\documentclass[
 aip,
 jmp,%
 amsmath,amssymb,
numerical,%
]{revtex4-1}

\pdfoutput=1

\usepackage{graphicx}
\usepackage{dcolumn}
\usepackage{bm}
\usepackage{subfigure,amsmath,subeqnarray,fancybox,multirow,hhline,amsmath}

\newcommand{\pdiff}[3][]{\dfrac{\partial^{#1} #2}{\partial {#3}^{#1}}}

\begin{document}

\preprint{AIP/123-QED}

\title{Coalescence of Liquid Drops: Different Models Versus Experiment}

\author{J. E. Sprittles}
 \email{sprittles@maths.ox.ac.uk}
\affiliation{Mathematical Institute, University of Oxford, Oxford, OX1 3LB, U.K.}

\author{Y. D. Shikhmurzaev}%
 \email{yulii@for.mat.bham.ac.uk}
 \affiliation{School of Mathematics, University of Birmingham, Edgbaston, Birmingham, B15 2TT, U.K. }

\date{\today}

\begin{abstract}

The process of coalescence of two identical liquid drops is
simulated numerically in the framework of two essentially
different mathematical models, and the results are compared with
experimental data on the very early stages of the coalescence
process reported recently. The first model tested is the
`conventional' one, where it is assumed that coalescence as the
formation of a single body of fluid occurs by an instant
appearance of a liquid bridge smoothly connecting the two drops,
and the subsequent process is the evolution of this single body of
fluid driven by capillary forces. The second model under
investigation considers coalescence as a process where a section
of the free surface becomes trapped between the bulk phases as the
drops are pressed against each other, and it is the gradual
disappearance of this `internal interface' that leads to the
formation of a single body of fluid and the conventional model
taking over.  Using the full numerical solution of the problem in
the framework of each of the two models, we show that the recently
reported electrical measurements probing the very early stages of
the process are better described by the interface formation/disappearance model.
New theory-guided experiments are suggested that would help to
further elucidate the details of the coalescence phenomenon. As a
by-product of our research, the range of validity of different
`scaling laws' advanced as approximate solutions to the problem
formulated using the conventional model is established.

\end{abstract}


\maketitle

\section{Introduction}

The phenomenon of coalescence where two liquid volumes, in most
cases drops, merge to form a single body of fluid exhibits a range
of surprisingly complex behaviour that is important to understand
from both the theoretical viewpoint as well as with regard to a
large number of applications. The dynamics of coalescing drops is
central for a whole host of processes such as viscous sintering\citep{bellehumeur04},
emulsion stability\citep{dreher99}, spray cooling\citep{grissom81}, cloud formation\citep{kovetz69}, and, in
particular, a number of emerging micro- and nanofluidic
technologies\citep{quake05}. The latter include, for example, the 3D-printing
devices developed for the rapid fabrication of custom-made products
ranging from hearing aids through to electronic circuitry
\citep{derby10,singh10}.  In this technology, structures are built by
microdrops ejected from a printer; these drops subsequently come into contact with a surface containing both dry solid substrate as
well as liquid drops deposited earlier, so that being able to
predict the behaviour of drops as they undergo stages of both
spreading over a solid and coalescence is critical to improving the
overall quality of the finished product.

The spatio-temporal scales characterizing the coalescence process are extremely small, so that
resolving the key (initial) stages of the process experimentally is
very difficult. This is particularly the case in microfluidics where the process of coalescence as such is inseparable from the overall dynamics. This difficulty, and the associated cost of performing high-accuracy experiments, becomes a strong motivation for
developing a reliable theoretical description of this class of flows
which would be capable of taking one down to the scales inaccessible
for experiments and allow one, in particular, to map the parameter space
of interest to determine, say, critical points at which the flow
regime bifurcates.

From a fundamental perspective, the phenomenon of coalescence is a
particular case from a class of flows where the flow domain
undergoes a topological transition in a finite time, so that
studying this phenomenon might help to elucidate common features and
develop methods of quantitative modelling applicable to other flows
in this class.  Technically, coalescence is the process by which two
liquid volumes that at some initial moment touch at a point or along a line, i.e.\ have a
common boundary point or points, become one body of fluid, where (a) there are
only `internal' (bulk) and `boundary' points and (b) every two
internal points can be connected by a curve passing only through
internal points. Once the coalescence as defined above has taken
place, the subsequent process is simply the evolution of a single body of liquid and it can be described
in the standard way.

\begin{figure}
     \centering
\includegraphics[scale=0.8]{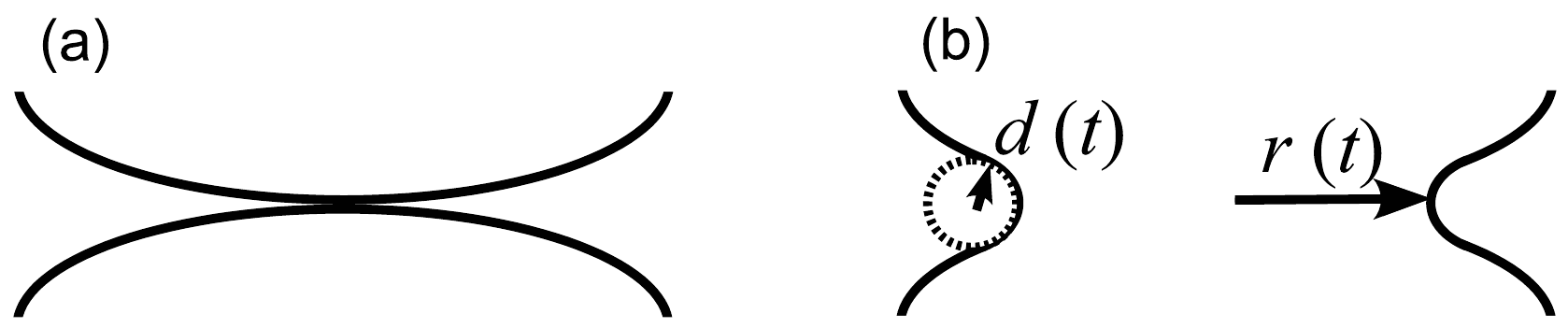}
 \caption{Sketch illustrating the scheme used in the conventional modelling of coalescence: the initial contact (a) is instantly followed by a finite-size `bridge' connecting the two fluid volumes (b), i.e.\ $r(0)=r_{min}>0$.  The subsequent evolution of the single body of fluid is driven by the capillary pressure, where the main contribution is due to the longitudinal curvature $1/d(t)$.}
 \label{F:bridge_evolution}
\end{figure}

In the conventional framework of fluid mechanics, the free surface
has to be smooth as otherwise, to compensate the action of the
surface tension on the singularities of the free-surface curvature,
one has to admit non-integrable singularities in the bulk-flow
parameters \citep{richardson68}. Therefore, when applied to the coalescence phenomenon, the conventional approach essentially by-passes the problem: it is assumed that
immediately after the two free surfaces
touch, there somehow appears a smooth liquid
bridge of a small but finite size connecting the two fluid volumes
(Figure~\ref{F:bridge_evolution}). In other words, the
coalescence, i.e.\ the formation of a single body of fluid, has already
taken place and the subsequent evolution of the free-surface shape
can be treated conventionally.  Hence, theoretical studies of
coalescence in the framework of conventional fluid mechanics
essentially boil down to a `backward analysis' of the process,
i.e.\ to considering what happens in the limit $t\to0+$ as the time
is rolled back to the initial singularity in order to uncover what
the early stages of the evolution of the free surface and the flow
parameters might be.

In the forthcoming subsections, we will describe how the development of new experimental techniques
and a new generation of experimental equipment, in particular the use
ultra high-speed optical cameras \citep{thoroddsen05} as well as
novel electrical methods \citep{paulsen11}, have made it possible to study
processes on the spatio-temporal scales that were previously unobtainable.  Therefore, this is a perfect opportunity for
a detailed comparison between theory and experiment in order to
probe the fundamental physics associated with this `singular' free-surface flow.

\section{Background}

\subsection{Plane 2D flows}

Much initial work on coalescence was motivated by Frenkel's 1945
paper on viscous sintering \citep{frenkel45} for inertialess viscous
flow with an inviscid dynamically passive gas in the exterior.
Later, consideration of the plane 2-dimensional flow of high viscosity liquids
led, in particular, in the works of
Hopper\citep{hopper84,hopper90,hopper93a,hopper93b} and
Richardson\citep{richardson92}, to an exact solution for coalescing
cylinders obtained using conformal mapping techniques. Notably,
as pointed out in \citep{eggers99}, in this solution, for small radii
of the liquid bridge one has that the radius of longitudinal
curvature $d(t) = O(r^3)$ as $r\to0+$
(Figure~\ref{F:bridge_evolution}), i.e.\ it is asymptotically even smaller than the
undisturbed distance between the free surfaces, which is of $O(r^2)$
as $r\to0+$. In other words, exact solutions obtained in the
framework of conventional fluid mechanics confirmed
that this formulation predicts that the free-surface curvature is
singular as $t\to0+$ and hence the conventional model is used beyond its limits of applicability through the initial
stage of the process. Correspondingly, the flow velocity in the exact solution is also
singular as $t\to0+$ and unphysically high for small $t$ and $r$.

\subsection{Scaling laws for axisymmetric flows}\label{scalings}

More recent works have been mainly concerned with deriving various
`scaling laws' for the radius of the liquid bridge $r(t)$, joining
two drops of initial radius $R$, as a function of time $t$
(Figure~\ref{F:bridge_evolution}). These scaling laws are obtained
by balancing the factors driving and resisting the fluid motion,
with the appropriate assumptions about how these factors can be
expressed quantitatively.

From a theoretical viewpoint, consideration of scaling laws is analogous to the approach of
Frenkel, as opposed to the rigorous fluid mechanical treatment of coalescence
initiated by Hopper and Richardson, since, as in Frenkel's
paper\citep{frenkel45}, to obtain the scaling laws, the solution of
the equations of fluid mechanics is found using some plausible
assumptions rather than being obtained directly. On the other hand, however, the
simple results obtained using the scaling laws approach, once tested
experimentally, can give an indication as to whether the rigorous
analysis of a given problem formulation is worth pursuing.

Analytic progress has been achieved by assuming that the process is
driven by surface tension $\sigma$ and opposed either by viscous or
inertial forces \citep{eggers99}. The driving force due to the surface tension is calculated by assuming that the mean curvature $\kappa$ of the
free surface is due primarily to the longitudinal curvature $1/d(t)$
(Figure~\ref{F:bridge_evolution}): $\kappa \propto 1/d(t)$. In the inertia-dominated case, it is assumed in \citep{eggers99} that $d(t)$ is determined by the initial free surface
shape, which for coalescing spheres gives $d\propto r^2/R$. As mentioned above, in the viscosity-dominated regime it is shown in \citep{eggers99} that when the surrounding gas is inviscid, one has $d(t)\propto r^3/R$.  In either case, one can calculate the surface tension force $\sigma\kappa(t)$ as a function
of time. In the situation where viscous forces dominate inertial
ones and hence are the main factor resisting the flow, a scale for
velocity is $U_{visc} = \sigma/\mu$ (so that the capillary number
$Ca=\mu U_{visc}/\sigma = 1$) and the corresponding time scale is
$T_{visc}=R\mu/\sigma$. Alternatively, if it is the inertial forces
that are the main factor resisting the motion, a scale for velocity
is $U_{inert} = (\sigma/\rho R)^{1/2}$ (so that the Weber number $We
= \rho U_{inert}^2 R/\sigma=1$) and the corresponding time scale is
$T_{inert}=(\rho R^3/\sigma)^{1/2}$. In the viscous case, the simplest
scaling is that the bridge radius evolves as $r/R \propto
t/T_{visc}$; however, in \citep{eggers99} it is shown that there is a
logarithmic correction to this term so that
\begin{equation}\label{viscous_scaling}
r/R = -C_{visc} \left(t/T_{visc}\right) \ln\left(t/T_{visc}\right),
\end{equation}
where $C_{visc}$ is a constant.  The limits of applicability of this scaling, based on the equivalence of the
two- and three-dimensional problems, is expected to hold\citep{eggers99} for $r/R<0.03$.

In \citep{eggers99}, it is suggested that when the Reynolds number,
based on $U_{visc}$ as the scale for velocity and the radius of the
bridge $r$ as the length scale, becomes of order one, $Re_r = \rho
\sigma r/\mu^2 \approx 1$, there will be a crossover point where the
dynamics switches from Stokesian to Eulerian, i.e.\ the main factor
resisting the motion is now inertia of the fluid. This crossover
point correspond to $r\approx\mu^2/(\rho\sigma)$ after which the
balance of the surface tension and inertia forces gives
\begin{equation}\label{inertial_scaling}
r/R = C_{inert}\left(t/T_{inert}\right)^{1/2},
\end{equation}
where $C_{inert}$ is a constant.  Notably, for water the crossover
from viscous to inertial scaling is predicted to occur at $r=14$~nm.

\subsection{Numerical simulations}

The use of computational simulation for what is, strictly speaking,
the evolution of the post-coalescence single body of fluid has
focussed both on the very early stages of the process as well as on
the global dynamics of the two drops \citep{duchemin03,eggers99,menchacarocha01,paulsen12}. In the early
stages, computations of the inviscid flow, using boundary integral
methods, have shown the formation of toroidal bubbles trapped inside
the drops as the two free surfaces reconnect themselves in front of
the bridge \citep{oguz89,duchemin03}. The appearance of these bubbles,
originally suggested in \citep{oguz89}, has been further investigated
in \citep{duchemin03} using inviscid boundary integral calculations,
and an attempt has been made to continue the simulation past the
toroidal bubble formation. It was shown that, despite the bubble
generation, the scaling (\ref{inertial_scaling}) still holds, with
the prefactor determined to be $C_{inert}=1.62$ for the period in
which bubble formation occurs ($r/R<0.035$).  However, as the
authors acknowledge, the computational approach for dealing with the
reconnection procedure is not entirely satisfactory, with the robust
simulation of such phenomena remaining an open problem.

Simulations of the entire post-coalescence process have been
performed to varying degrees of accuracy, dependent in many cases on
the computational power available at the time, in
\citep{jagota88,martinez95,menchacarocha01}.  A recurring question in
these studies was how to initialize the simulation.  For example, in
\citep{menchacarocha01}, it is assumed that the singular curvature at
the moment of touching of the two fluid volumes is immediately
smoothed out over a grid-size dependent region, so that as the grid
is refined, the radius of curvature decreases, i.e.\ the curvature
tends to the required singular initial condition. This behaviour is
reflected in Figure~9 of \citep{menchacarocha01}, showing the bridge
radius versus time, where changing the grid resolution changes the
results considerably, i.e., as expected, the solution is
mesh-dependent.  A similar approach is used in \citep{paulsen12};
however, there, the results from the simulation are only plotted
when the ``transients from the initial conditions have decayed''
(see Figure~3 of this paper), so that it is difficult to observe the
influence of the initial conditions.  Due to an inability to resolve
multiscale phenomena computationally, until now, no studies have
considered in detail both the very initial stages of coalescence
alongside the global dynamics of the drops.

\subsection{Experimental data}

\begin{figure}
     \centering
\includegraphics[scale=0.8]{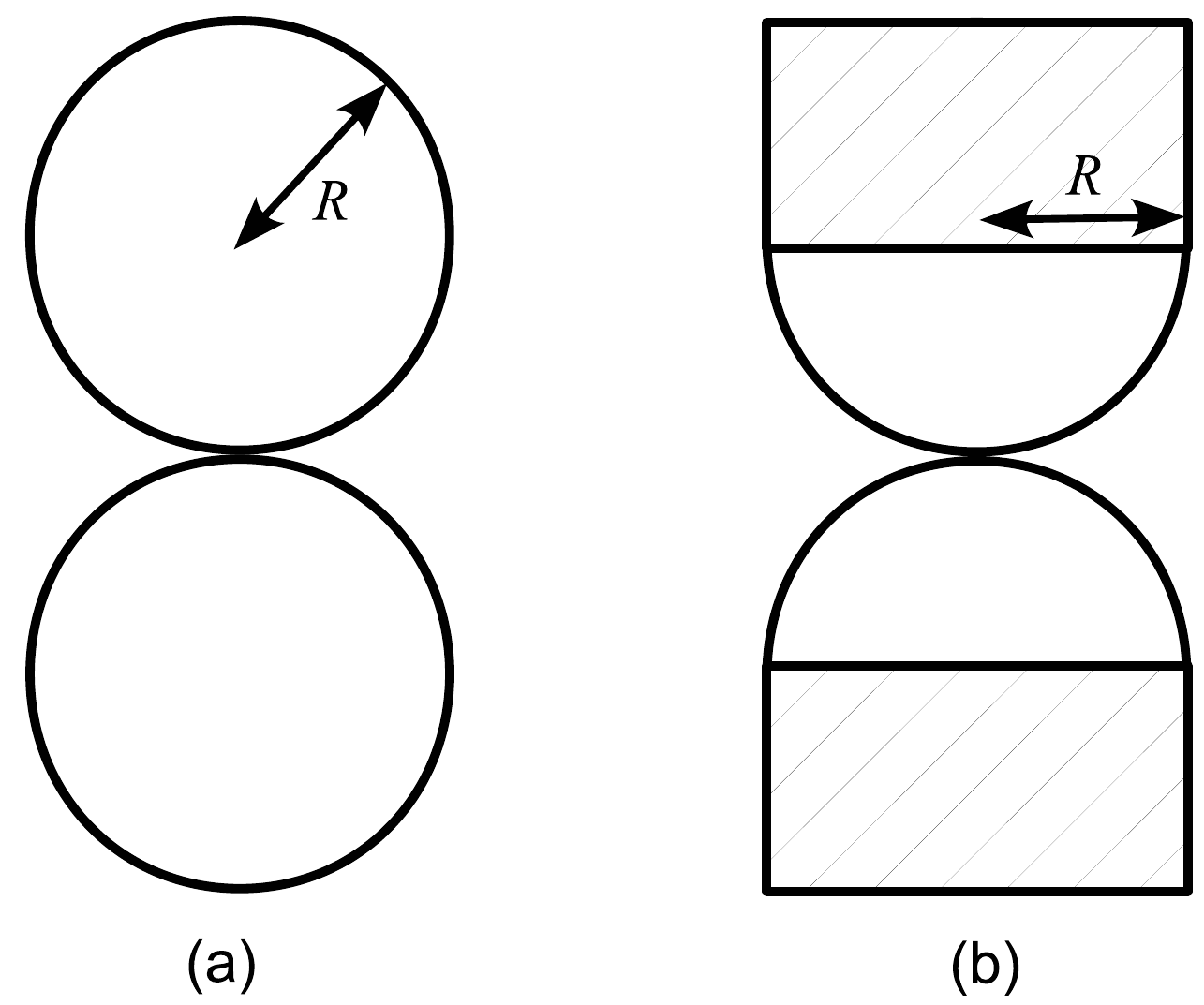}
\caption{\label{F:drop_geometries} (a): Sketch of the benchmark problem of the coalescence of two free identical liquid spheres. (b): typical experimental setup where drops are grown from capillaries until they begin to coalesce.}
\end{figure}

Several experimental studies have probed the dynamics of
coalescing drops.  The study of coalescing free liquid drops (Figure~\ref{F:drop_geometries}a) is rather complicated, as it is
difficult to control and monitor the movement of the drops with the
required precision. Therefore, since coalescence as such is a local
process, a common experimental setup is based on using supported
hemispherical drops, with one drop sitting on a substrate,
or being grown from a capillary tube, and the other, a pendent drop,
being grown from a capillary above
(Figure~\ref{F:drop_geometries}b). As coalescence is initiated, the
bridge radius is then measured as a function of time either
optically or using some indirect methods. To date the most
exhaustive study of coalescence, using the aforementioned
experimental setup, has been carried out by Thoroddsen and
co-workers\citep{thoroddsen05}, who investigated a range of
viscosities and drop sizes, with the shapes of the drop monitored
using ultra high-speed cameras capable of capturing up to one million
frames per second. Similar experiments have been reported in
\citep{aarts05} and \citep{wu04} with the same setup.

As shown in \citep{thoroddsen05}, after correcting the initial shapes
of the drops to account for the influence of gravity, the inviscid
scaling law (\ref{inertial_scaling}) appears to be in good agreement
with experimental results for the initial stages of the
coalescence of drops of low viscosity ($\mu<10$~mPa~s) fluid.  It is
found in \citep{aarts05,wu04} that the prefactor $C_{inert}=1.62$ predicted
in \citep{duchemin03} is considerably higher than all the values
obtained experimentally, which for hemispherical drops are seen to
be around $C_{inert}=0.8$. Also, no toroidal bubbles have been
observed\footnote{It should be pointed out here that the
measurements are taken for relatively large bridge radii, so that a
comparison with the inertial scaling (\ref{inertial_scaling}) is
valid, but the theory of \citep{duchemin03} is well past its limits
of applicability, so that one cannot expect good agreement for the
proposed prefactor.}. At intermediate viscosities
($40$~mPa~s$~<\mu<220$~mPa~s), it is found in \citep{thoroddsen05}
that neither the inertial nor viscous scalings are able to fit the
data whilst at the highest viscosity ($\mu=493$~mPa~s) a region of
linear growth of the bridge radius with time is observed. In both
\citep{thoroddsen05} and \citep{aarts05}, linear growth in the initial
stages shows no signs of the logarithmic correction as in
equation (\ref{viscous_scaling}).  Instead, the scaling $r/R = B
t/T_{visc}$ is shown to fit the data best, where $B$ is the coefficient of proportionality. Notably, the value of the constant $B$ is
seen to be a factor of two smaller in \citep{aarts05} than in
\citep{thoroddsen05}.

Recently, a new experimental technique has been developed to study
the coalescence phenomenon at spatio-temporal scales inaccessible to
optical measurements.  In \citep{case08,case09,paulsen11}, an
electrical method, extending the techniques utilized in
\citep{burton04} to study drop pinch-off dynamics, has been used to
measure the radius of the bridge connecting two coalescing drops of an electrically conducting
liquid down to time scales of $\sim10$~ns, giving at least two
orders of magnitude better resolution than optical techniques. In
doing so, it is shown that the initial radius of contact is very
small, as suggested in \citep{aarts05}, so that there is no evidence
for the initial area of $\sim100$~$\mu$m suggested in
\citep{thoroddsen05}. It is noted that the method loses accuracy
towards the end of the process ($t>400$~$\mu$s for water), but that
in this range optical experiments are available and reliable, so
that by using both electrical measurements alongside optical ones,
it is possible to obtain accurate measurements over the entire range
of bridge radii (see Figure~\ref{F:230cP_cl} below where we do
precisely this).

In \citep{case08,case09}, it is found that for low viscosity fluids a
new regime exists for $t<10$~$\mu$s which is inconsistent with the
assumption that the inertial scaling, equation
(\ref{inertial_scaling}), will kick-in almost instantaneously for
such liquids. In \citep{paulsen11}, the same electrical method is
used to measure the influence of viscosity on the coalescence
dynamics, with other parameters (surface tension, density, drop
size) almost constant, and similar behaviour is observed over two
orders of magnitude variation in the viscosity of water-glycerol
mixtures, with, as before, the cross-over time between different
flow regimes being vastly different from what the combination of scaling
laws predicts. It is suggested in \citep{paulsen11} that this is because the cross-over from regimes is based on the
Reynolds number whose length scale is taken to be the bridge
radius, whereas, in fact it should be based on the undisturbed free surface height at a given radius, which is proportional to $r^2$ as opposed to $r$,
giving a much later cross-over time, as observed experimentally.


Thus, although in experiments one can observe some of the general
trends following from the scalings (\ref{viscous_scaling}) and
(\ref{inertial_scaling}), experimental studies have been unable to
validate these scaling laws. At low viscosities, the prefactor
obtained in \citep{duchemin03} for small bridge radii has not been
confirmed; at intermediate viscosities, neither inertial nor viscous
forces can be neglected so that both scaling laws become
inapplicable; whilst at the highest viscosities, logarithmic
corrections have not been observed and different experiments give
different values of the prefactor to a linear power law, which has
not been predicted by theory.  It should also be pointed out that
when using power laws, there is no guarantee that the prefactor
which fits the experimental data is necessarily the one that would
be obtained from solving the full problem formulation.  Thus, it is
clear that full-scale computational simulation of this class of
flows is called for.  Such a simulation will allow one to accurately
compare theoretical predictions with experimental data and hence,
first of all, show whether or not the model itself accounts for all the key
physics involved in the coalescence process. As a by-product, the
simulation will be able to test the validity of the scalings
(\ref{viscous_scaling}) and (\ref{inertial_scaling}) by comparing
them to the exact solution.

\subsection{Coalescence as an interface formation/disappearance process}

In order to study coalescence over a range of viscosities for a
sustained period of time and to test the
mathematical model of the phenomenon, as opposed to different
approximations, against experiments we need to use computational
methods that are capable of solving the full Navier-Stokes equations
with the required accuracy. This would allow one not only to account
in full for the effects of inertia, capillarity and viscosity and
hence make the comparison of the conventional model with experiments
conclusive; it will also make it possible to incorporate and test
against experiments the `extra' physics that carries the system
through the topological transition, which is, technically, what
coalescence actually is and what is not considered in the conventional model.

\begin{figure}
     \centering
\includegraphics[scale=0.8]{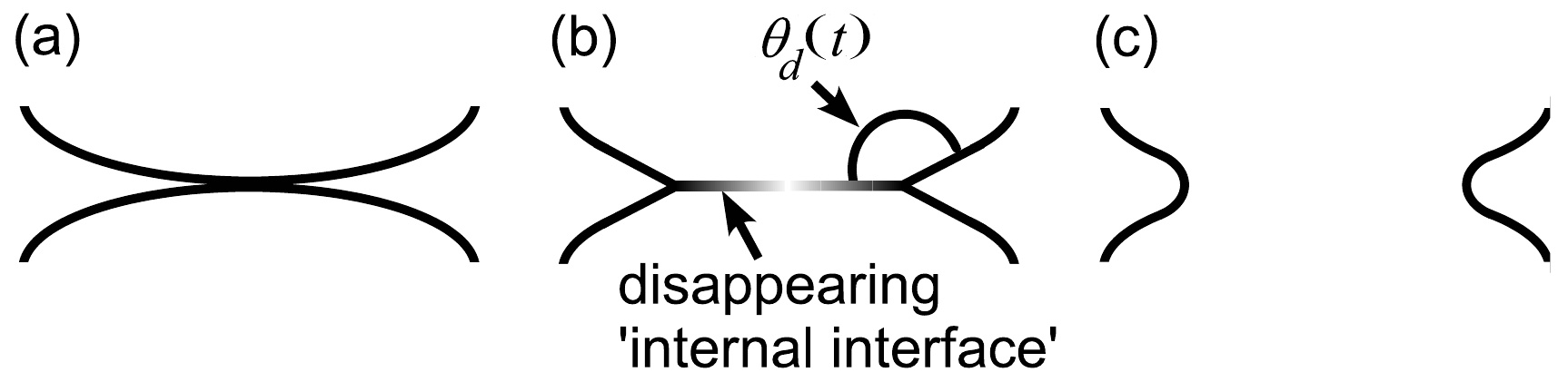}
\caption{\label{F:ifm} Sketch illustrating the scheme used in the interface formation/disappeance theory: the initial contact point (a) is followed by a fraction of the free surface being `trapped' between the bulk phases, forming a gradually disappearing `internal interface' (b), and, as the `internal interface' disappears and the `contact angle' $\theta_d$, being initially equal to $180^\circ$, relaxes to its `equilibrium' value of $90^\circ$, the conventional mechanism takes over (c).  The interface formation/disappearance model provides boundary conditions on interfaces, which are modelled as zero-thickness `surface phases'; these interfaces, including the `internal interface' in (b), are shown as finite-width layers for graphical purposes only.}
\end{figure}

As pointed out in the Introduction, the conventional fluid mechanics
model essentially deals with the post-coalescence process, i.e.\ the
evolution of a single body of fluid that the coalescence phenomenon
has produced, and, as the limit $t\to0+$ is taken, gives rise to
unphysical singularities. This suggests that some `additional'
physics, not accounted for in the conventional model, takes the
system through the topological change, and the conventional physics
takes over when the liquid bridge between the two drops already has
a finite size determined by this `additional' physics. The first
study identifying this `additional' physics, which was aimed at embedding
coalescence into the general physical framework as a particular case
of a more general physical phenomenon, has been reported in
\citep{shik00}. It has been shown that coalescence is in fact a
particular case of the interface formation/disappearance process: as the two
drops are pressed against each other, a section of their free
surfaces becomes trapped between the bulk phases
(Figure~\ref{F:ifm}). As this trapped interface gradually (albeit, in physical
terms, very quickly) loses its surface properties (such as the
surface tension), the angle $\theta_d$ (Figure~\ref{F:ifm}) formed
by each of the free surfaces of the drops with the `internal
interface' sandwiched between the two drops goes to $90^\circ$, so
that eventually a bridge of a finite physically-determined radius
emerges and the conventional model takes over. The outlined physics
allows for the existence of a non-smooth free surface without
unphysical singularities in the flow field since the surface
tensions acting on the line where the free-surface curvature is
singular are balanced not by the bulk stress, as in \citep{richardson68}, but by the (residual) surface tension in the
`internal' interface. The existence of such non-smooth free surfaces has
been confirmed experimentally \citep{joseph91} and has already been described theoretically using the above
approach \citep{shik00,shik05a}.

The approach outlined above removes the unphysical singularities in
the mathematical description of the coalescence process and allows
one to treat it in a regular way, as just one of many fluid
mechanics phenomena. The developed model (which came to be known as
`interface formation model' or, for brevity, IFM) unifies the
mathematical modelling of such seemingly different phenomena as
coalescence \citep{shik00}, breakup of liquid threads \citep{shik00,shik05b} and
free films \citep{shik05c}, as well as dynamic wetting \citep{shik93,shik97,shik06,shik97a}; an
exposition of the fundamentals of the theory of capillary flows with
forming/disappearing interfaces can be found in \citep{shik07}.

Applying the interface formation model to coalescence phenomena
results in a new perspective on the problem.  Instead of thinking of
coalescence as the process by which \emph{one} deformed body
evolves, which is how equations (\ref{viscous_scaling}) and
(\ref{inertial_scaling}) were derived, it is thought of as the
process by which \emph{two} drops evolve into a single entity.
Specifically, just after the drops first meet, an internal interface
divides them, which allows an angle to be sustained in the free
surface, and the coalescence process is thought of as the time it
takes for this internal divide between the two drops to disappear,
and hence for the free surface to become smooth.  A characteristic
time of this process is the surface tension relaxation time, and,
given that this parameter's value is expected to be proportional to viscosity, it
is likely that, for high viscosity fluids, such as the $58000$~mPa~s silicon oil used in
\citep{paulsen12}, direct experimental evidence for this model, in particular
the angle in the free surface at a finite time after the drops first
come into contact, may be possible to observe in the optical range.

The local asymptotic analysis carried out in \citep{shik00,shik07} has shown that the singularities inherent in the
conventional treatment of the early stage of coalescence are
removed, but, to validate the theory experimentally, a global solution
must be found. The obstacle here is that the interface formation
model introduces a new class of problems where boundary conditions
for the Navier-Stokes equations are themselves differential
equations along {\it a priori\/} unknown interfaces, and this class
of problems poses formidable difficulties even for numerical
treatment. The decisive breakthrough in this direction has been made
recently as a regular framework for computing this kind of problems
has been developed\citep{sprittles_jcp}.  This advance together with
the development of the aforementioned novel experimental techniques,
which can probe the coalescence process on the spatio-temporal
scales well beyond the reach of previous studies, make a full
comparison between theory and experiment possible for the first
time.

\subsection{Outline of the paper}

The aim of this paper is to address whether the conventional model
or the interface formation model are able to describe experimental
results which give the bridge radius as a function of time over a
range of viscosities.  To do so, in Section~\ref{model} we present
the problem formulations for both models and, notably, list the equations of
interface formation, with a very brief description and references for detail.  In
Section~\ref{numerics}, the computational tool, which was originally
devised to describe dynamic wetting flows, is briefly described and
references are given to the publications where detailed benchmarking
and mesh-independence tests have been reported. In
Section~\ref{previous}, simulations from this code, for both low and
high viscosity liquids, obtained using the conventional model, are shown to be in agreement with previous
benchmark computational results. Besides validating the code, this allows us to consider the accuracy of the
scaling laws proposed in various limits. Then, in
Section~\ref{experiments}, the predictions of both the conventional model and the interface
formation model are compared to
experiments conducted in both \citep{thoroddsen05} and \citep{paulsen11}.  This allows us to assess which of these models
describes the underlying physics of the coalescence phenomenon.
Next, in Section~\ref{viscosity}, a comparison is made to
experiments in \citep{paulsen11} over a range of viscosities, in
order to ascertain how well the models are able to capture the
observed drop behaviour.  In subsections A and B of Section~\ref{theory_driven}, we propose
a theory-guided test case which could potentially bring the differences between
the two models' predictions into the optical range.  Concluding
remarks in Section~\ref{conclusion} summarize the main results and point out some open issues for future research.


\section{Modelling of Coalescence Phenomena}\label{model}

Consider the axisymmetric coalescence of two drops whose motion
takes place in the $(r,z)$-plane of a cylindrical coordinate
system.  The liquid is incompressible and Newtonian with constant
density $\rho$ and viscosity $\mu$, and the drops are surrounded
by an inviscid dynamically passive gas of a constant pressure $p_g$. To
non-dimensionalize the system of equations for the bulk variables,
we use the drop radius $R$ as the characteristic length scale;
$U_{visc}=\sigma/\mu$ as the scale for velocities (so that $Ca=\mu
U_{visc}/\sigma = 1$), where $\sigma$ is the equilibrium surface
tension of the free surface; $T_{visc}=R/U_{visc} = \mu R/\sigma$
as the time scale; and $\sigma/R$ as the scale for pressure. Then,
the continuity and momentum balance equations take the form
\begin{equation}\label{ns}
\nabla\cdot\mathbf{u} = 0,\qquad Re~\left[\pdiff{\mathbf{u}}{t} + \mathbf{u}\cdot\nabla\mathbf{u}\right] = -\nabla p + \nabla^2\mathbf{u} + Bo~\mathbf{g},
\end{equation}
where $t$ is time, $\mathbf{u}$ and $p$ are the liquid's velocity
and pressure, and $\mathbf{g}$ is the gravitational force density, which in the nondimensional formulation is a unit vector in the negative $z$-direction.
The non-dimensional parameters are the Reynolds number $Re=\rho
\sigma R/\mu^2$ and the Bond number $Bo=\rho g R^2/\sigma$.  To
simplify the computations, we shall assume that gravitational forces
are negligible $Bo=0$, so that, for two identical drops of radius
$R$, the process can be regarded as symmetric with respect to the
plane touching the two drops at the moment of their initial contact,
and we can consider the flow in one drop, using the symmetry
conditions at the symmetry plane $z=0$. The point in the
$(r,z)$-plane at which the free surface meets the plane of symmetry
will be referred to as the `contact line', since, as we will show
below, there is a certain analogy between the process of dynamic
wetting and the coalescence phenomenon, where, in the present case,
the drop (for definiteness, the one above $z=0$) `spreads' over the
plane of symmetry (see Figure~\ref{F:ifm}). For the same reason, the
angle $\theta_d$ between the free surface and the symmetry plane
$z=0$ will be referred to as the `contact angle', so that, in the
analogy with dynamic wetting, the `equilibrium' contact angle is
$90^\circ$.

The effect of neglecting gravity is estimated in the Appendix, where
we show that, as one would expect, gravity influences only the late
stages of the drops' evolution, i.e.\ the global geometry of the
flow, where it is important whether the drops are spherical or
hemispherical. In the present paper, we are interested primarily in the local
process where coalescence as such takes place, and this process can
be studied without taking gravity into account.

The boundary conditions to equations (\ref{ns}) will be given by
two different models.  First, we give the conventional model
formulation routinely used for studying free-surface flows, and
then we will present the interface formation model, which, until
now, has not been used in full to describe this class of flows.

\subsection{Conventional modelling}

The standard boundary conditions used in fluid dynamics of
free-surface flows are the kinematic condition, stating that the
fluid particles forming the free surface stay on the free surface at
all time, and the conditions of balance of tangential and normal forces acting on
an element of the free surface from the two bulk phases and from the
neighbouring surface elements:
\begin{equation}\label{ckin}
\pdiff{f}{t} + \mathbf{u}\cdot\nabla f = 0,
\end{equation}
\begin{equation}\label{cstress}
 \mathbf{n}\cdot\left[\nabla\mathbf{u}+\left(\nabla\mathbf{u}\right)^T\right]
 \cdot\left(\mathbf{I}-\mathbf{n}\mathbf{n}\right) =\mathbf{0},
\end{equation}
\begin{equation}
 \label{cnormstress}
 p_g-p+\mathbf{n}\cdot\left[\nabla\mathbf{u}+
 \left(\nabla\mathbf{u}\right)^T\right]\cdot
 \mathbf{n}=\nabla\cdot\mathbf{n}.
\end{equation}
Here $f(r,z,t)=0$ describes the \emph{a priori} unknown free surface, with
the inward normal $\mathbf{n} = \nabla f/|\nabla f|$; $\mathbf{I}$
is the metric tensor of the coordinate system, so that the
convolution of a vector with the tensor
$(\mathbf{I}-\mathbf{n}\mathbf{n})$ extracts the component of this
vector parallel to the surface with the normal $\mathbf{n}$ (in what
follows, for brevity, we will mark these components with a subscript
$\parallel$, so that $\mathbf{u}_{||} = \mathbf{u}\cdot(\mathbf{I}-\mathbf{n}\mathbf{n})$).

At the plane of symmetry $z=0$, one has the standard symmetry
conditions of impermeability and zero tangential stress,
\begin{equation}\label{cnormal}
\mathbf{u}\cdot\mathbf{n}_s = 0,~~~\mathbf{n}_s\cdot\left[\nabla\mathbf{u}+\left(\nabla\mathbf{u}\right)^T\right]\cdot
\left(\mathbf{I}-\mathbf{n}_s\mathbf{n}_s\right) =\mathbf{0},
\end{equation}
where $\mathbf{n}_s$ is the unit normal to the plane of symmetry.  One also has the condition that the free surface is smooth, i.e.\
$\theta_d\equiv\pi/2$, or, in terms of the normals $\mathbf{n}$ and
$\mathbf{n}_s$ to the free surface and the plane of symmetry,
respectively, $\mathbf{n}\cdot\mathbf{n}_s=0$.

We will consider an axisymmetric flow, and on the axis of symmetry
the standard impermeability and zero tangential stress condition apply:
\begin{equation}\label{axis_bulk}
 \mathbf{u}\cdot\mathbf{n}_a=0,
 \qquad
 \mathbf{n}_a\cdot\left[\nabla\mathbf{u}
 +\left(\nabla\mathbf{u}\right)^T\right]\cdot
 (\mathbf{I}-\mathbf{n}_a\mathbf{n}_a)=\mathbf{0},
 \end{equation}
where $\mathbf{n}_a$ is the unit normal to the axis of symmetry in
the $(r,z)$-plane.

With regard to the overall drop geometry, there
are two cases (Figure~\ref{F:drop_geometries}). In the case of the
coalescence of free spherical drop, one needs the symmetry condition
on the free-surface shape, namely that the free surface is smooth at
the axis of symmetry,
\begin{equation}\label{axis_shape}
 \mathbf{n}\cdot\mathbf{n}_a=0,\quad\hbox{for}\quad f(0,z,t)=0,~~t\ge0.
\end{equation}
In the case of hemispherical drops pinned to the solid support, we
need the condition that the coordinates of the free surface are
prescribed where the free surface meets the solid:
\begin{equation}\label{pinned_shape}
 f(1,1,t)=0 \qquad (t\ge0).
\end{equation}

To complete the formulation, one needs the initial conditions, which
we will discuss and specify below.

\subsection{The interface formation/disappearance model}

The interface formation/disappearance model formulates the boundary
conditions that generalize (\ref{ckin})--(\ref{cnormal}) to account
for situations in which the interfaces are forming or disappearing.
In these cases, the interfaces have dynamic interfacial properties,
and, in particular, the surface tension is no longer a constant; it
varies as the interface is forming/disappearing, and this creates
spatial gradients of the surface tension which give rise to the
Marangoni flow in the bulk. The equations of the interface formation
model consider interfaces as two-dimensional `surface phases'
characterized, besides the surface tension, by the
surface density $\rho^s$ and the surface velocity $\mathbf{v}^s$
with which the surface density is transported. The normal to the
interface component of $\mathbf{v}^s$ can differ from the normal
component of the bulk velocity $\mathbf{u}$ evaluated at the
interface as there can be mass exchange between the surface and bulk
phases.

The details of the interface formation model can be found
elsewhere\citep{shik07}, so that here we will give the necessary
equations in the dimensionless form, using as characteristic scales
for $\rho^s$, $\mathbf{v}^s$ and $\sigma$ the surface density
corresponding to zero surface tension $\rho^s_{(0)}$, the same
velocity scale as used in the bulk $\sigma/\mu$, and the equilibrium
surface tension of the liquid-gas interface $\sigma = \sigma_{1e}$,
respectively. In what follows, subscripts 1 and 2 will refer,
respectively, to the surface variables on the free surface and on
the plane of symmetry $z=0$, which will be regarded as a gradually
disappearing `internal interface' trapped between the two coalescing
drops as they are pressed against each other.  Notably, the plane of symmetry $z=0$ actually cuts
the internal interface into two symmetric halves and we consider the upper half of this interface which, for brevity, is referred to as the `internal interface'.

On the liquid-gas free surface, we have
\begin{equation}\label{kin}
\pdiff{f}{t} + \mathbf{v}^s_1\cdot\nabla f = 0,
\end{equation}
\begin{equation}\label{stress-norm}
 p_g-p+
 \mathbf{n}\cdot\left[\nabla\mathbf{u}+\left(\nabla\mathbf{u}\right)^T\right]\cdot
\mathbf{n}=\sigma_1\nabla\cdot\mathbf{n},
\end{equation}
\begin{equation}\label{stress-tang}
\mathbf{n}\cdot\left[\nabla\mathbf{u}+\left(\nabla\mathbf{u}\right)^T\right]\cdot
\left(\mathbf{I}-\mathbf{n}\mathbf{n}\right) +\nabla\sigma_1 =\mathbf{0},
\end{equation}
\begin{equation}\label{mass1b}
 \left(\mathbf{u}- \mathbf{v}^s_{1}\right)\cdot\mathbf{n}
 = Q\left(\rho^{s}_1-\rho^{s}_{1e}\right),
\end{equation}
\begin{equation}\label{mass1i}
 \epsilon\left[\pdiff{\rho^{s}_{1}}{t} + \nabla\cdot\left(\rho^{s}_{1}\mathbf{v}^{s}_{1}\right)\right] = - \left(\rho^{s}_{1}-\rho^{s}_{1e}\right),
\end{equation}
\begin{equation}\label{others1a}
4\bar{\beta}\left(\mathbf{v}^{s}_{1||}-\mathbf{u}_{||}\right)
=\left(1+4A\right)\nabla\sigma_1,
\end{equation}
\begin{equation}\label{others1b}
 \sigma_{1}=\lambda(1-\rho^{s}_{1}),
\end{equation}
where the following nondimensional parameters have been introduced:
$Q=\rho^s_{(0)}/(\rho\sigma\tau_{\mu})$, $\epsilon =
\sigma\tau_{\mu}/R$, $\bar{\beta} = \beta R/\mu$, $A = \alpha\beta$,
$\rho^s_{1e} = (\rho^s_{1e})_{dim}/\rho^s_{(0)}$, $\lambda = \gamma
\rho^s_{(0)}/\sigma_{1e}$. Here, we have used the experimentally
ascertained result \citep{blake02} that, for a class of fluids
commonly used in experiments, the characteristic relaxation time of
the interface $\tau$ is linearly proportional to the liquid's
viscosity, with coefficient of proportionality $\tau_{\mu}$, so that
$\tau = \tau_{\mu}\mu$.

Our assumption of symmetry between the two coalescing drops means
that the position of the `trapped' `internal interface' is known
\emph{a priori}, so that the normal stress condition, which in the general
case is used to find the interface's shape, is not required, and we
have the following equations:
\begin{equation}\label{kin2}
\mathbf{v}^s_2\cdot\mathbf{n} = 0,
\end{equation}
\begin{equation}\label{stress2}
\mathbf{n}\cdot\left[\nabla\mathbf{u}+\left(\nabla\mathbf{u}\right)^T\right]\cdot
\left(\mathbf{I}-\mathbf{n}\mathbf{n}\right) +\nabla\sigma_2 =\mathbf{0}
\end{equation}
\begin{equation}\label{mass2}
\left(\mathbf{u}- \mathbf{v}^s_{2}\right)\cdot\mathbf{n} = Q\left(\rho^{s}_2-1\right),\qquad \epsilon\left[\pdiff{\rho^{s}_{2}}{t} + \nabla\cdot\left(\rho^{s}_{2}\mathbf{v}^{s}_{2}\right)\right] = - \left(\rho^{s}_{2}-1\right),
\end{equation}
\begin{equation}\label{others2}
4\bar{\beta}\left(\mathbf{v}^{s}_{2||}-\mathbf{u}_{||}\right)=\left(1+4A\right)\nabla\sigma_2,
\qquad\sigma_{2}=\lambda(1-\rho^{s}_{2}).
\end{equation}
As one can see, these equations are the same as
(\ref{kin})--(\ref{others1b}) with $\rho^s_{1e}=1$.  This means that in
equilibrium the `internal interface' vanishes, no longer having the
surface tension and mass exchange with the `bulk', which are the
only factors that distinguish it as a special `surface phase'.

Although the boundary conditions of the interface formation model
have been explained in detail elsewhere \citep{shik07}, it seems
reasonable to briefly recapitulate their physical meaning. On the
free surface, besides the standard kinematic condition (\ref{kin})
and also standard conditions on the normal and tangential stress
(\ref{stress-norm}) and (\ref{stress-tang}), where the latter
includes the Marangoni effect due to the (potentially) spatially
nonuniform surface tension, one has the conditions describing the
mass exchange between the interface and the bulk (\ref{mass1b}),
(\ref{mass1i}), the equation describing how the difference between
the tangential components of the surface velocity and the bulk
velocity evaluated at the interface is related to the surface
tension gradient (\ref{others1a}), and the surface equation of state
(\ref{others1b}). The conditions on the internal interface are a
simplification of the conditions on the free surface due to the fact
that the shape of this interface is known ($z=0$), so that the normal-stress boundary condition, which applied to the
entire internal interface, i.e. the upper and lower halves put together,
is automatically satisfied, due to the symmetry of the problem with respect
to the $z=0$ plane, and is hence not needed, and the kinematic boundary
condition simplifies to (\ref{kin2}). In the case of a problem not
symmetric with respect to the plane $z=0$ both of these conditions
should be used in their full form.

Estimates for the phenomenological material constants $\alpha$,
$\beta$, $\gamma$, $\rho^s_{(0)}$ and $\tau$ have been
obtained by comparing the theory to experiments in dynamic wetting,
e.g.\ in \citep{blake02}, but could equally well have been taken from
any other process involving the formation or disappearance of
interfaces.

Boundary conditions (\ref{kin})--(\ref{others2}) are themselves
differential equations along the interfaces and therefore are in need of
boundary conditions at the boundaries of the interfaces, i.e.\ at
the contact line where the free surface meets the internal
interface, at the axis of symmetry (if free drops are considered) or
the solid boundary (in the case of pinned drops). At the contact
line, one has the continuity of surface mass flux and balance of
horizontal projection of forces due to surface tensions acting on
the contact line:
\begin{equation*}
 \rho^s_1\left(\mathbf{v}^s_{1||}
 -\mathbf{U}_{c}\right)\cdot\mathbf{m}_1
 +\rho^s_2\left(\mathbf{v}^s_{2||}
 -\mathbf{U}_{c}\right)\cdot\mathbf{m}_2=0,
\end{equation*}
\begin{equation}\label{cl}
 \sigma_{2}+\sigma_{1}\cos\theta_d=0.
\end{equation}
Here $\mathbf{m}_{i}$ are the unit vectors normal to the contact
line and inwardly tangential to the free surface ($i=1$) and the plane of symmetry ($i=2$); $\mathbf{U}_c$
is the velocity of the contact line (which is, obviously, directed
horizontally). Equation (\ref{cl}) is the well-known Young's
equation\citep{young05} that introduces and determines the contact
angle in the processes of dynamic wetting.  The present model
essentially considers coalescence as the process where the two drops
`spread' over their common boundary which gradually loses its
`surface' properties, and the contact angle tends to its
`equilibrium value' of $90^\circ$, where one will have the familiar
smooth free surface, whose evolution can be described by the
conventional model.

For the bulk velocity $\mathbf{u}$ one again has (\ref{axis_bulk})
on the axis of symmetry and conditions (\ref{axis_shape}) or
(\ref{pinned_shape}) for the free surface. Additionally, at the axis
of symmetry (in the case of free drops) or at the solid surface (in
the case of the drops pinned to the solid), the
boundary condition is the absence of a surface mass source/sink, so
that one has
\begin{equation}\label{on_v^s}
\mathbf{v}^s\cdot\mathbf{k}=0,
\end{equation}
where $\mathbf{k}=\mathbf{n}_a$ for the free drops and a unit vector
tangential to the free surface in the case of pinned drops.

Notably, at leading order in the limit $\epsilon\rightarrow 0$,
which is associated with taking to zero the ratio of the
characteristic length scale of interface formation $U\tau$~$(=
\sigma\tau_{\mu}$) to that of the bulk flow $R$, the interface
formation model reduces to the standard model. In simple terms: one
can see that for $\epsilon=0$ equation (\ref{mass1b}) and the second
equation in (\ref{mass2}) immediately give $\rho^s_1 = \rho^s_{1e}$
and $\rho^s_{2}=1$, i.e.\ the interfaces are in equilibrium, so that
$\sigma_1=1$ and $\sigma_2 = 0$, which results in the standard
stress-balance and kinematic equations on the free surface, the
absence of an internal interface, and, from the Young equation
(\ref{cl}), an instantaneously smooth free surface
$\theta_d=90^\circ$.

\subsection{Initial Conditions}

The initial conditions for the conventional model and the interface
formation model are essentially different as they represent how the
two models view the onset of coalescence. In the conventional model,
it is assumed that, after coming into contact, the two drops
instantaneously produce a smooth free surface, i.e.\ they
immediately coalesce and round the corner enforced by the drops'
initial configuration at the moment of touching.  Therefore, besides
prescribing the fluid's initial velocity, which we will assume to be
zero,
\begin{equation}\label{ic-u}
 \mathbf{u}=\mathbf{0}\qquad\hbox{at }t=0,
\end{equation}
we need to specify the initial shape as having,
near the origin, a tiny bridge whose free surface crosses the plane of symmetry at the
right angle. The free-surface shape far away from the origin (i.e.\ from the point of the
initial contact) is the undisturbed spherical (or hemispherical) drop. The initial radius of the bridge $r_{min}$ is a
parameter whose influence is to be investigated, although it is known \emph{a priori} that the limit $r_{min}\rightarrow 0$ gives rise to a singularity. For both a
spherical and a hemispherical drop, the free surface below the
drop's centre is conventionally prescribed as the one given by
Hopper's solution \citep{hopper84}, that is the analytic
two-dimensional solution for Stokes flow, whose parametric form is
\begin{eqnarray} \notag
 r(\theta) = \sqrt{2}\left[(1-m^2)(1+m^2)^{-1/2}
 (1+2m\cos\left(2\theta\right)+m^2)^{-1}\right](1+m)\cos\theta, \\  \label{ic2}
 z(\theta) =
 \sqrt{2}\left[(1-m^2)(1+m^2)^{-1/2}(1+2m\cos\left(2\theta\right)+m^2)^{-1}\right]
 (1-m)\sin\theta,
\end{eqnarray}
for $0<\theta<\theta_u$, where $m$ is chosen so that $r(0)=r_{min}$
and $\theta_u$ is chosen so that $r(\theta_u)=z(\theta_u)=1$.
Notably, for $r_{min}\to0$ we have $m\to1$  and $r^2+(z-1)^2=1$,
i.e.\ the drop's profile is a semicircle of unit radius which touches the plane of
symmetry at the origin.

The interface formation model does not presume an instant
coalescence, so that, after the two drops touch and then establish a
nonzero area of contact, (a) there is still an internal interface
between them, and hence coalescence as the formation of a single
body of fluid is only starting, and (b) the free surface is not
smooth, as the initial angle of contact of $180^\circ$ is only
starting its evolution towards $90^\circ$, i.e.\ a smooth interface.
For both a spherical and a hemispherical drop, the free surface
below the drop's centre can be prescribed as
\begin{equation}\label{ic1}
 (r-r_{min})^2+(z-z_c)^2 = z_c^2,
 \end{equation}
where $z_c=\hbox{$\frac{1}{2}$}(1+(1-r_{min})^2)/2$, so that if
there is no base, i.e.\ $r_{min}=0$, one has $z_c=1$, i.e.
$r^2+(z-1)^2=1$, which is a circle of radius $1$ centred at $(0,1)$ that coincides with the shape obtained from (\ref{ic2}) in the same limit. Importantly, for the interface formation model the limit $r_{min}\rightarrow0$ does not give rise to a singularity.

In addition to the free-surface shape given by (\ref{ic1}) and the
flow field in the bulk, by (\ref{ic-u}), we need to specify the initial
state of the interfaces, which will be given by
\begin{equation}\label{irho}
 \rho^s_1 =\rho^s_2= \rho^s_{1e},\qquad (t=0),
\end{equation}
These conditions in (\ref{irho}) describe the fact that (a) the free surface is
initially in equilibrium, and (b) the part of the free surface that
has been sandwiched between the two drops and becomes an `internal'
interface initially possesses the equilibrium properties of the
free-surface, since it can equilibrate to its new environment in a
finite time. Then, for $t>0$, the internal interface will start to relax towards its equilibrium state, which in turn will drive the free surface away from its initial (equilibrium) state, so that in the early stages of the coalescence phenomenon both interface will be out of equilibrium and will, in particular, deviate from the initial values given in (\ref{irho}). Notably, the assumption that all the interfaces are
unchanged from their pre-coalescence state is consistent with an
initial contact angle of $\theta_d=180^\circ$, which follows from
the Young equation (\ref{cl}) for $\sigma_1=\sigma_2=1$, i.e.\ when
$\rho^s_1 =\rho^s_2 = \rho^s_{1e}$.

\section{A Computational Framework for Free-Surface Flows with Dynamic Interfacial Effects}\label{numerics}

A finite-element-based computational platform for simulating
free-surface flows with dynamic interfacial effects has been
developed in \citep{sprittles11c,sprittles_jcp} and originally
applied to microfluidic dynamic wetting processes, which are the
most complex case of these flows. The ability of the developed framework to
simulate flows involving strong deformations of a drop has already
been confirmed in \citep{sprittles_pof}, where the predictions of the
code are shown to be in excellent agreement with previous literature
for the benchmark test-case of a freely oscillating liquid drop.  In
\citep{sprittles_jcp}, the interface formation model was
incorporated in full into the framework and allowed the simulation
of microfluidic phenomena such as capillary rise, showing excellent
agreement with experiments, and, in \citep{sprittles_pof}, the impact and spreading of microdrops on surfaces of varying wettability. The exposition in \citep{sprittles_jcp}
together with the preceding paper \citep{sprittles11c} provide a
detailed step-by-step guide to the development of the code, allowing
one to reproduce all results, as well as curves for benchmark
calculations and a demonstration of the platform's capabilities.
Therefore, here it is necessary only to point out a few aspects of
the computations.

The code is based on the finite element method and uses an arbitrary
Lagrangian-Eulerian mesh design \citep{kistler83,heil04,wilson06} to
allow the free surface to be accurately represented whilst bulk
nodes remain free to move. For the drop geometry, the mesh is based on the bipolar
coordinate system, and is graded to allow for extremely small
elements near the contact line and progressively larger elements in
the bulk of the liquid.  This ensures that all the physically-determined
smallest scales near the contact line are well resolved whilst the problem is
still computationally tractable.  The conditions on the mesh needed
to resolve the scales associated with the interface formation in
dynamic wetting problems are given in \citep{sprittles_jcp}. However,
for the coalescence phenomenon, even smaller elements are required to
capture the free-surface shape associated with the {\it
conventional\/} model.  Indeed, the initial free-surface shape given
by (\ref{ic2}) requires that the free surface bends near the plane
of symmetry $z=0$ to meet this boundary perpendicularly at
$r=r_{min}$. The radius of curvature of the free
surface where it meets $z=0$ is of $O(r_{min}^3)$ and for
$r_{min}=10^{-4}$, used in our computations, one has the radius of curvature $\sim10^{-12}$, i.e.\
extremely small and many orders of magnitude smaller than the length
scales associated with the interface formation dynamics. Here the model is used beyond
its area of applicability, as in the derivation of the capillary pressure due to the free-surface
curvature it is assumed that the radius of curvature is much larger than the physical
thickness of the interface, which is modelled as a geometric surface of zero thickness.  However, the conventional model dictates that this is the scale which needs to be resolved, so that in order to provide mesh-independent solutions from this model the elements near
the contact line have to be exceptionally small. On such scales, it is
somewhat surprising that, even with the huge amount of care taken,
we have been able to produce mesh-independent converged solutions.
Any further reduction of $r_{min}$ for the conventional model has
been seen to be impossible. To capture dynamics on this scale would
require one to `zoom in' on the coalescence event, which will
initially be isolated from the global dynamics, and then stitch this
solution to a global result at a later time, i.e. essentially to
mimic numerically the technique of matched asymptotic expansions.

The result of our spatial discretization is a system of non-linear
differential algebraic equations of index two \citep{lotstedt86}
which are solved using the second-order backward differentiation
formula, whose application to the Navier-Stokes equations is
described in detail in \citep{gresho2}, using a time step which
automatically adapts during a simulation to capture the appropriate
temporal scale at that instant.


\section{Benchmark Simulations}\label{previous}

In order to compare our computations for the conventional model to
the numerical results presented in \citep{paulsen12}, we consider the
coalescence of liquid spheres of radius $R=1$~mm, density
$\rho=970$~kg~m$^{-3}$, surface tension $\sigma_{1e}=20$~mN~m$^{-1}$
for viscosities $\mu=1$~mPa~s and $\mu=58000$~mPa~s.  For these
parameters, the Reynolds numbers are $Re=1.9\times10^4$ and
$Re=5.8\times10^{-6}$, respectively, which allows us to investigate
both the inertia-dominated and viscosity-dominated regimes.

\begin{figure}
     \centering
\includegraphics[scale=0.35]{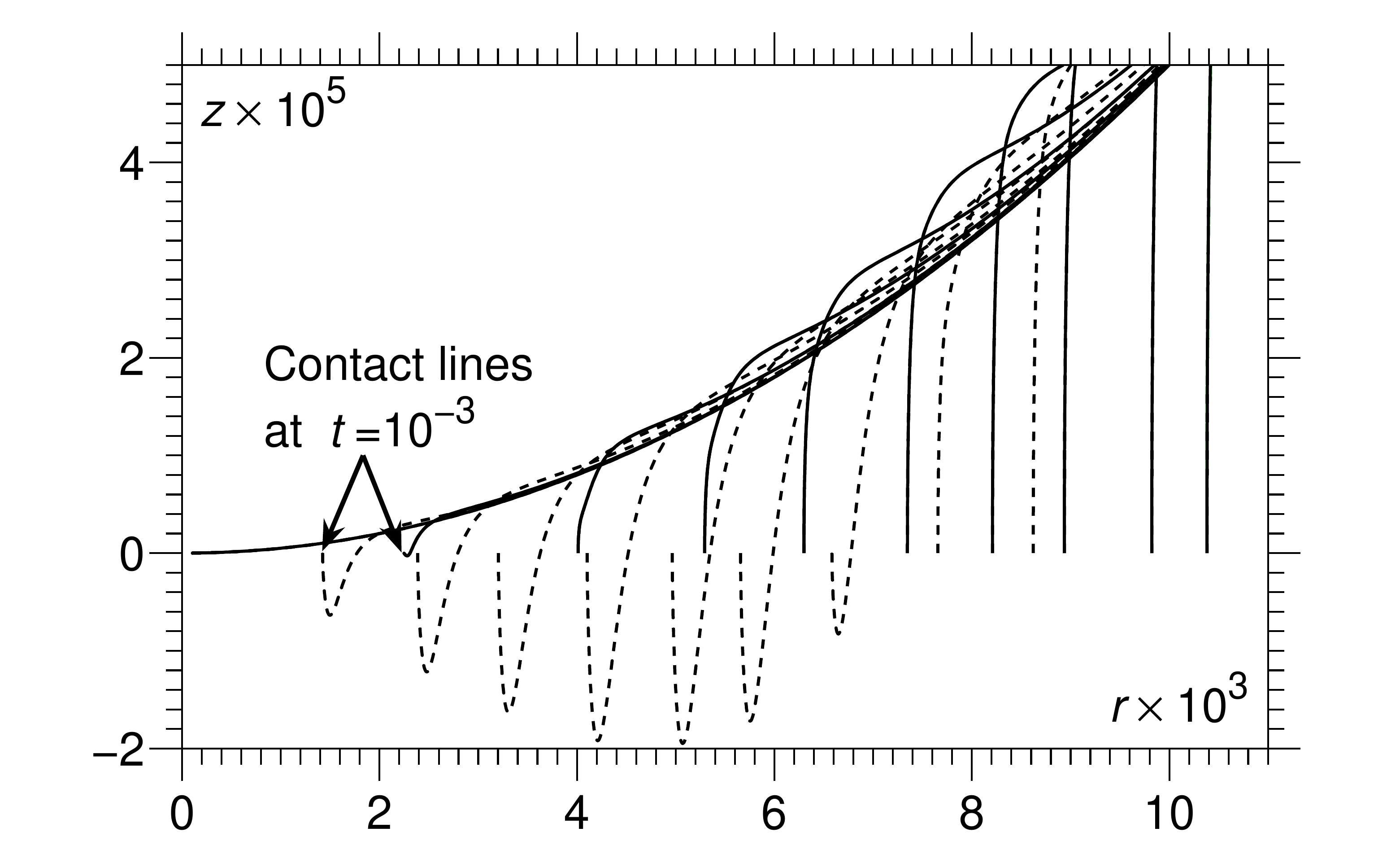}
\caption{\label{F:1cp_cush} Free-surface profiles obtained using the conventional model for the coalescence of two free drops with $Re=1.9\times10^4$ at intervals of $\triangle t = 10^{-3}$.  Dashed lines: the computed solution in which the free surface is allowed to freely pierce the plane of symmetry ($z=0$).  Solid lines: solution when the free surface is prevented, as it is henceforth, from crossing the symmetry plane.}
\end{figure}

Before doing so, we must make some comments regarding the
computation of the very initial stages of coalescence.  In
particular, in some simulations, for both the conventional and the
interface formation models, we have observed the tendency towards the formation of
toroidal bubbles, as the disturbance to the free surface,
initiated by the coalescence event, leads to capillary waves which
come into contact with the plane of symmetry (i.e., for the two
drops, into contact with each other), in front of the propagating
contact line.  This effect is essentially the same as the one reported in \citep{duchemin03,oguz89} and in our computations only occurs
for low-viscosity liquids. It is particularly severe for the
conventional model's computations, where the contact angle
variation, from $180^\circ$ at the moment of touching to $90^\circ$
when the computations start, creates a greater disturbance of the initial (equilibrium) free-surface shape and hence causes
larger free surface waves than those produced by the interface
formation model when the contact line begins to move.
Computationally, as only one drop in this symmetric system is
considered, there is nothing to stop the free surface piercing the
$z=0$ plane of symmetry, and in Figure~\ref{F:1cp_cush} the dashed
curves show the profiles obtained if no special treatment is
provided, for computations of the $Re=1.9\times10^4$ liquid using
the conventional model, i.e.\ the worst case scenario.

Physically, if the two free surfaces reconnect instantaneously upon
coming into contact, i.e.\ begin to coalesce, then the simulation should be
continued with a trapped toroidal bubble and a multiply-connected
domain.  However, as the capillary waves propagating along the free surfaces of the two drops try to reconnect, the viscosity of the gas in the narrow gap between them can no longer be neglected since the gas will be acting as a lubricant preventing the free surfaces from touching.  In any case, at present, accounting for the dynamics of a trail
of toroidal bubbles deposited behind an advancing free surface is beyond
developed computational methods.  An alternative approach, is to
assume that, as the free surfaces of the two drops try to touch ahead of
the contact line, they do not coalesce immediately, i.e.\ remain
\emph{free} surfaces for the short time that they are in contact, as
then the capillary waves propagate further and these free surfaces
separate.  This approach may well mimic the reality, as one has to drain the air film between the two converging surfaces before coalescence can occur, which could explain why there is yet to be any experimental validation of the existence of the toroidal bubbles. The profiles obtained using this approach are shown as the
solid lines in Figure~\ref{F:1cp_cush}, and it is this approach that
we use henceforth in the situations where the free surface touches
or tries to pierce the plane of symmetry.

\begin{figure}
     \centering
\includegraphics[scale=0.35]{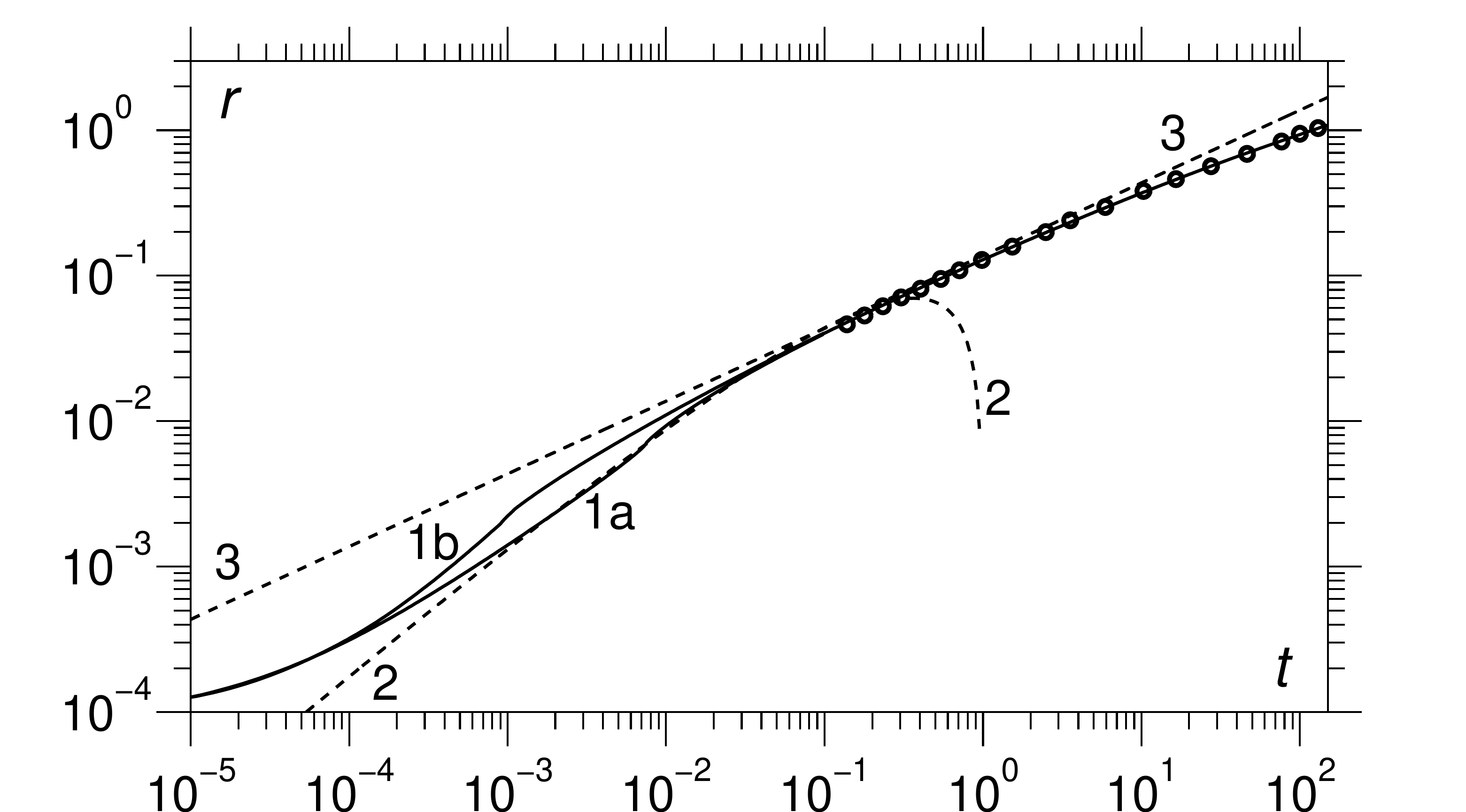}
\includegraphics[scale=0.35]{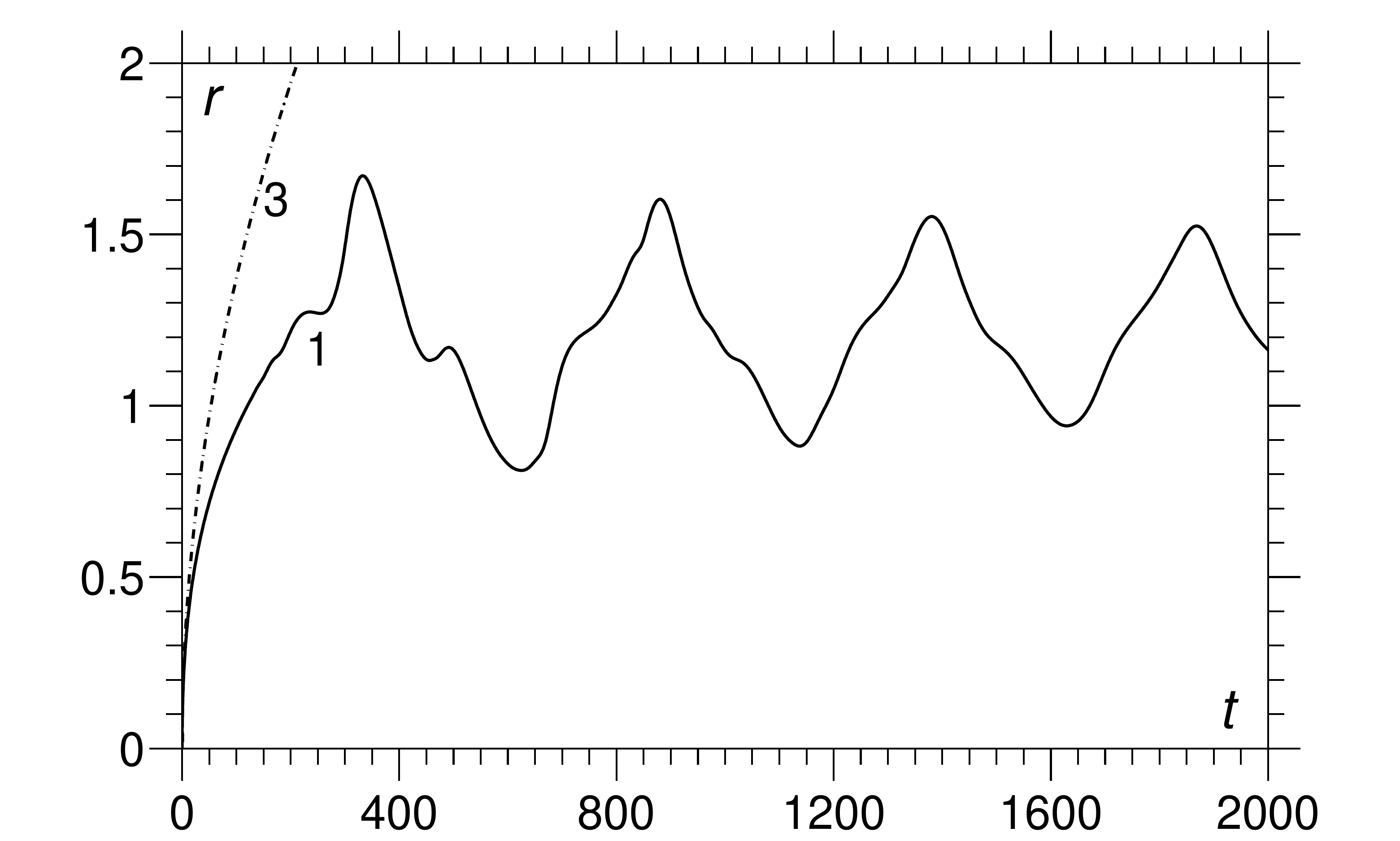}
 \caption{\label{F:1cP_cl} Bridge radius as a function of time obtained using the conventional model and scaling laws (\ref{viscous_scaling}) and (\ref{inertial_scaling}). Curve 1a: the free surface is allowed to pierce the plane of symmetry; curve 1b: simulations where piercing of the plane of symmetry was not allowed; curve 2: best fit ($C_{visc}=0.19$) of the scaling law (\ref{viscous_scaling}); curve 3: the inviscid scaling law (\ref{inertial_scaling}) with $C_{inert}=1.62$; circles: the numerical solution obtained in \citep{paulsen12} for the same problem. After the initial stages curves 1a and 1b are graphically indistinguishable and so the label curve 1 is used.}
\end{figure}

In Figure~\ref{F:1cP_cl}, one can see that the difference between the two approaches, i.e.\
between allowing the free surface to freely pierce the plane of
symmetry and using the plane of symmetry as a geometric constraint,
is visible but small, with the second approach (curve 1b), where the
free surface is unable to pass through the symmetry plane,
predicting a slightly faster evolution of the bridge radius than
when the penetration of the plane is allowed (curve 1a). This
phenomenon clearly deserves more attention, and the development of
more advanced computational techniques, but in what follows we use
the method proposed above and note that the specific treatment does
not appear to have a significant influence on the bridge radius,
certainly compared to the error bars in the experiments shown in
Section~\ref{experiments} (see for example Figure~\ref{F:3p3cP_cl})
and only affects the lowest viscosity liquid drops.

The $\log$-$\log$ plot in Figure~\ref{F:1cP_cl} shows the radius of
the liquid bridge connecting the two coalescing drops as a function
of time. Henceforth, $r$ refers to the radius of the free surface at
the plane of symmetry, i.e.\ $r=r(0,t)$.  The curves shown in
Figure~\ref{F:1cP_cl} have been computed using either of the two
approaches to deal with the capillary waves piercing through the
plane of symmetry: both curves 1a and 1b are graphically
indistinguishable from the corresponding numerical results obtained in
\citep{paulsen12}, so that circles had to be used to highlight the
region, roughly $0.1<t<100$, for which a comparison was available.

Our results also give an opportunity to compare the full numerical
solution we obtained to the scaling laws given by equations
(\ref{viscous_scaling}) and (\ref{inertial_scaling}) described in
Section~\ref{scalings}.  As one can see in Figure~\ref{F:1cP_cl}, both scaling laws,
(\ref{viscous_scaling}) and (\ref{inertial_scaling}), provide a good
approximation of the conventional model's solution over a
considerable period of time. As expected, the
viscosity-versus-capillarity scaling law (curve 2), with
$C_{visc}=0.19$ in equation (\ref{viscous_scaling}), provides a good
approximation for early time, until roughly $t=0.1$. The
inertia-versus-capillarity scaling law (curve 3), with
$C_{inert}=1.62$ in equation (\ref{inertial_scaling}) taken from
\citep{duchemin03}, despite being used well outside its limits of
applicability, agrees fairly well with our numerical solution from
roughly $t\approx0.1$ until approximately $t\approx10$, at which
point the (non-local) influence of the drop's overall geometry
becomes pronounced.

Of particular interest is that our simulations show the
$r\sim t\ln t$ behaviour predicted in \citep{eggers99}, which, as far
as we are aware, has never previously been observed in either
experiments or simulations.  In \citep{eggers99}, it is claimed that
the viscosity-versus-capillarity scaling law is only valid when the
Reynolds number $Re_r$ based on the bridge radius is less than one,
i.e.\ $Re_r\equiv r\,Re<1$, which corresponds to $r<10^{-4}$ for our
values of parameters. However, we observe that the scaling law
approximates the actual solution up until almost $r=10^{-1}$, i.e.\
well outside its apparent limits of applicability. This is in
agreement with the conclusions in \citep{paulsen11}, where it is
claimed that, to ascertain the limits of applicability of the
scaling law, one should use the Reynolds number $Re_h$ based on the
undisturbed height of the free surface at a given radius, as opposed to the bridge radius itself.  Given that
$h\propto r^2$, we have the condition $Re_h = r^2 Re<1$, which
suggests that the scaling law is valid until $r<10^{-2}$, which is
far closer to what we see. The above regime is followed by the
inertial one after which something close to a $t^{1/2}$ scaling is
observed.

\begin{figure}
     \centering
     \begin{minipage}[l]{.99\textwidth}
\subfigure[$t=0$]{\includegraphics[scale=0.19]{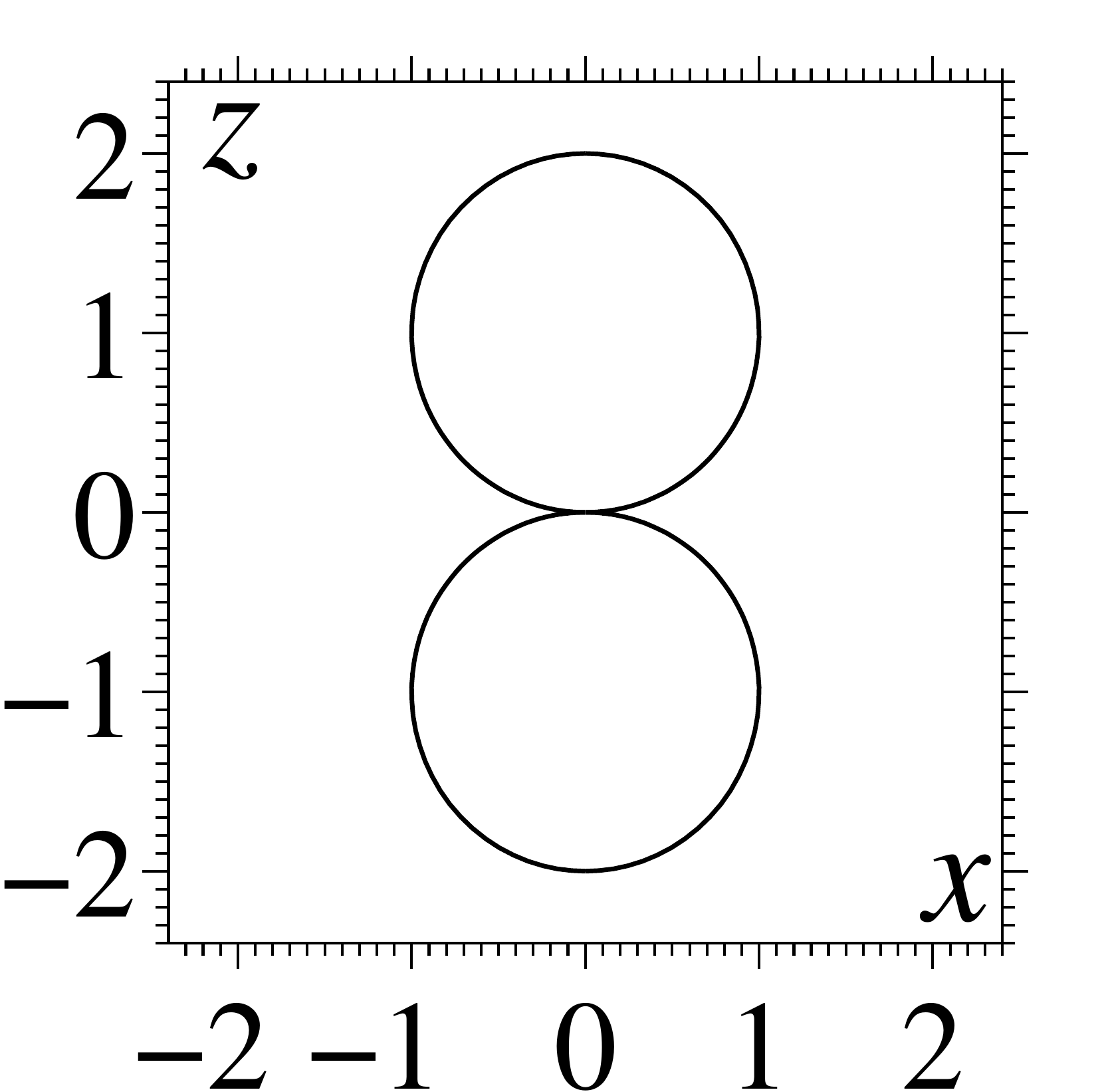}}
\subfigure[$t=10$]{\includegraphics[scale=0.19]{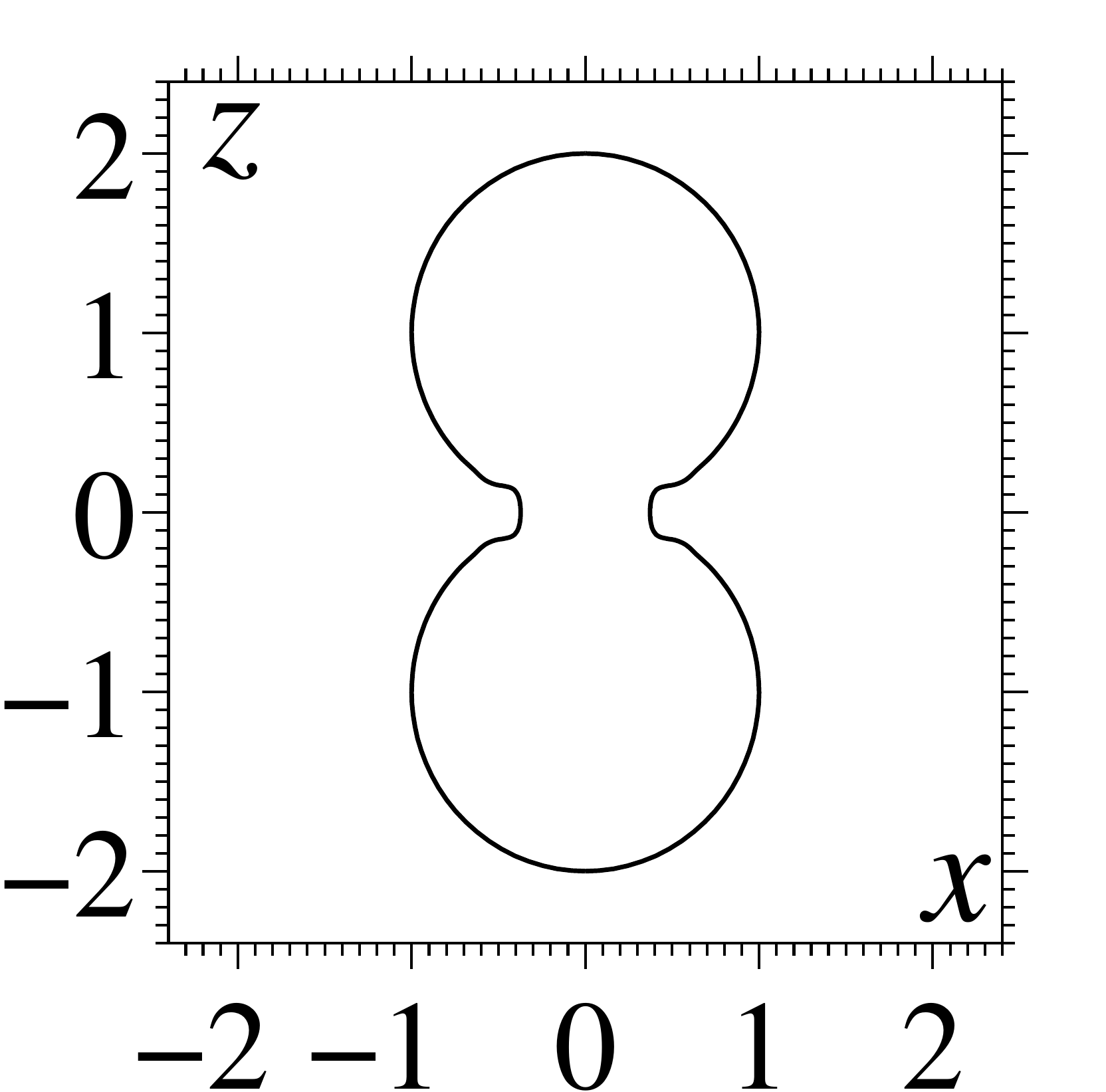}}
\subfigure[$t=20$]{\includegraphics[scale=0.19]{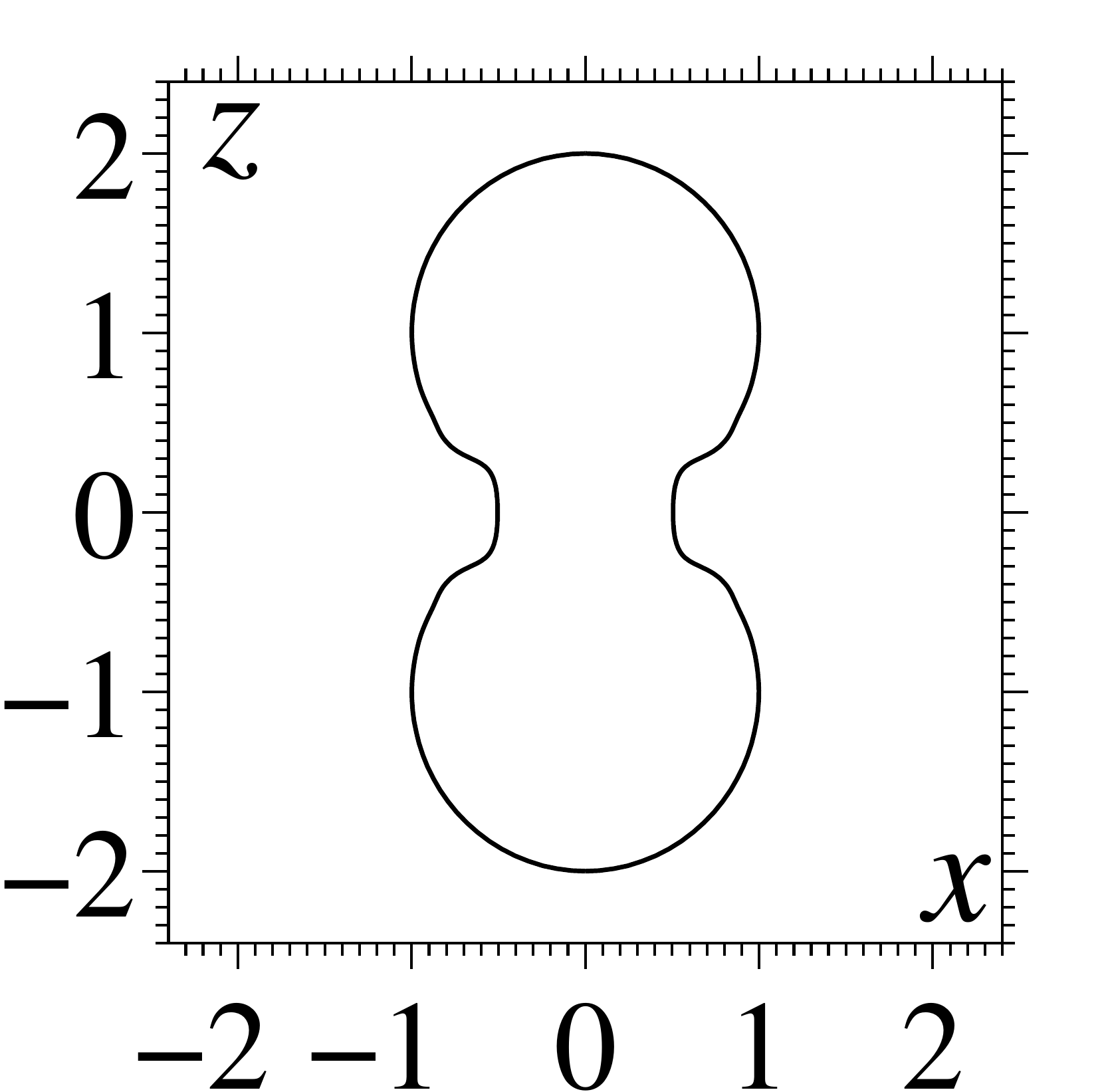}}
\subfigure[$t=50$]{\includegraphics[scale=0.19]{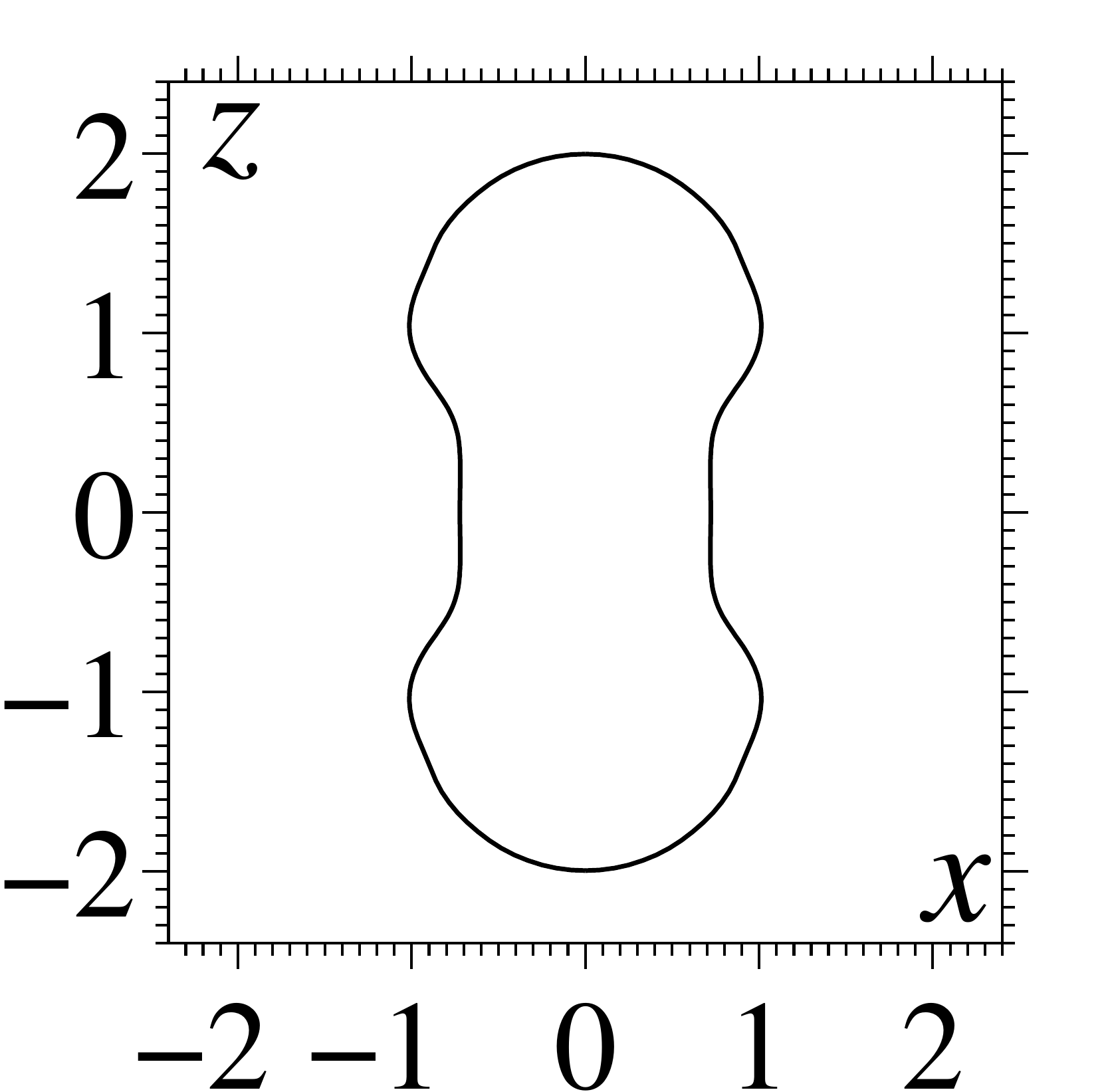}}
    \end{minipage}
     \begin{minipage}[l]{.99\textwidth}
\subfigure[$t=100$]{\includegraphics[scale=0.19]{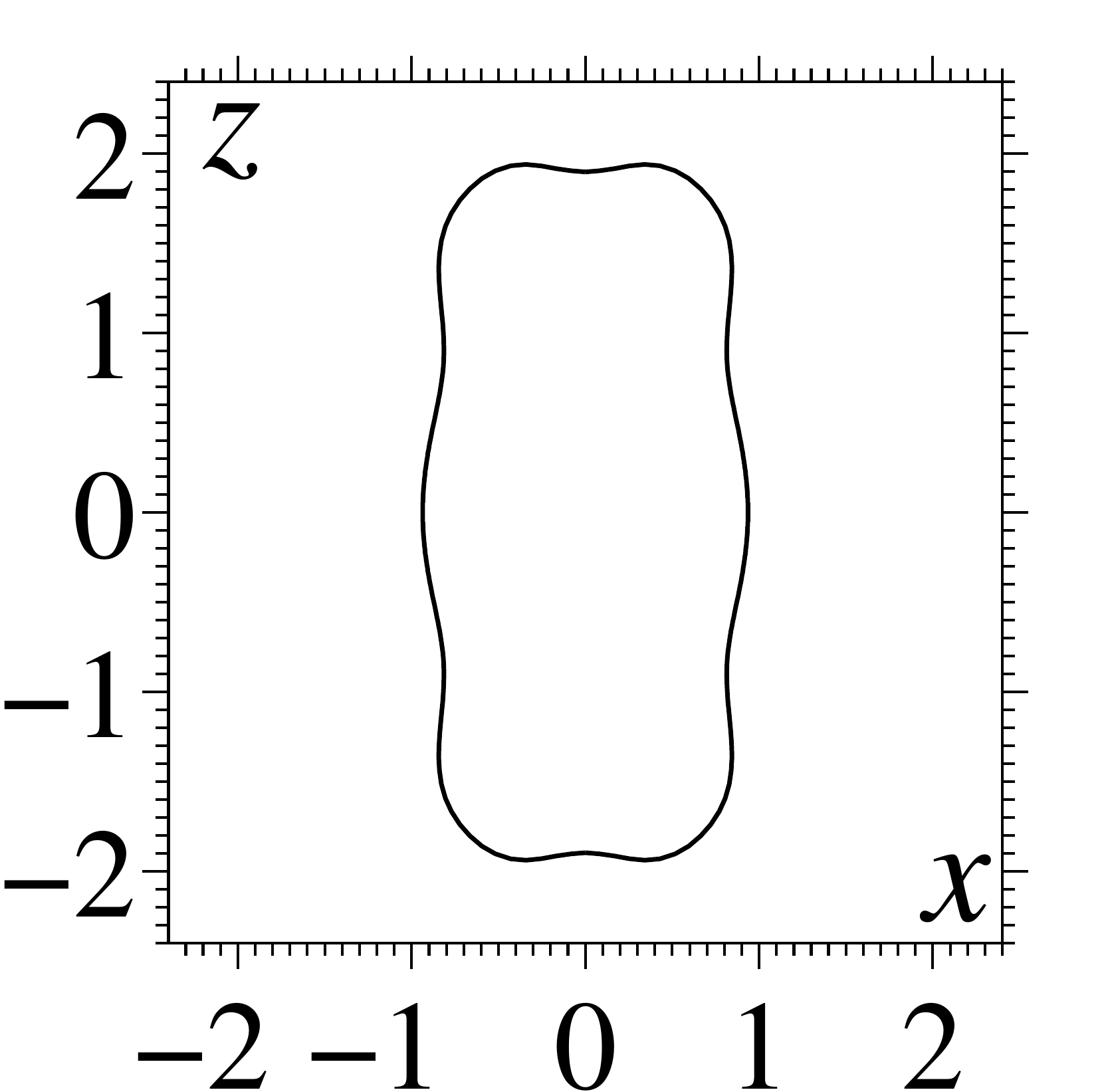}}
\subfigure[$t=150$]{\includegraphics[scale=0.19]{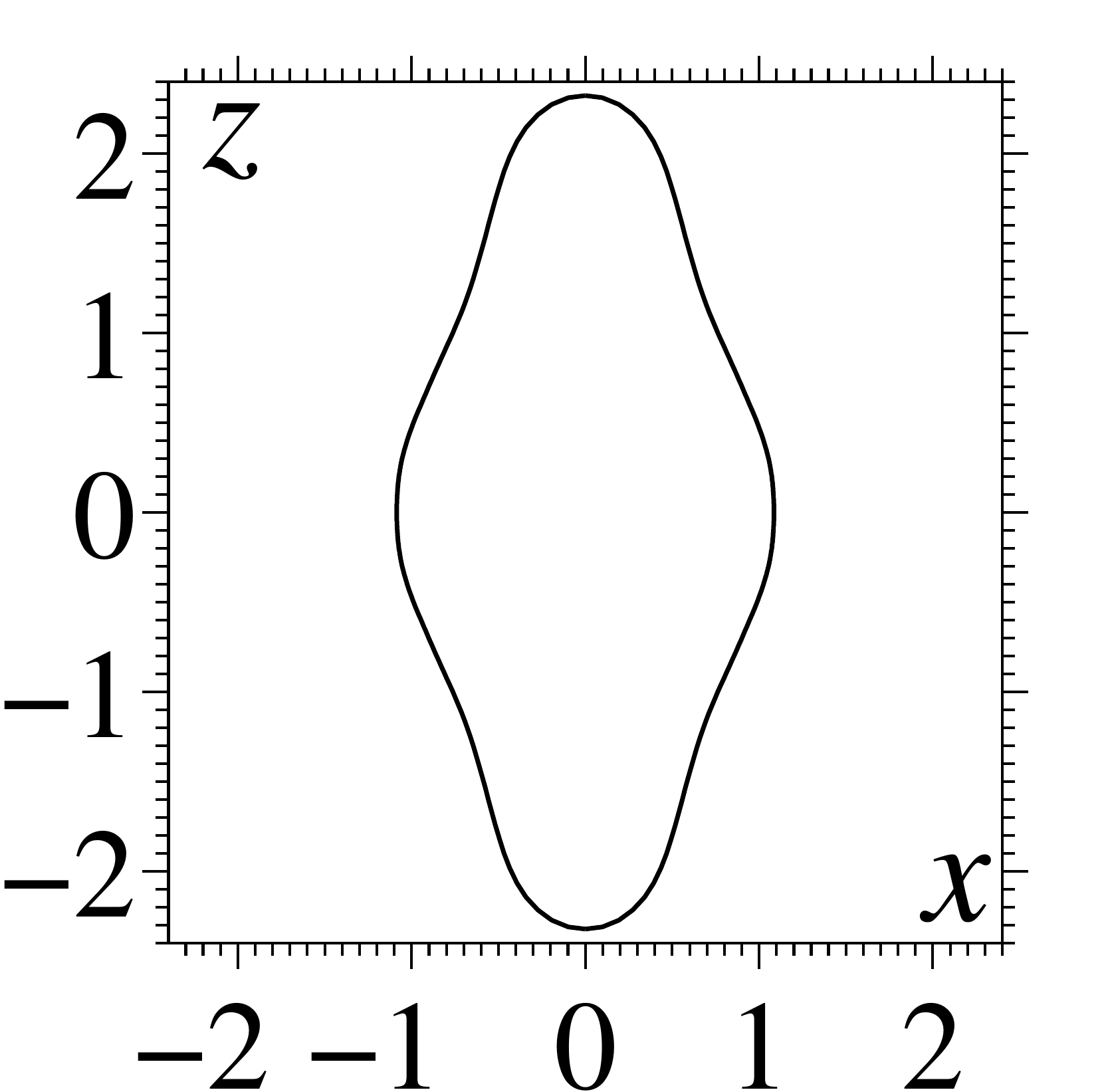}}
\subfigure[$t=200$]{\includegraphics[scale=0.19]{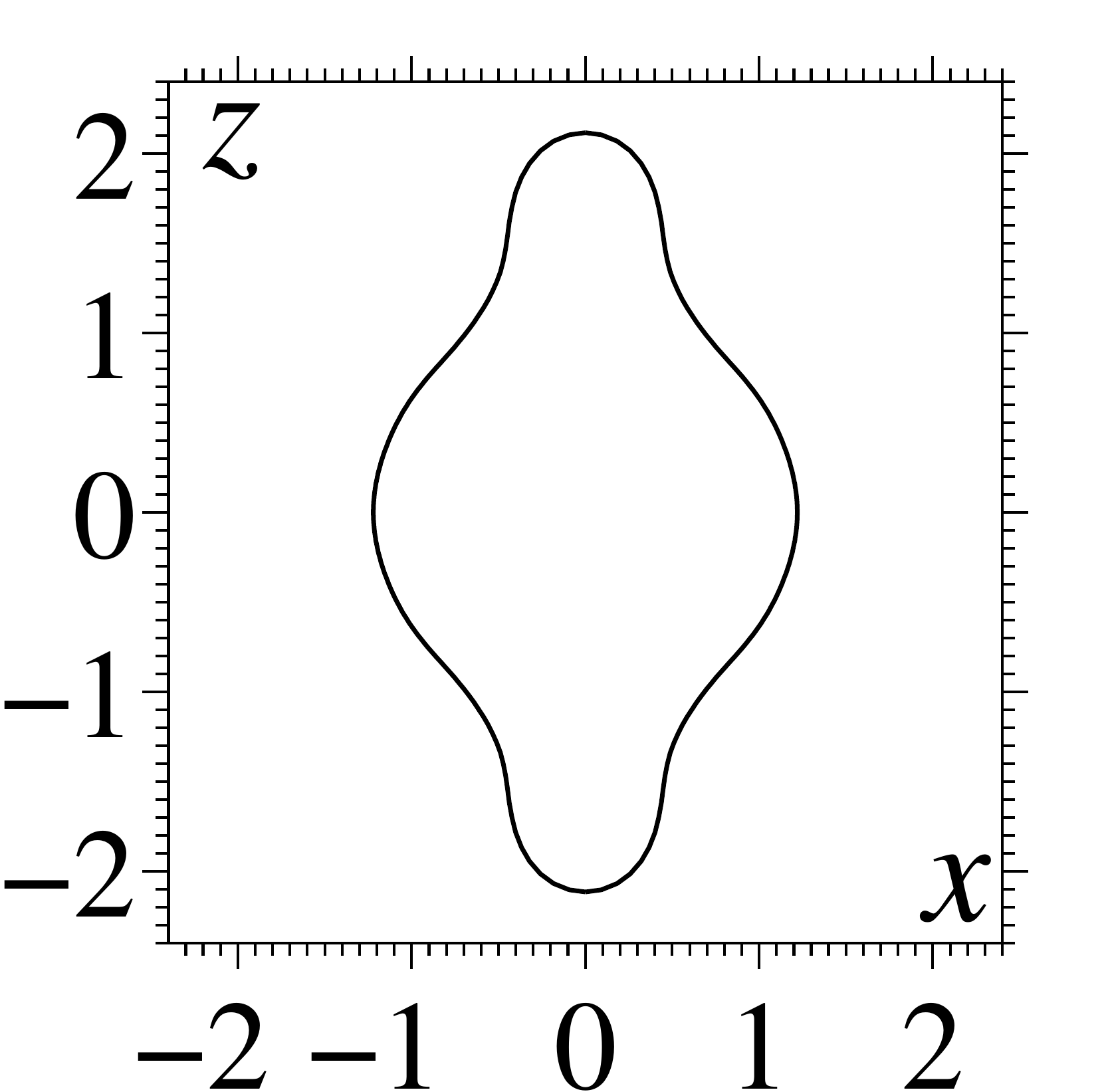}}
\subfigure[$t=250$]{\includegraphics[scale=0.19]{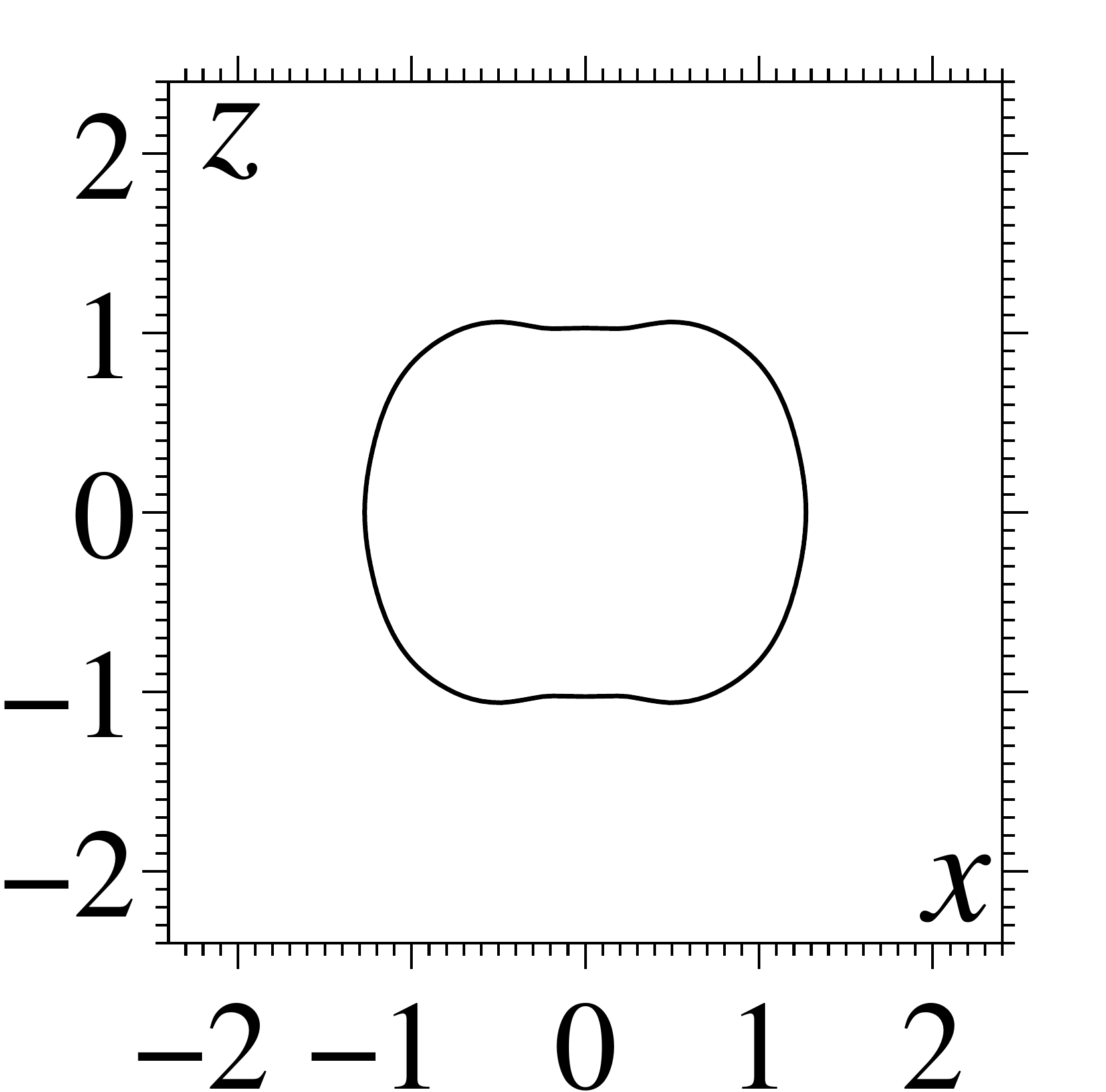}}
    \end{minipage}
    \begin{minipage}[r]{.99\textwidth}
\subfigure[$t=300$]{\includegraphics[scale=0.19]{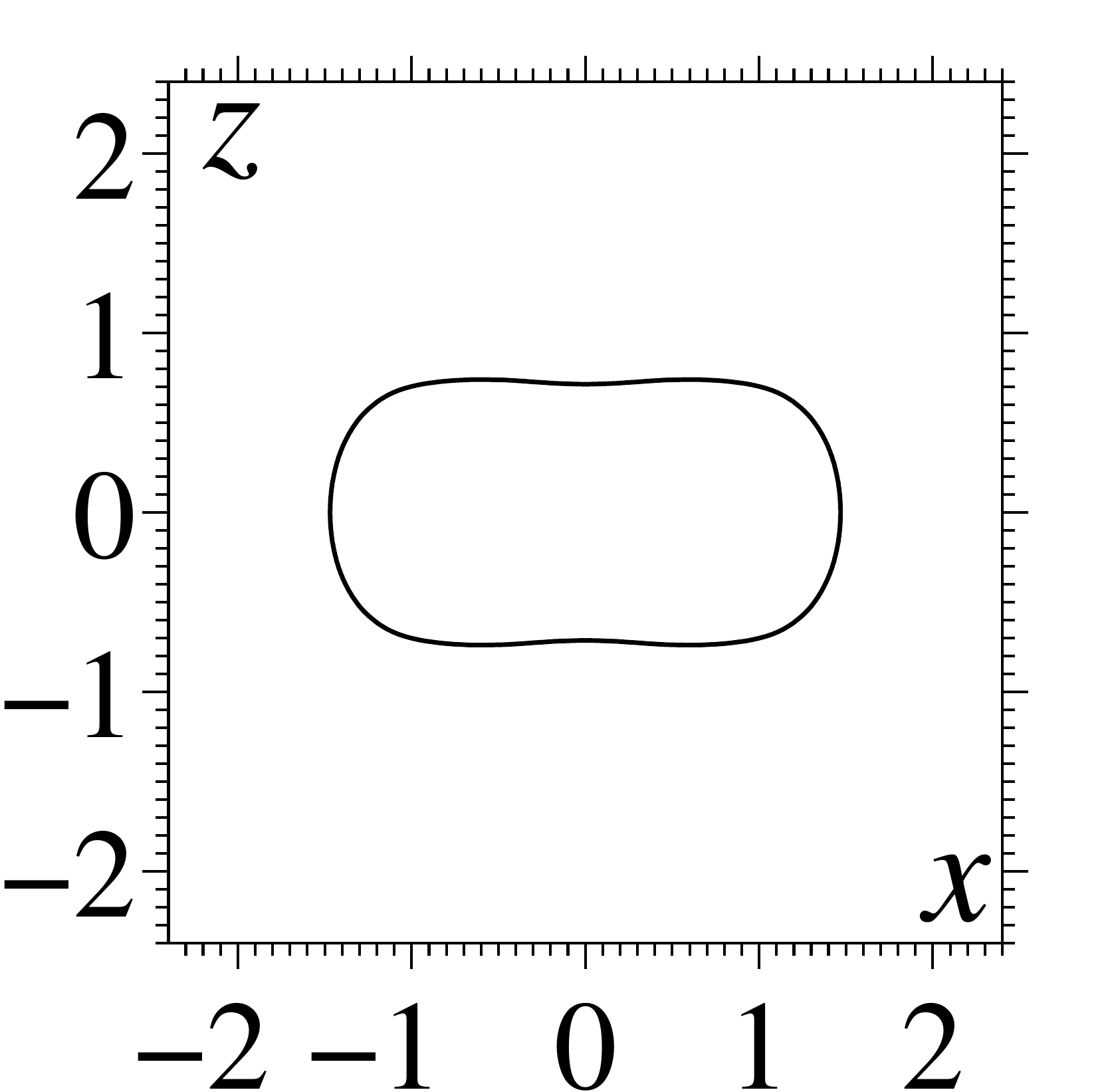}}
\subfigure[$t=350$]{\includegraphics[scale=0.19]{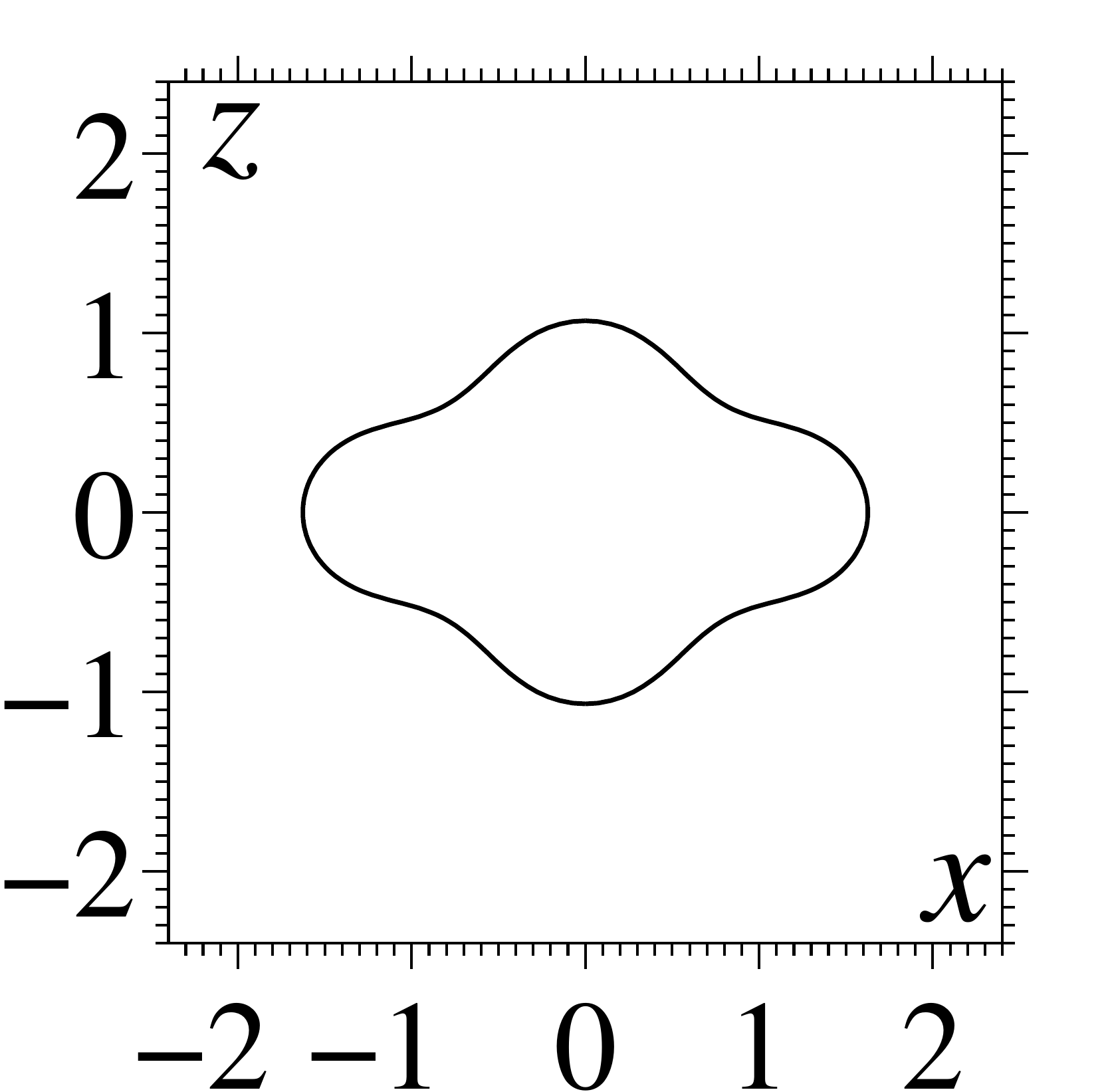}}
\subfigure[$t=450$]{\includegraphics[scale=0.19]{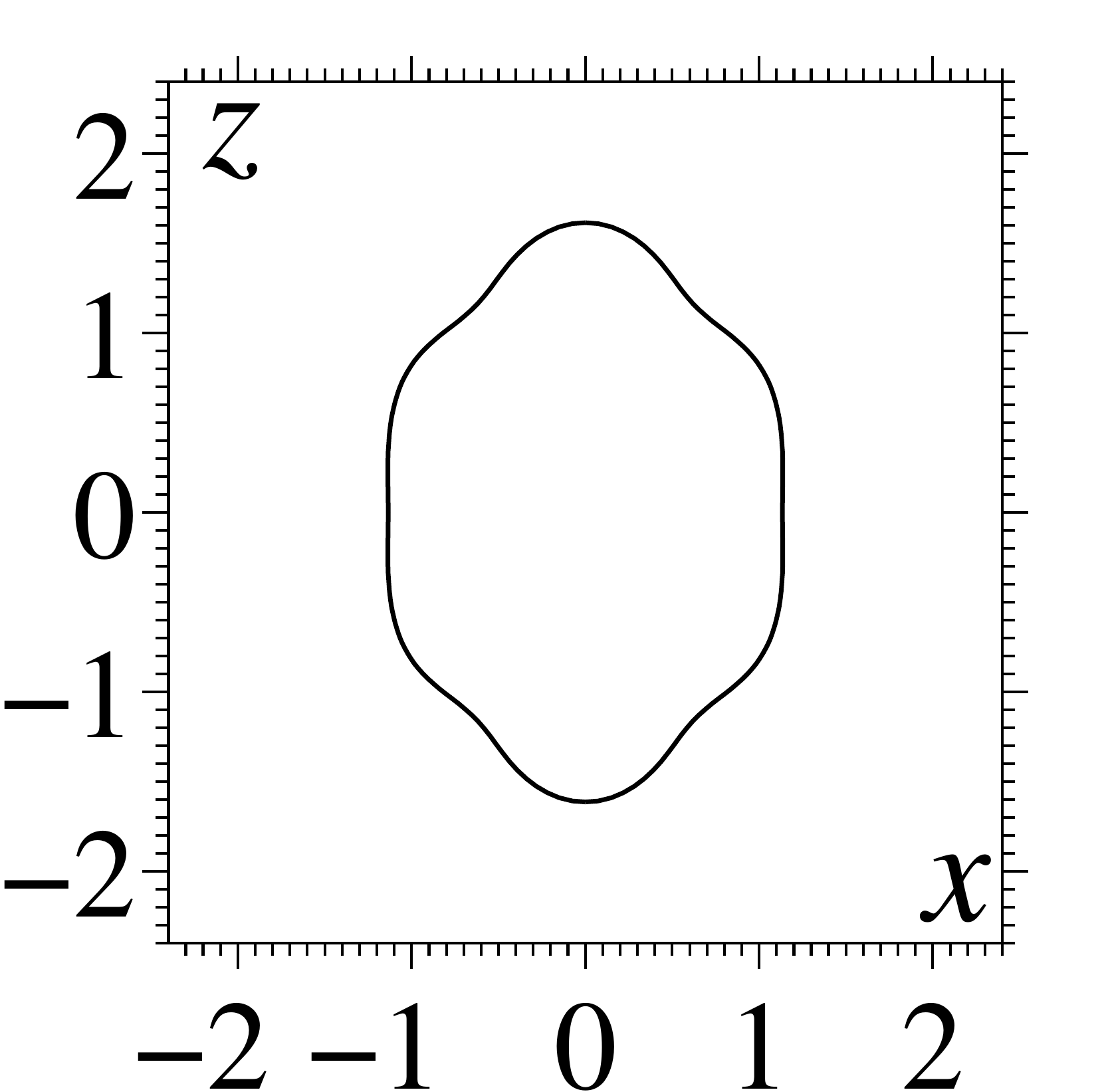}}
\subfigure[$t=550$]{\includegraphics[scale=0.19]{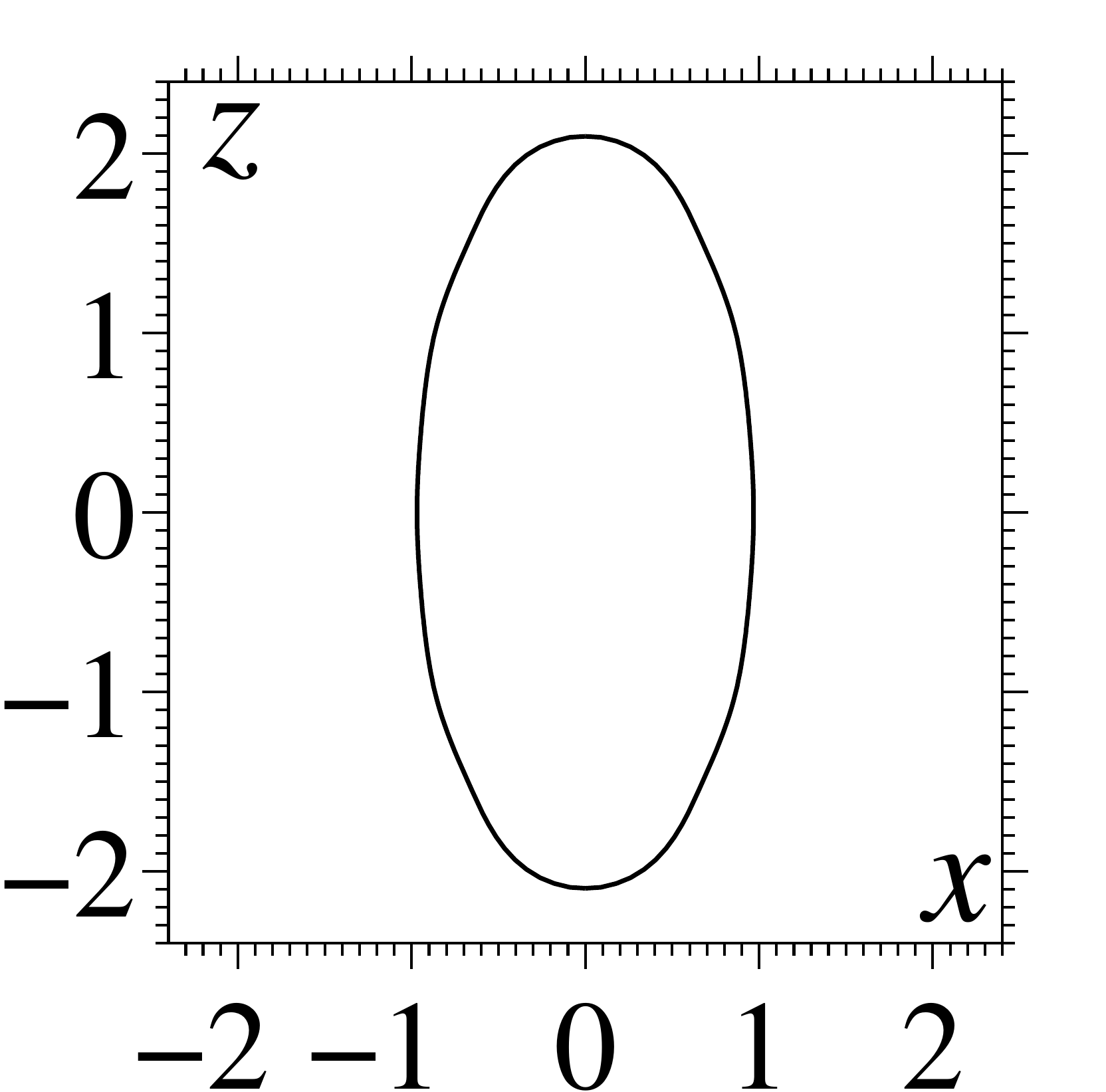}}
     \end{minipage}
\caption{Coalescence of two low-viscosity free liquid
drops with $Re=1.9\times10^4$. \label{F:1cP_evo}}
\end{figure}

In Figure~\ref{F:1cP_evo}, we show the results
obtained using the conventional model for the global dynamics of
the coalescence of low-viscosity ($Re=1.9\times10^4$) spherical drops,
as those considered in \citep{basaran92}. In this figure, we use the
Cartesian $x$-coordinate instead of $r$ to give the full profile of
the drops rather than just a half of it ($r\ge0$).  As one can see
from Figure~\ref{F:1cP_cl}, as well as from the shape of the free surface in the
last image in Figure~\ref{F:1cP_evo} at $t=550$, the free surface continues to evolve for $t>550$, but this period is
concerned with the free oscillation of a single liquid drop (see for
example \citep{basaran92}), as opposed to the coalescence event which
we are interested in here.

\begin{figure}
     \centering
     \begin{minipage}[l]{.99\textwidth}
\subfigure[$t=0$]{\includegraphics[scale=0.19]{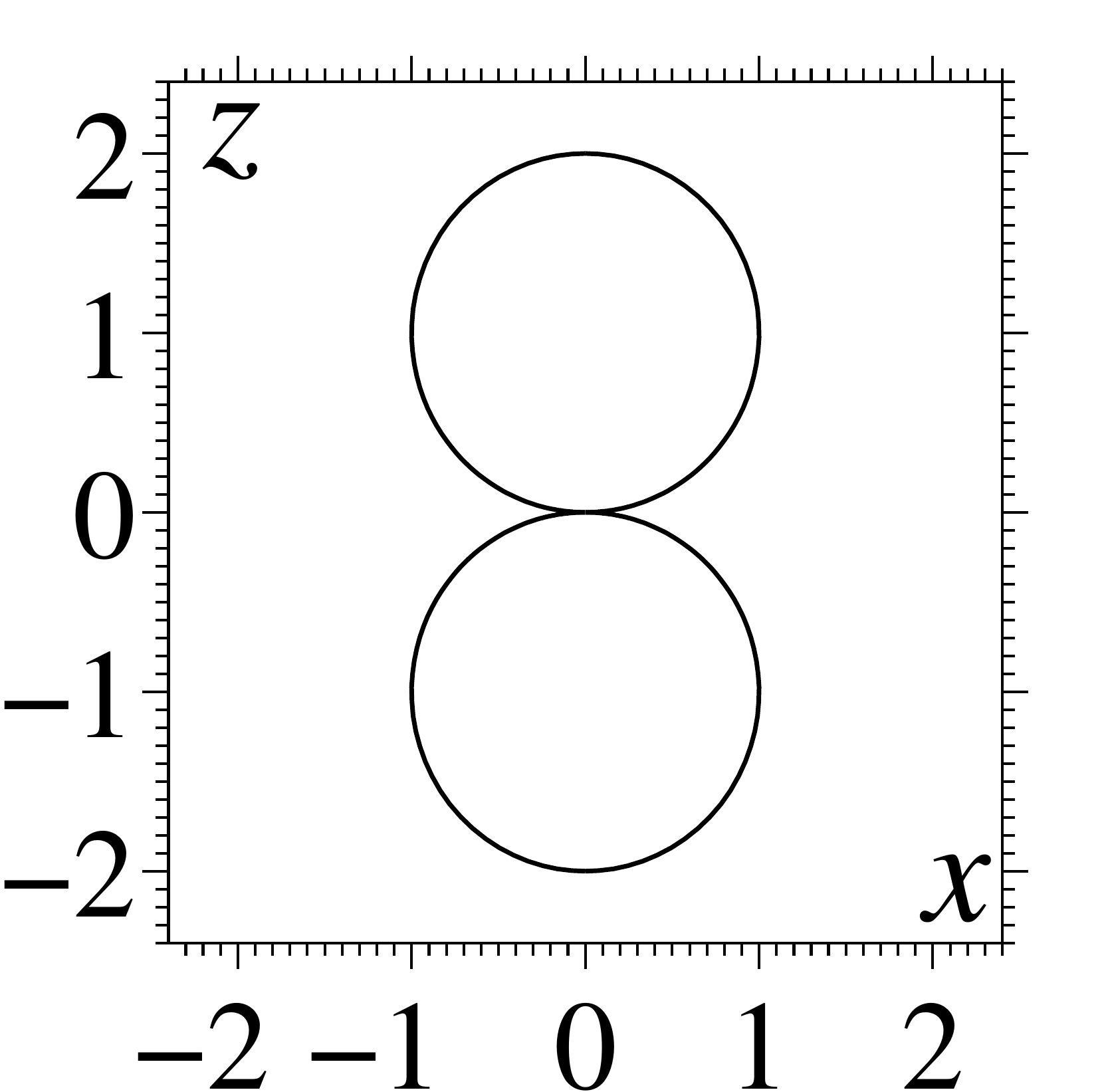}}
\subfigure[$t=0.2$]{\includegraphics[scale=0.19]{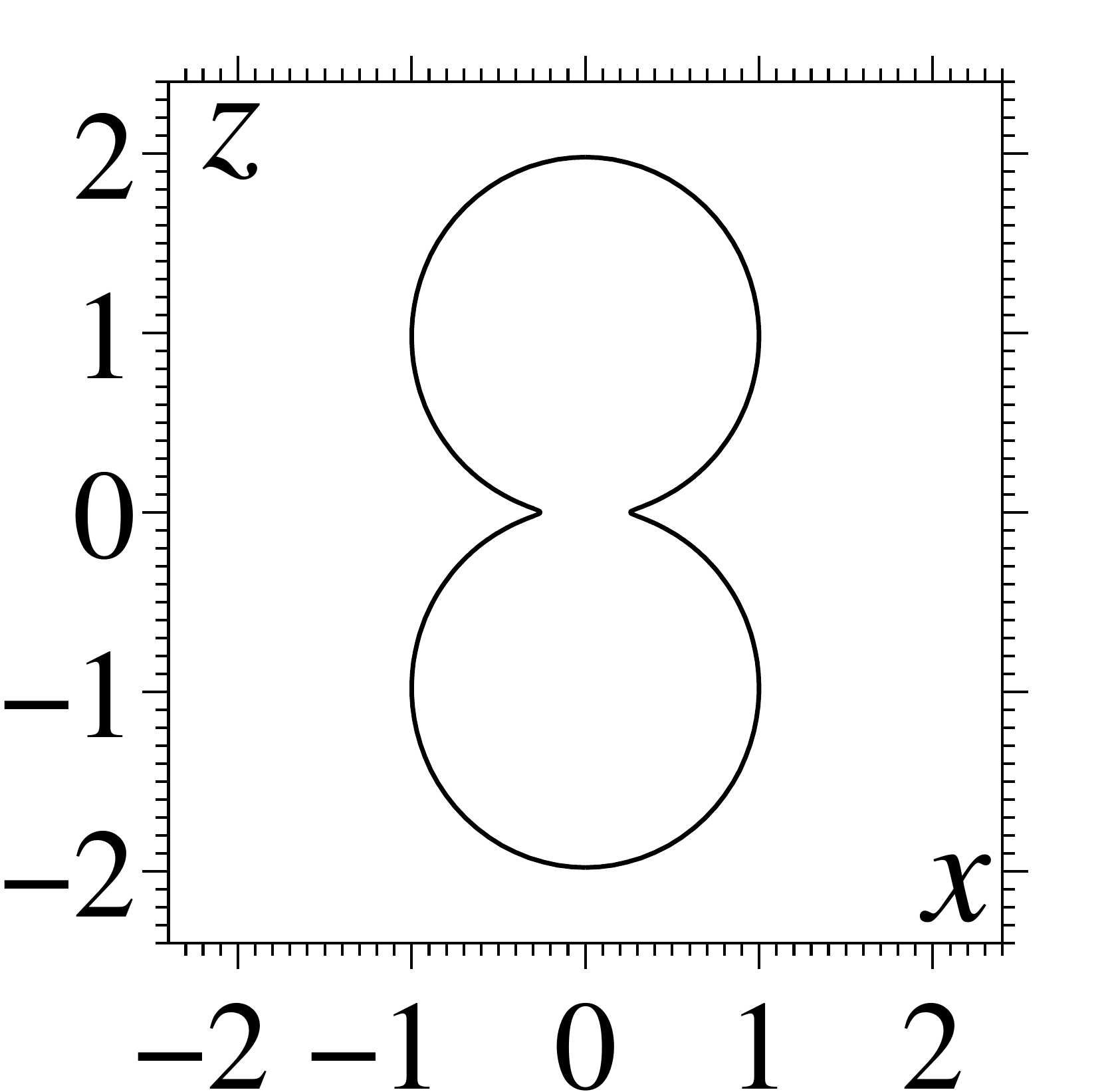}}
\subfigure[$t=0.5$]{\includegraphics[scale=0.19]{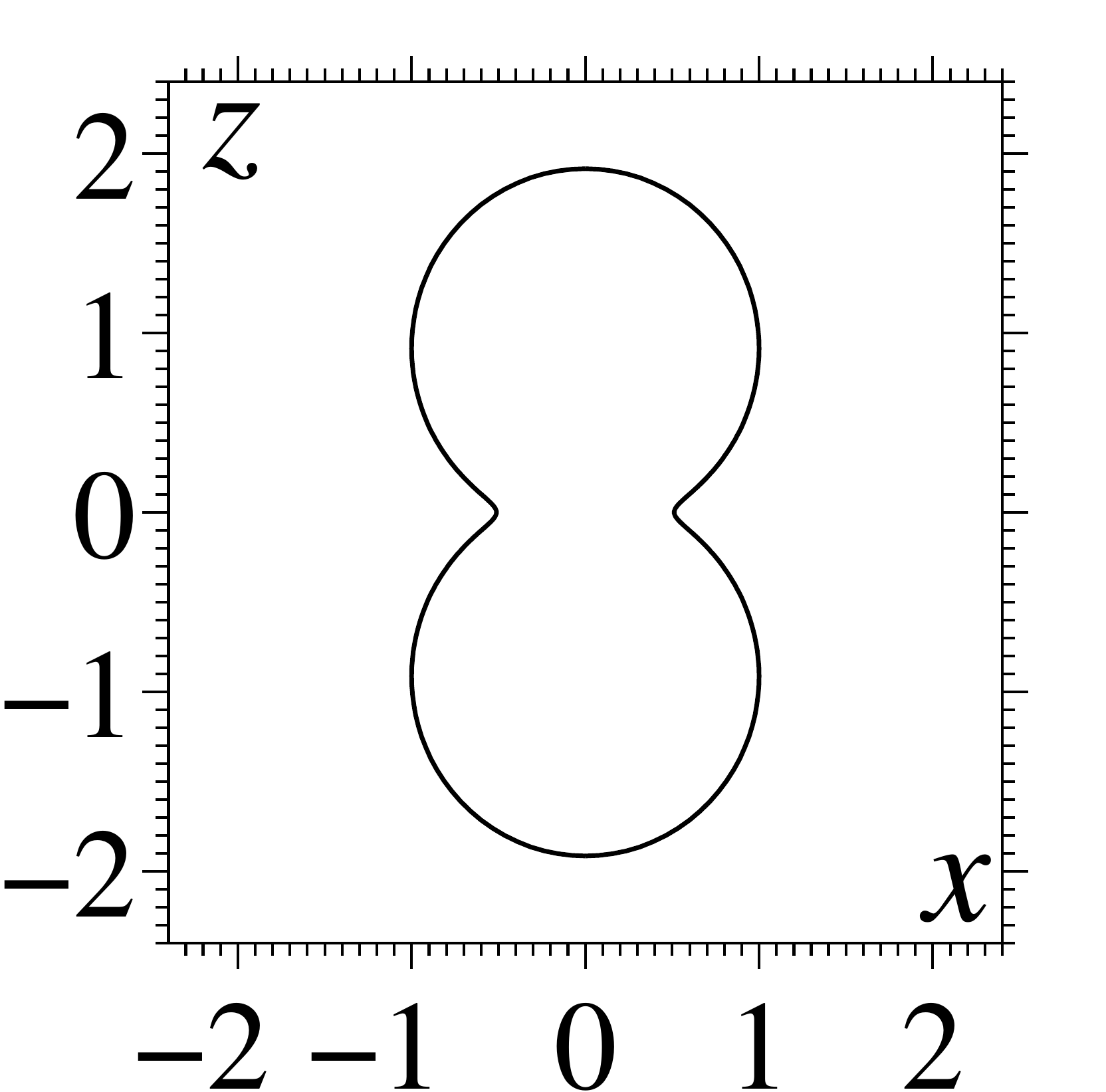}}
\subfigure[$t=1$]{\includegraphics[scale=0.19]{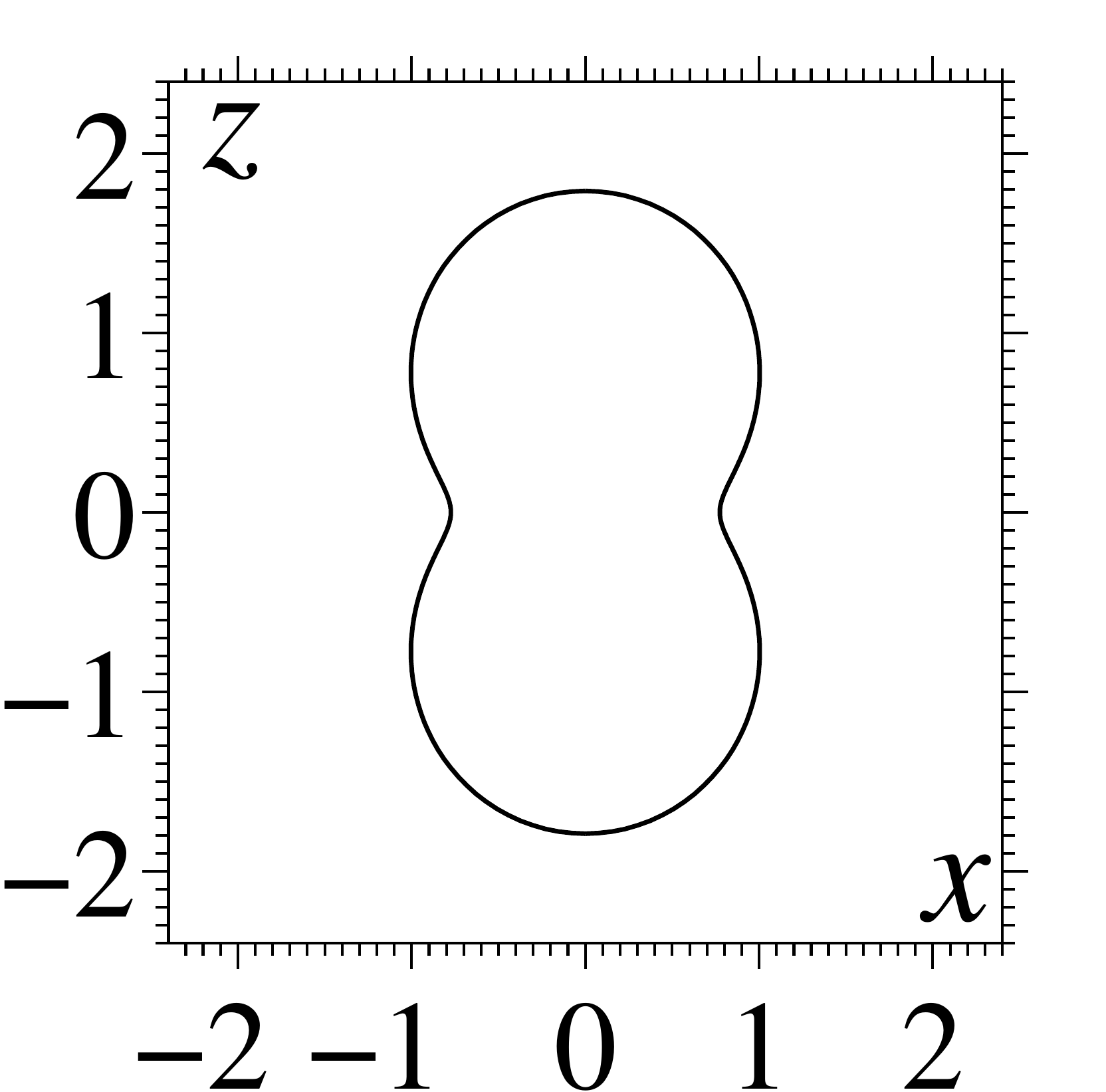}}
    \end{minipage}
     \begin{minipage}[l]{.99\textwidth}
\subfigure[$t=1.5$]{\includegraphics[scale=0.19]{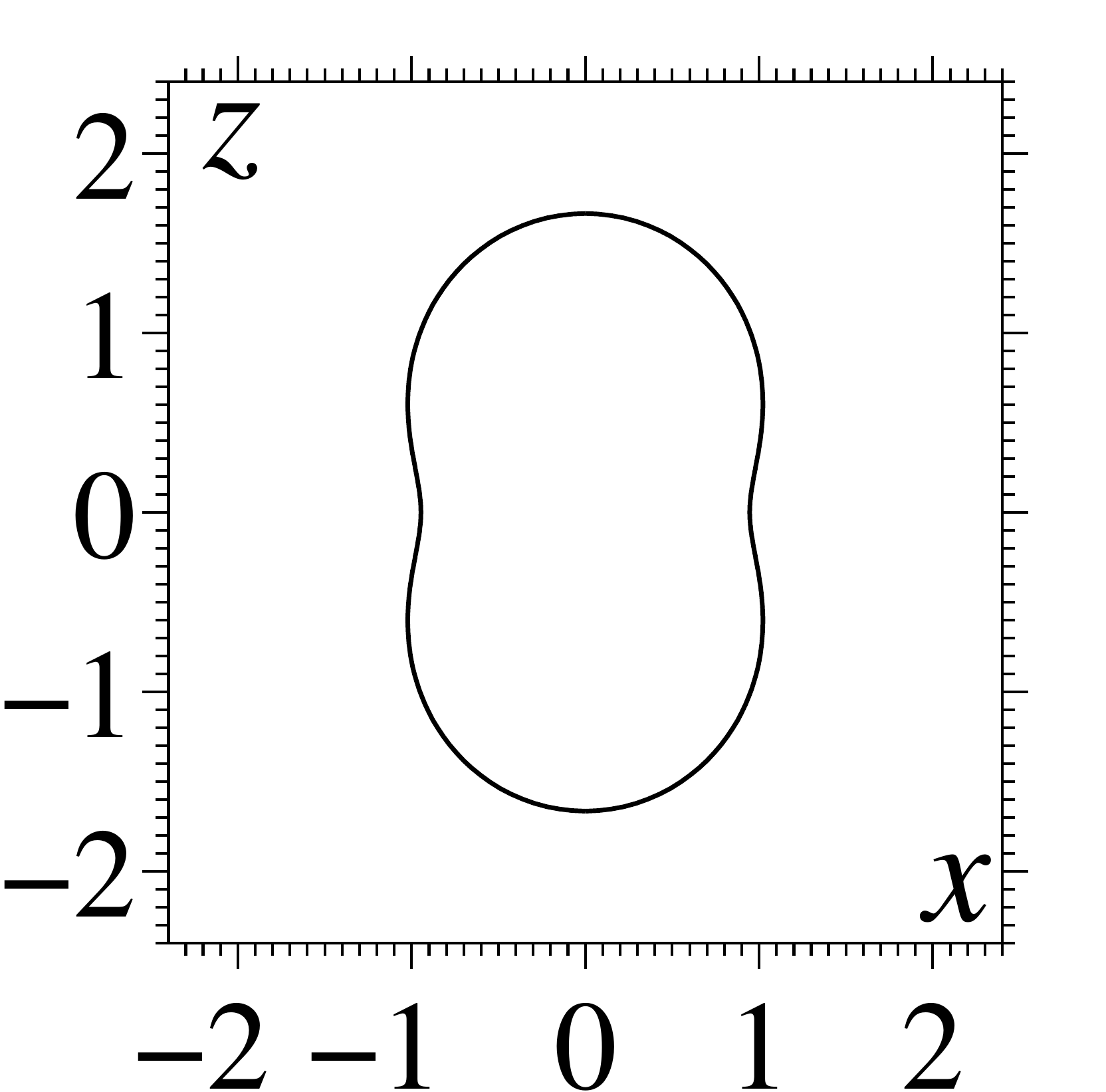}}
\subfigure[$t=2$]{\includegraphics[scale=0.19]{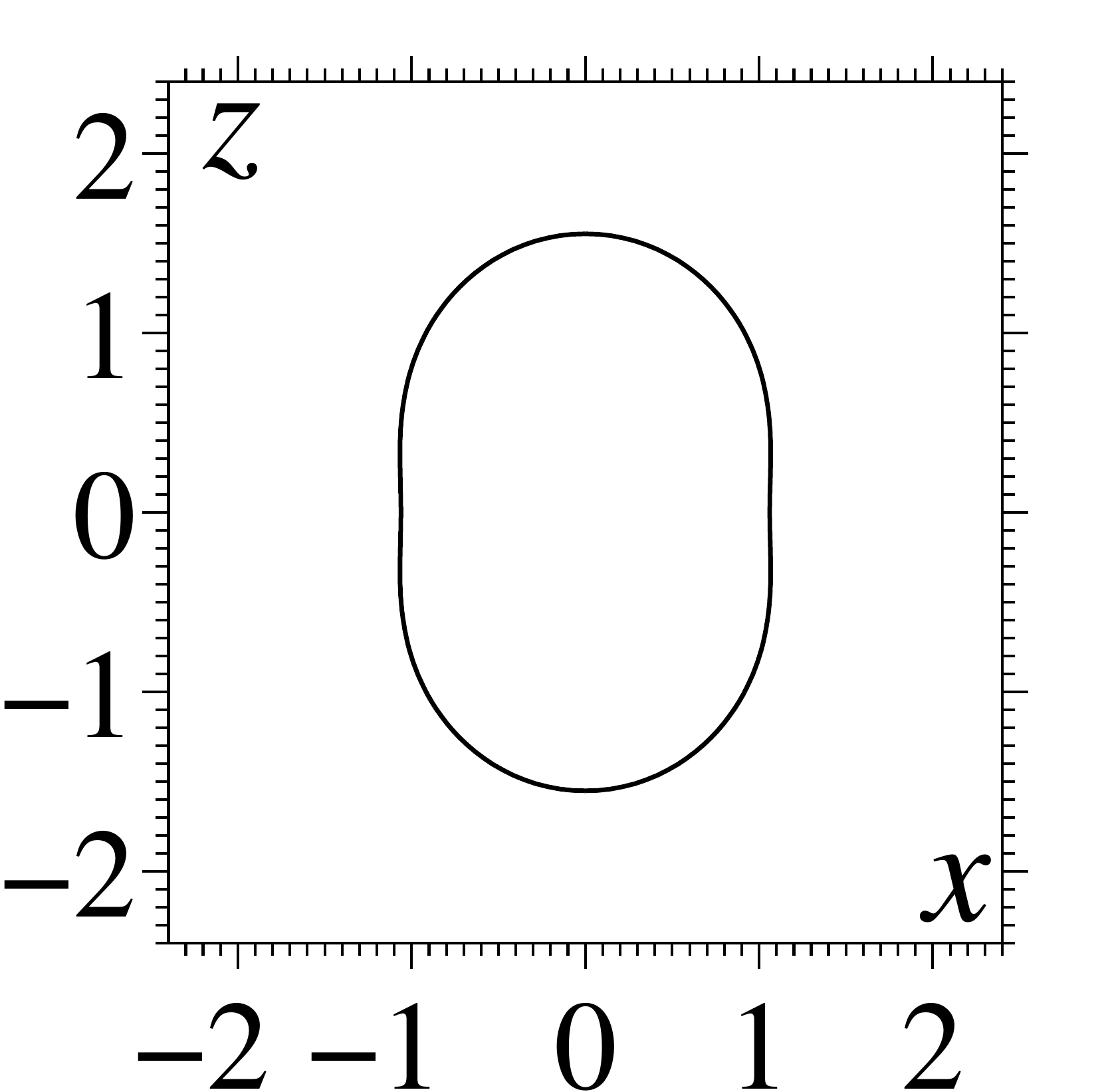}}
\subfigure[$t=3$]{\includegraphics[scale=0.19]{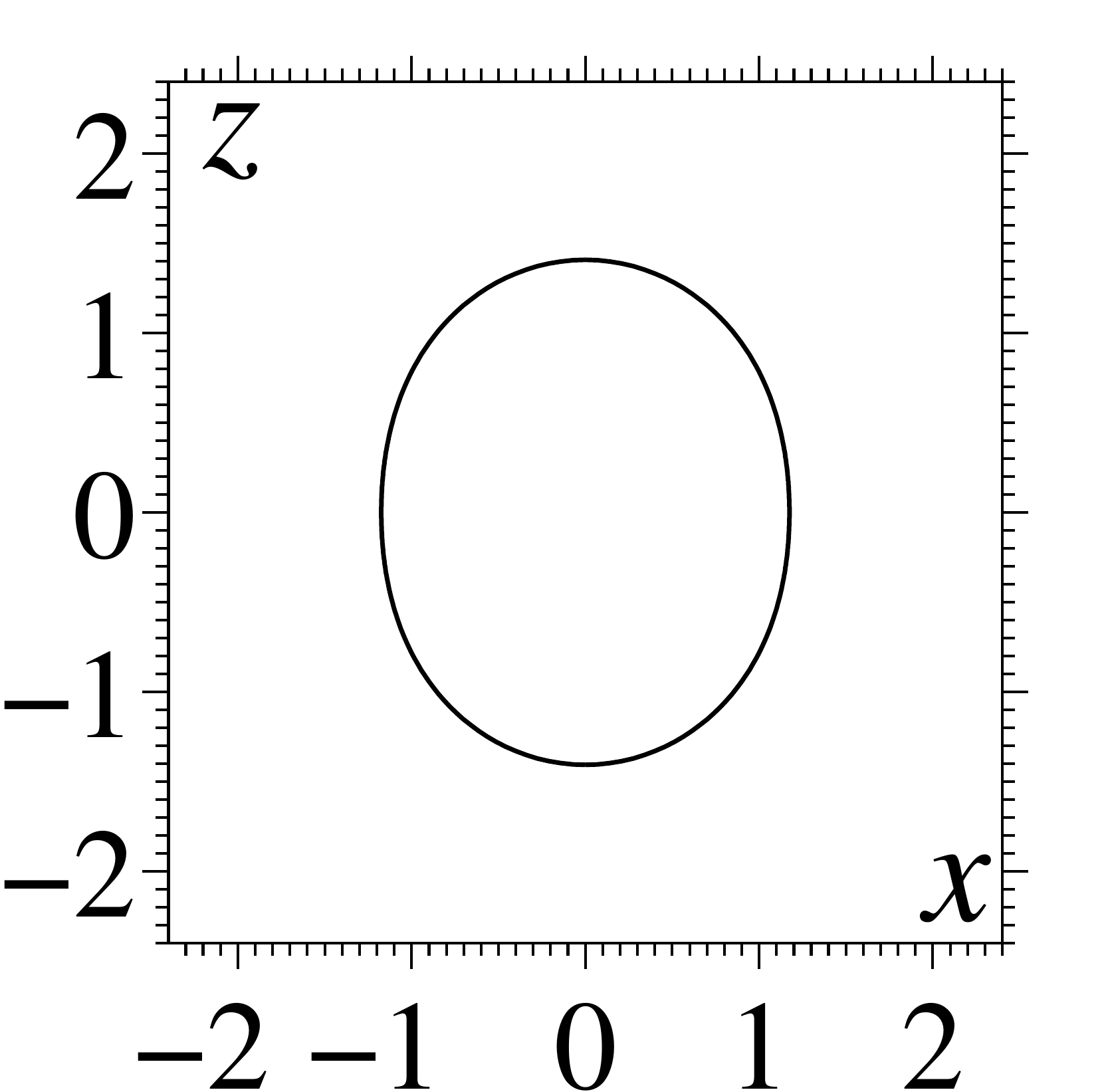}}
\subfigure[$t=4.5$]{\includegraphics[scale=0.19]{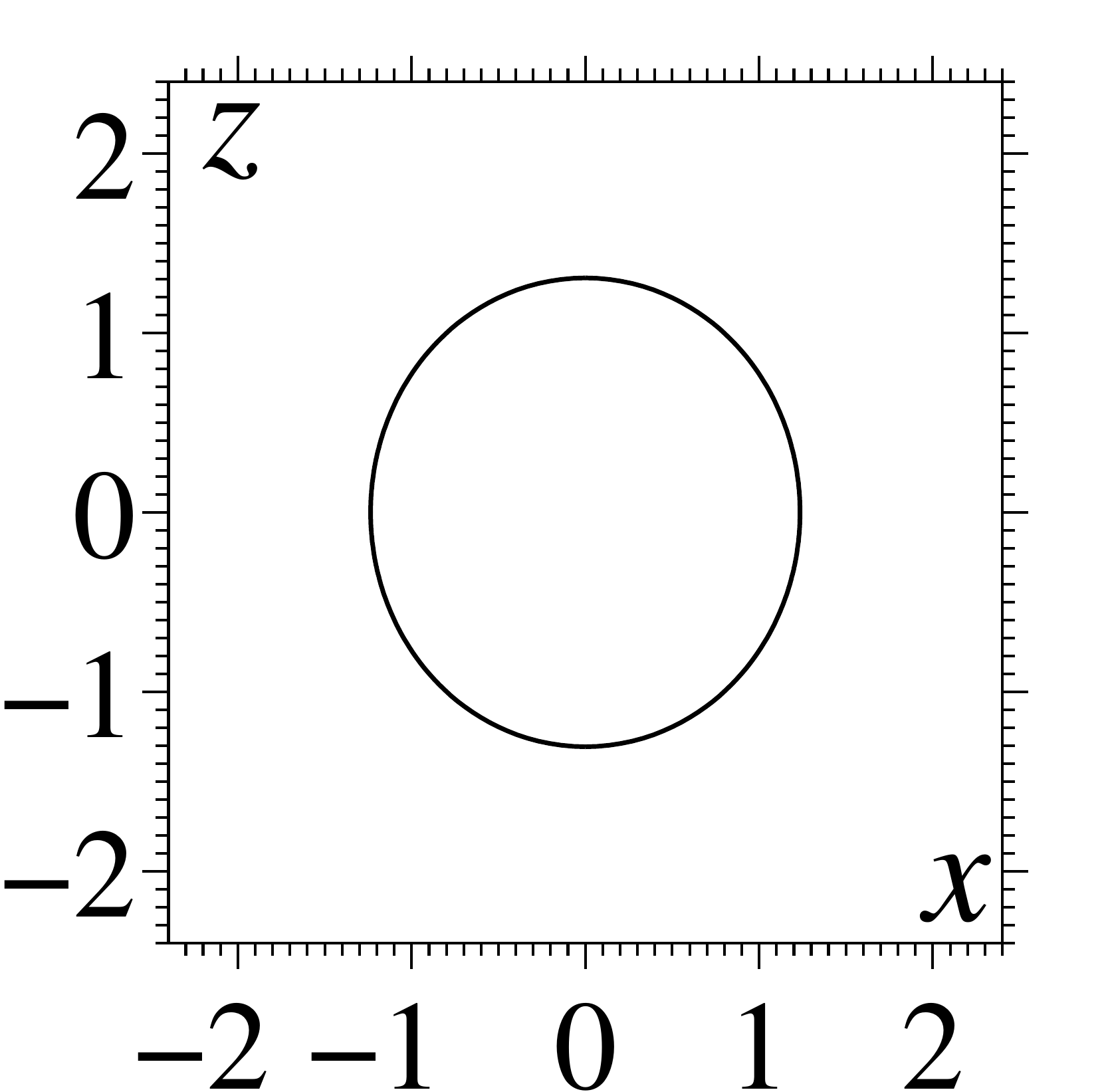}}
    \end{minipage}
 \caption{Coalescence of two high viscosity free liquid
 drops with $Re=5.8\times10^{-6}$.\label{F:58000cP_evo}}
\end{figure}

\begin{figure}
     \centering
\includegraphics[scale=0.35]{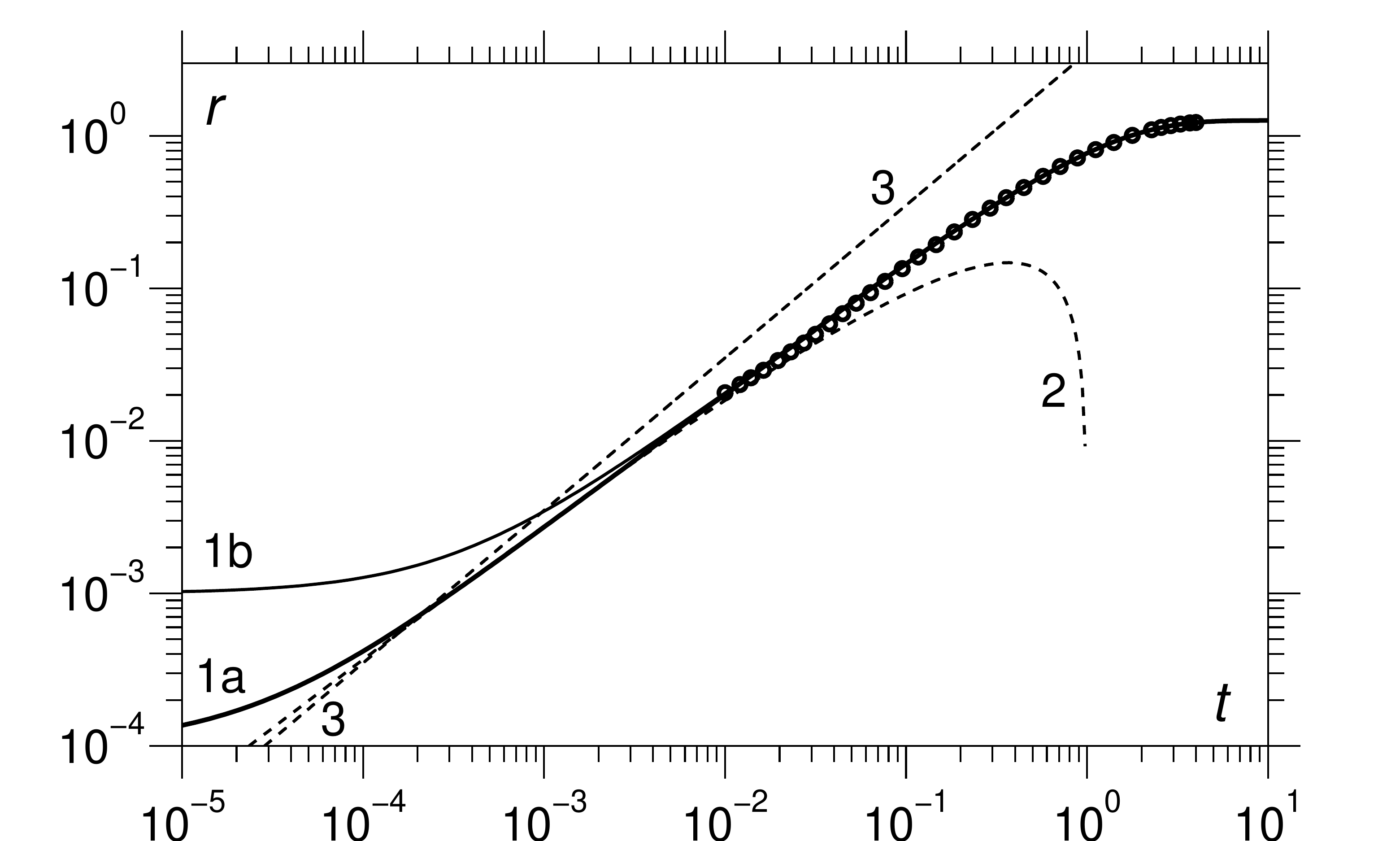}
\caption{\label{F:58000cP_cl}
 Bridge radius as a function of time
obtained using the conventional model  and scaling laws (\ref{viscous_scaling}) and (\ref{inertial_scaling}). Curve 1a:~$r_{min}=10^{-4}$; curve 1b:~$r_{min}=10^{-3}$; curve 2: best
 fit ($C_{visc}=0.4$) of the scaling law (\ref{viscous_scaling}); curve 3: best fit linear curve ($r=3.5t$); circles: the numerical solution obtained in \citep{paulsen12}
 for the same problem.}
\end{figure}

The coalescence of the two high viscosity drops
($Re=5.8\times10^{-6}$) is shown in Figure~\ref{F:58000cP_evo},
where one can observe that, as one would expect, the drops coalesce
without any oscillations. The $\log$-$\log$ plot in
Figure~\ref{F:58000cP_cl} confirms that our code is giving results
in agreement with previous computations that used the conventional
model \citep{paulsen12}. This has been tested for both
$r_{min}=10^{-4}$, curve~1a, as well as $r_{min}=10^{-3}$, curve~1b,
and we can see that both curves converge well before reaching the
circles which correspond to the results of \citep{paulsen12}. It is
interesting to note that the curves converge on to curve~2 obtained
from the viscosity-versus-capillarity scaling law, equation
(\ref{viscous_scaling}) with $C_{visc} = 0.4$, after a time of
$O(r_{min})$, i.e.\ the effect of the finite initial radius is lost
after a (dimensionless) time $r_{min}$, which is generally very short in the cases
we consider for the conventional model.  Our estimate above, based
on the extended period in which the scaling law held for a low
viscosity fluid, suggested that this law will be valid until
$Re_h=r^2 Re<1$, which in this case gives $r<10^2$, i.e.\ for the
entire period of motion. This cannot be the case as the scaling law
blows up before reaching such radii, see Figure~\ref{F:58000cP_evo},
and in fact we see once again that the scaling law agrees with the
simulation up until almost $r=t=10^{-1}$. Notably, the $t\ln t$
behaviour approximates the simulation better than the best linear
fit, curve~3, in contrast to experimental results \citep{thoroddsen05,aarts05} which suggest that the linear fit is a better one.  No
$t^{1/2}$ scaling, as predicted by the inertia-versus-capillarity
scaling law, is seen, as one would expect, and we have therefore
omitted this case from the plot.

Having confirmed that, for the conventional model, our framework is
giving results that are in agreement with previous studies into
coalescence, and having used this model to study the limitations of
the scaling laws proposed in the literature, we can now turn to a
direct comparison of the two theories, the conventional model and
the interface formation model, to recently published experimental
data.

\section{Comparison of different models to experiment}\label{experiments}

In this section, we will compare the predictions of both the
conventional model and the interface formation model with
experiments reported in \citep{thoroddsen05} and \citep{paulsen11}. In
both experimental setups, drops are formed from two nozzles and
slowly brought together until coalescence occurs.  In what follows,
we will initially consider the drops to be hemispheres of radius
$R=2$~mm, pinned at the nozzle edge from which they emanate (see
Figure~\ref{F:drop_geometries}). In the Appendix, the influence of
gravity and of the far-field flow geometry are quantified and shown
to be negligible for the initial stages of coalescence which we are
interested in, so that, for example, altering the length of the
capillary, or its inlet conditions will have no influence on our
forthcoming conclusions.

As in the experiments in \citep{paulsen11}, we consider the dynamics of water-glycerol
mixtures of density $\rho=1200$~kg~m${}^{-3}$ and surface tension
with air of $\sigma_{1e}=65$~mN~m${}^{-1}$ for a range of
viscosities $\mu=3.3,~48,~230$~mPa~s, which are chosen as some of the cases
where $\sigma_{1e}$ and $\rho$ vary least\footnote{We can see this
from the data provided to us by Dr J.D.~Paulsen, Dr J.C.~Burton and Professor S.R.~Nagel, which was published
in \citep{paulsen11}.}, giving $Re=1.4\times10^4,~68,~2.9$.  The dependence of the interface formation model's parameters on surface tension and drop radius are
\begin{align}\label{base_parameters}
Q=q_1\sigma^{-1},\quad \epsilon = q_2\sigma R^{-1} ,\quad \bar{\beta} = q_3 R,\quad A = 1,\quad \rho^s_{1e}=0.6,\quad \lambda = (1-\rho^s_{1e})^{-1},
\end{align}
and estimates for the dimensional constants, the $q$'s, for water-glycerol mixtures have been obtained from experiments on dynamic wetting\citep{blake02} as $q_1 = 3\times10^{-4}$~N~m$^{-1}$, $q_2=7\times10^{-6}$~N$^{-1}$~m$^2$ and $q_3=5\times10^{8}$~m$^{-1}$.

Fortuitously, at the highest viscosity we can also compare our
results to those in \citep{thoroddsen05} where the same liquid
mixture was used \footnote{Despite the drops in \citep{thoroddsen05}
having a different radius, $R=1.5$~mm, and very slightly different
viscosity, $\mu=220$~mPa~s, our simulations show that at such high
viscosity these alterations can be scaled out by an appropriate
choice of viscous time-scale, as we use.}.  Furthermore, at the
highest viscosity there are no complications from toroidal bubbles
and the viscosity ratio between the liquid and surrounding air is
large, so that all influences on the coalescence dynamics additional
to those considered, such as the dynamics of the gas, are
negligible.  In other words, this is the perfect test case for a
comparison between the conventional model, the interface formation
model and experimental data.

\begin{figure}
     \centering
     \begin{minipage}[l]{.99\textwidth}
\subfigure[t=0]{\includegraphics[scale=0.19]{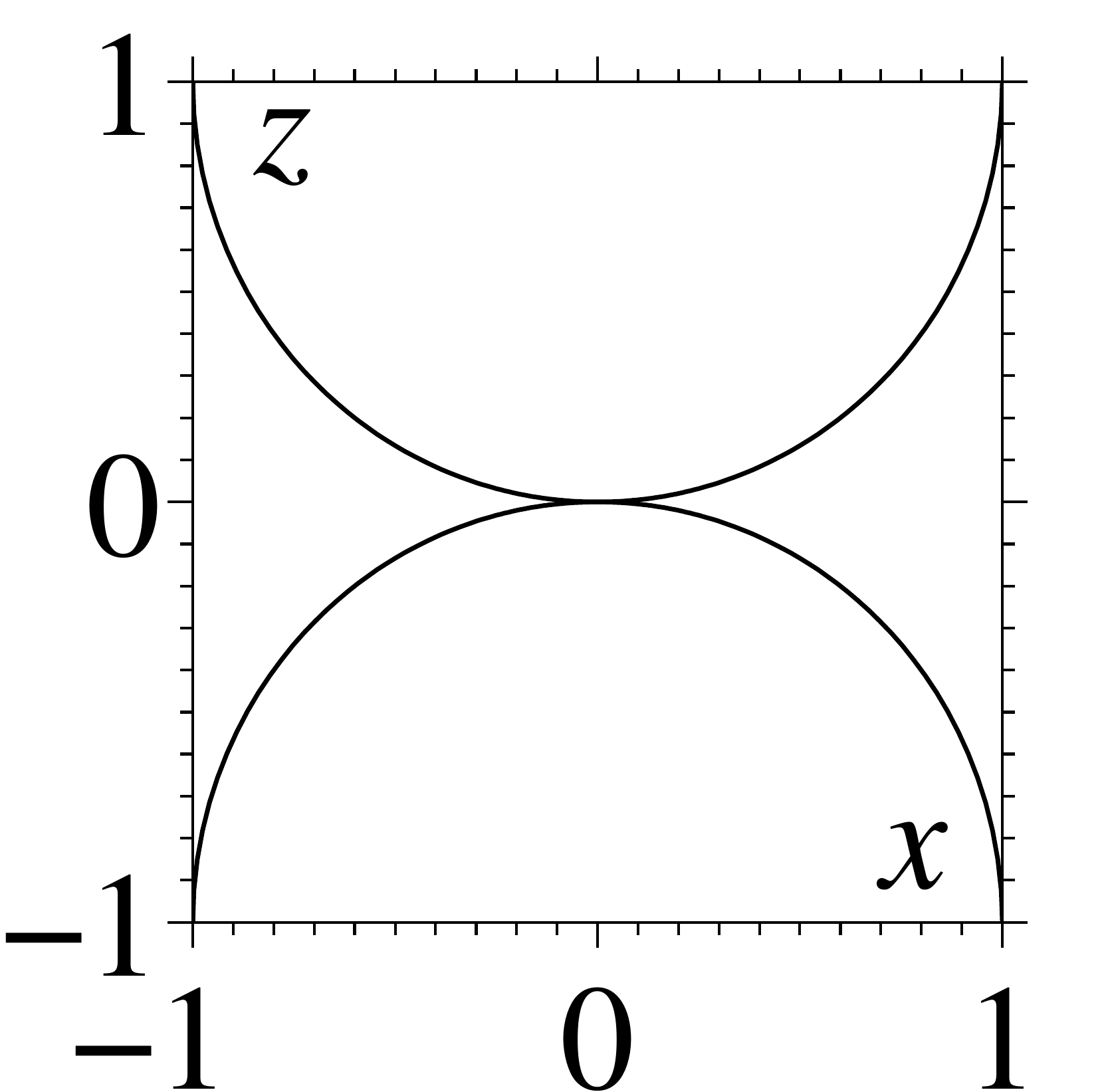}}
\subfigure[t=0.5]{\includegraphics[scale=0.19]{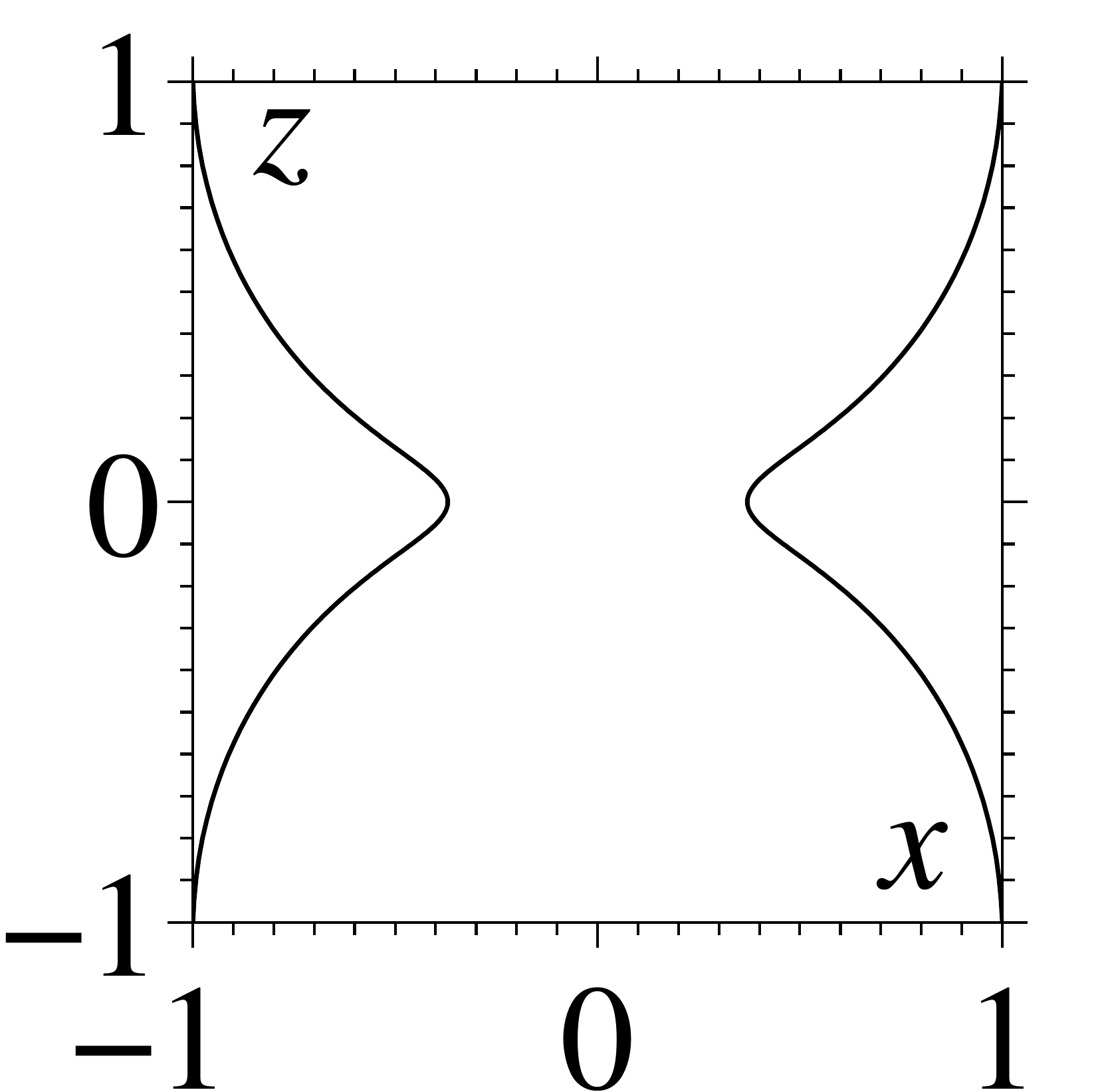}}
\subfigure[t=1]{\includegraphics[scale=0.19]{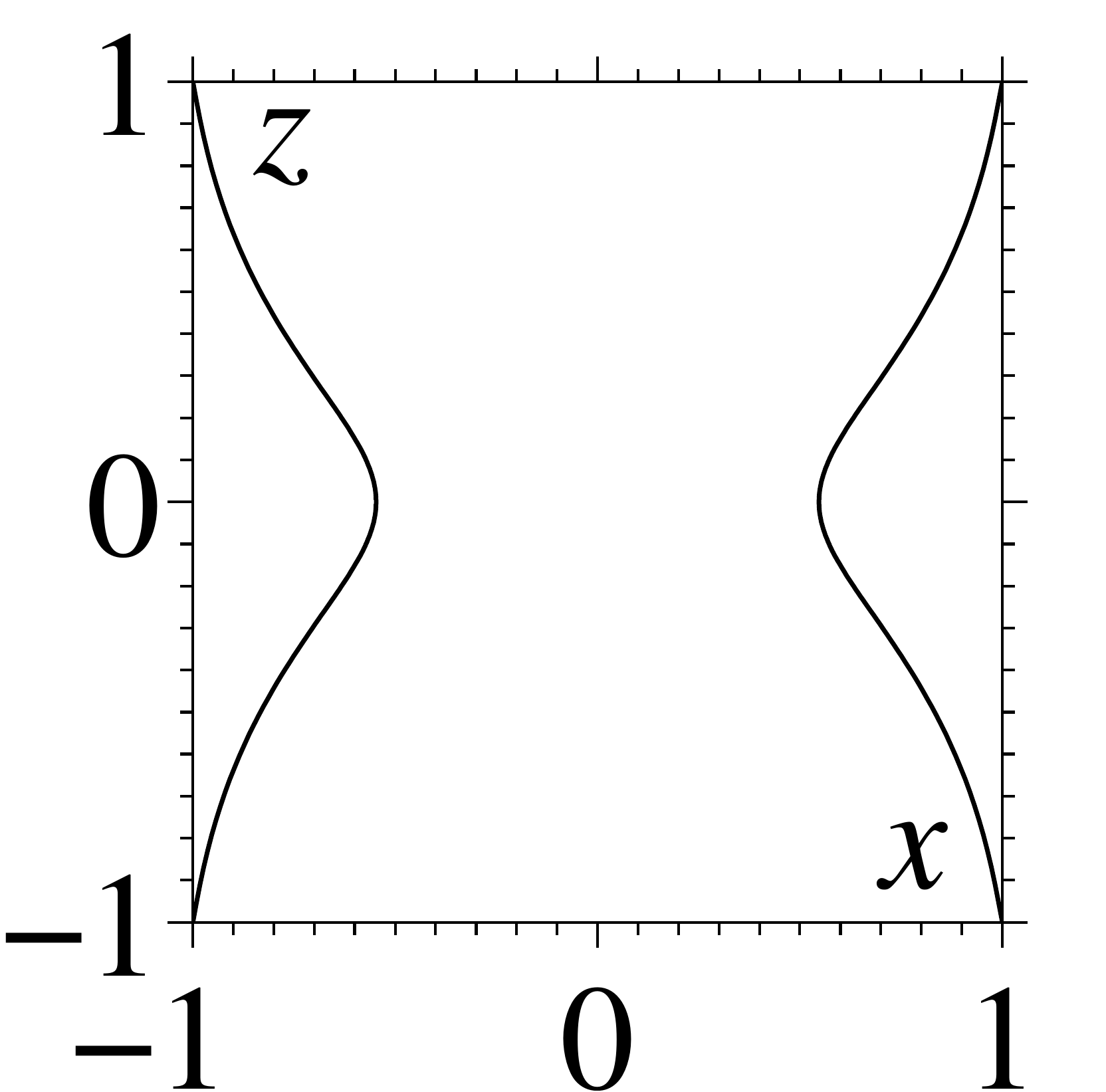}}
\subfigure[t=4]{\includegraphics[scale=0.19]{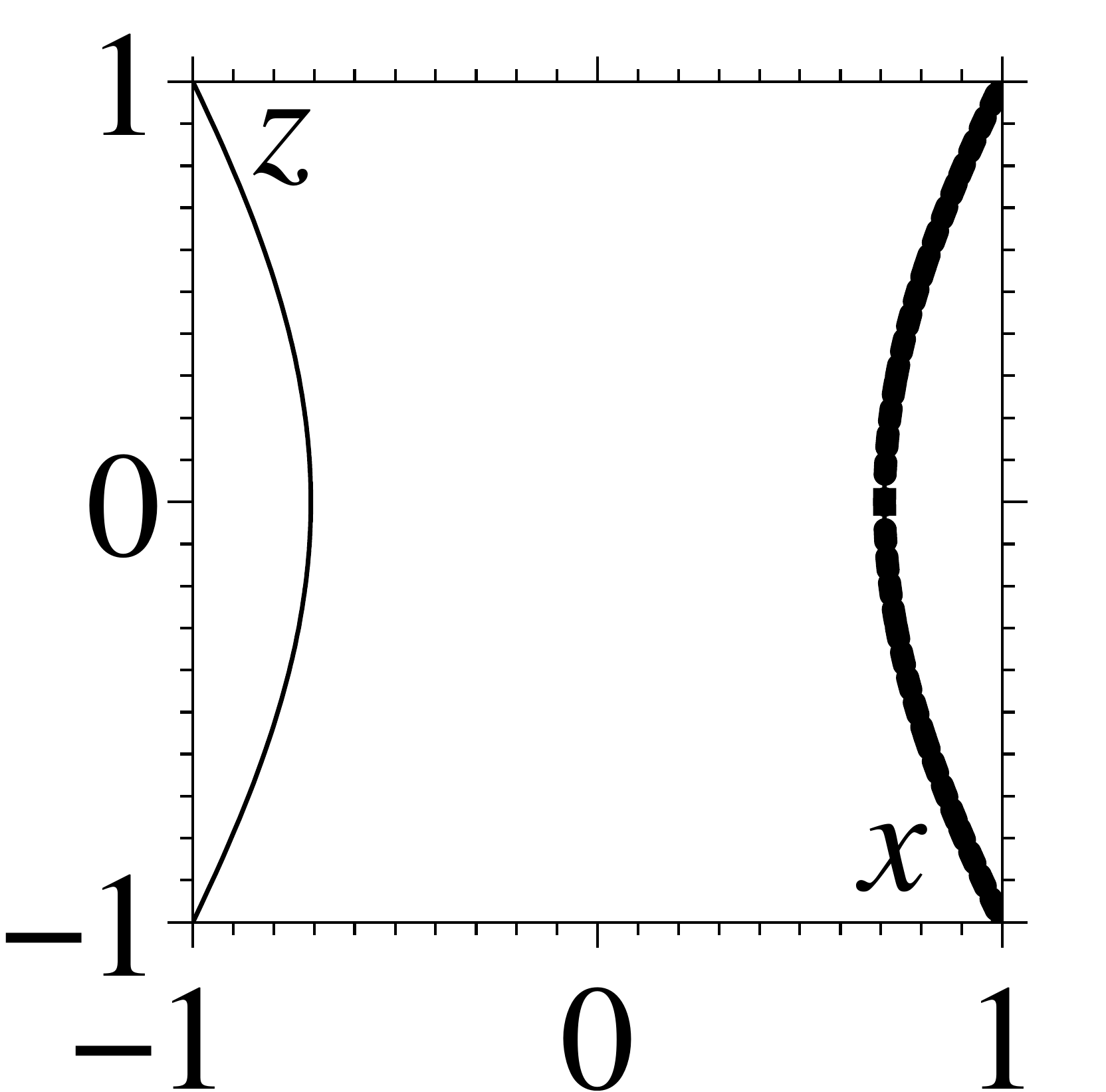}}
    \end{minipage}
\caption{Coalescence of two pinned hemispheres with $Re=2.9$, with the independently obtained equilibrium position shown by a dashed line. \label{F:230cP_evo}}
\end{figure}

Notably, in contrast to the coalescence of two free liquid drops,
where the final stage of the process is one spherical drop of the
combined volume, the equilibrium shape of the two coalescing
hemispheres pinned at the capillary edge is no longer analytically
calculable. So, a simple code was written to solve for the static
equilibrium shapes of the drops using the approach outlined in
\citep{fordham48}. In Figure~\ref{F:230cP_evo}, snapshots from the
coalescence event are shown and, critically, it can be seen that our
simulations predict the correct equilibrium shape.  On this scale,
there is seemingly little difference between the two models'
predictions, as one would expect given that the two equilibrium
shapes are the same. To access verifiable differences between the
models and to compare the results with the experiments, we now
consider the initial stages of the coalescence process.

\begin{figure}
     \centering
\includegraphics[scale=0.35]{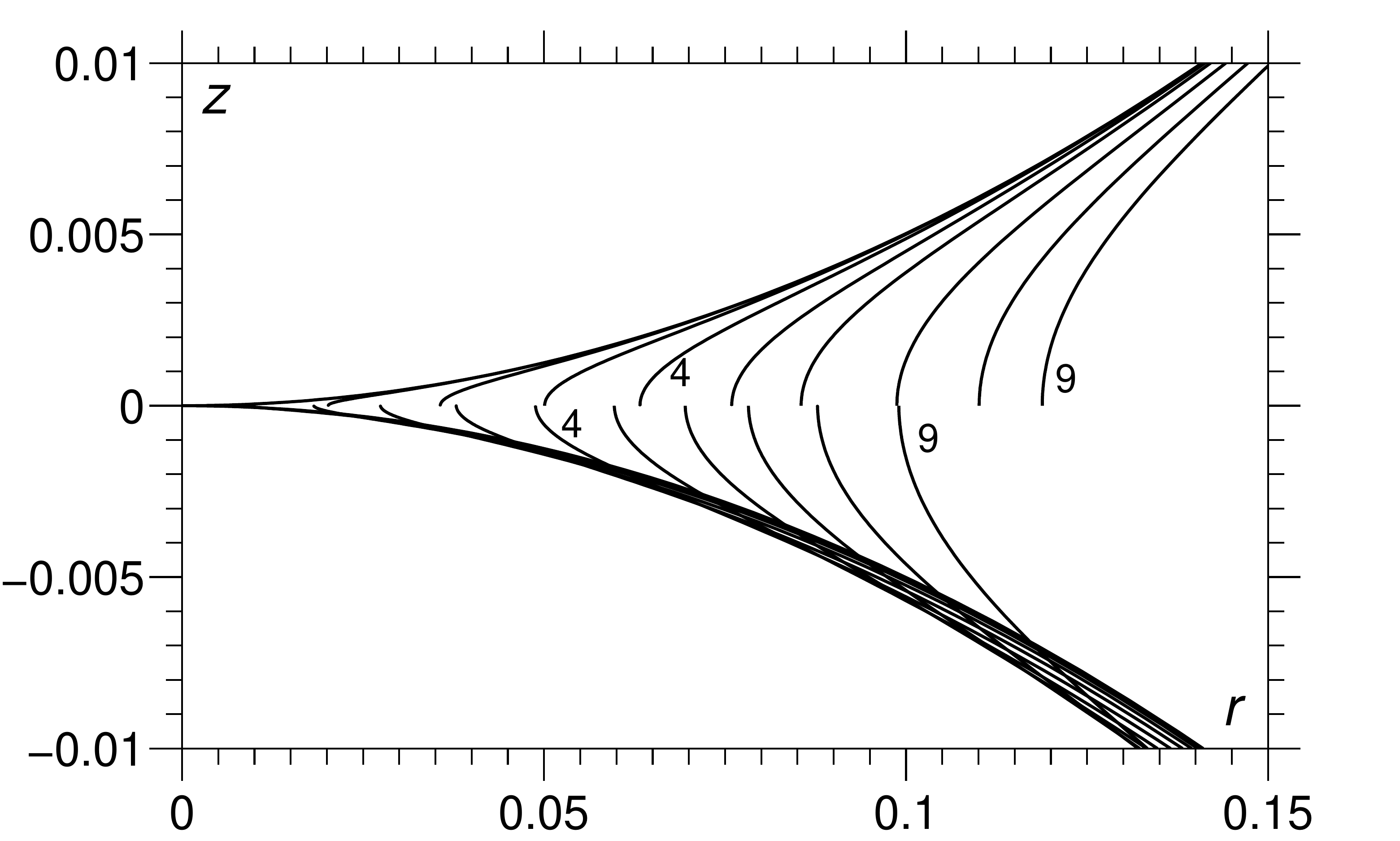}
\includegraphics[scale=0.35]{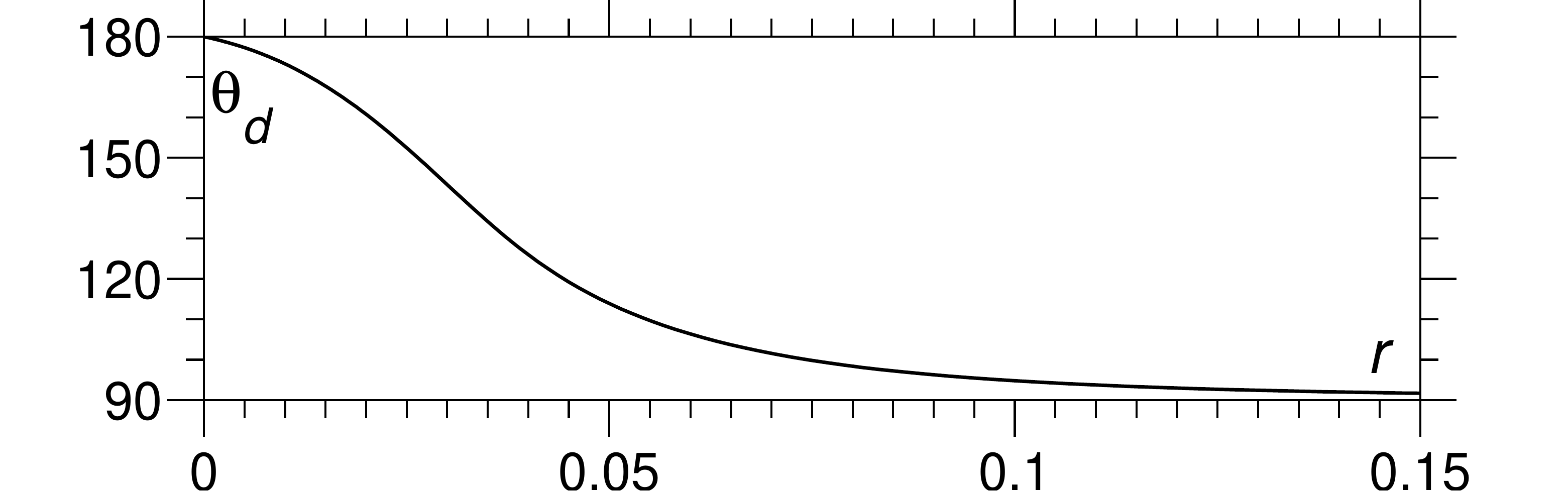}
\caption{\label{F:230cP_fs} Top: Comparison of the free surface profiles for the initial stages of coalescence for the highest viscosity fluid ($Re=2.9$) obtained  using the conventional model (upper curves) and the interface formation model (lower curves).  Snapshots are taken every $\triangle t = 10^{-2}$ so that curves 4 are at $t=0.04$ and curves 9 are at $t=0.09$.  Bottom: Contact angle at which the free surface meets the plane of symmetry for the interface formation model.}
\end{figure}

In Figure~\ref{F:230cP_fs}, we show the free-surface profiles
obtained from our simulations using the two different models. In the
initial stages of coalescence, one can see that the conventional
model (upper curves) predicts a faster motion than that given by the
interface formation model (lower curves). As can also be seen from
Figure~\ref{F:230cP_fs}, the contact angle predicted by the
interface formation model takes a finite time to evolve and
establish the smooth free surface.  This time period is associated
with $\theta_d>90^\circ$, and only towards the end of the evolution of the free
surface shown in the figure the contact angle approaches $90^\circ$,
indicating that the physics embodied in the conventional model can
take over. This gradual evolution of the contact angle results in a
slower motion in the initial stages than that predicted by the
conventional model where, as we know, the initial velocity, driven
by a region of extremely high curvature and hence high capillary
pressure, is huge. As we shall see, this difference between the
models' predictions will reduce as time from the onset of the
process passes and the two drops evolve towards the same equilibrium
position. Therefore, it is the initial stages of the evolution, such
as those shown in Figure~\ref{F:230cP_fs}, that the discrepancies
between theory and experiment will be most easily picked up.

\begin{figure}
     \centering
\includegraphics[scale=0.35]{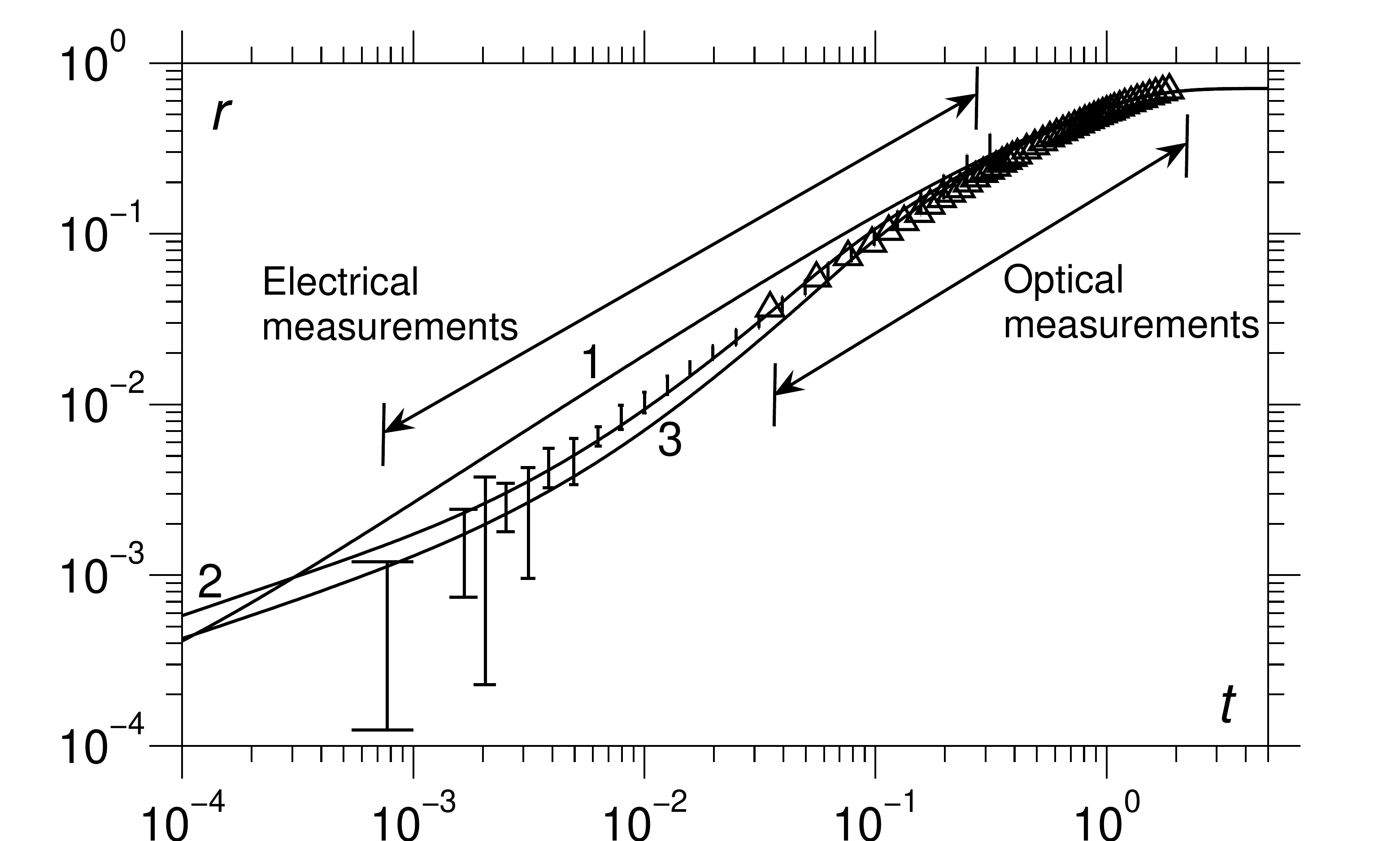}
\caption{\label{F:230cP_cl} Bridge radius as a function of time for viscosity $\mu=220$~mPa~s ($Re=2.9$) obtained using different models compared to experiments from \citep{paulsen11} (with error bars) and Figure~17b in \citep{thoroddsen05} (triangles). Curve 1: the conventional model; curve 2: the interface formation model with $\rho^s_{1e}=0.2$ and other parameters from (\ref{base_parameters}); curve 3: the interface formation model with parameters from (\ref{base_parameters}).}
\end{figure}

The bridge radius as a function of time for the highest viscosity
($Re=2.9$) is given in Figure~\ref{F:230cP_cl}, which shows a comparison
between the two models' predictions and the experimental data from
in \citep{thoroddsen05} and \citep{paulsen11}.  In
particular, the initial time of coalescence in the optical
experiment of \citep{thoroddsen05}, which is known to be uncertain,
is chosen such that one has an overlap with the data of the
electrical experiments of \citep{paulsen11}, where the initial time
was more accurately determined.

It is immediately apparent that the bridge radius predicted by the
conventional model overshoots the experimental values of both
studies for a considerable amount of time.  For the interface
formation model, using parameters (\ref{base_parameters}), we obtain
curve $3$. Alteration of any of these parameters is seen to
result in a worse agreement with experiment with the exception of
the parameter $\rho^s_{1e}$, which is the equilibrium surface
density on the free surface. Decreasing its value to
$\rho^s_{1e}=0.2$ gives curve $2$, which goes through all of the
error bars and data points except for the very first one. As one
would expect, all the curves coincide as the equilibrium position is
approached and both agree with the optical experiments in these
final stages of evolution.

\begin{figure}
 \centering
 \includegraphics[scale=0.35]{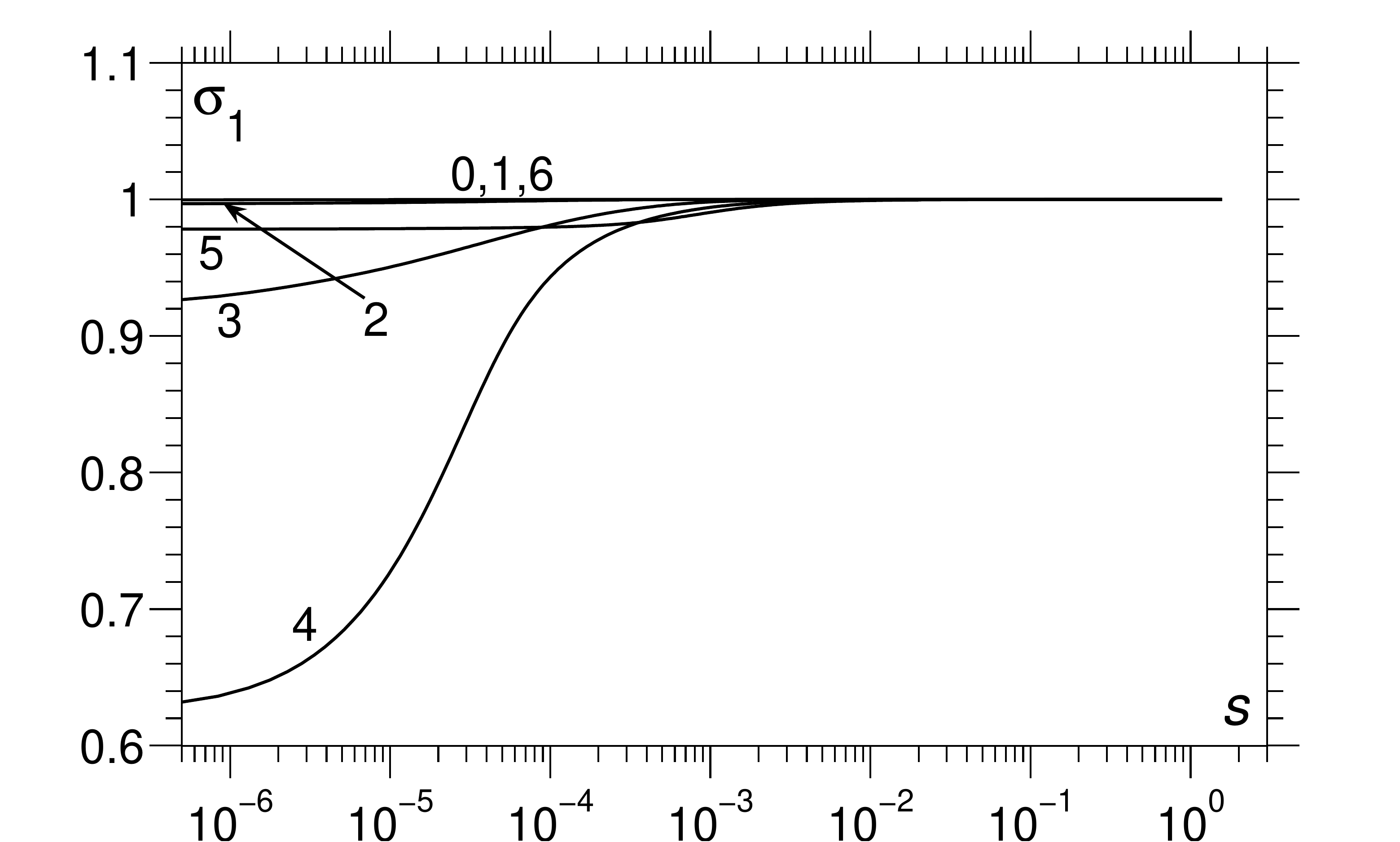}
 \includegraphics[scale=0.35]{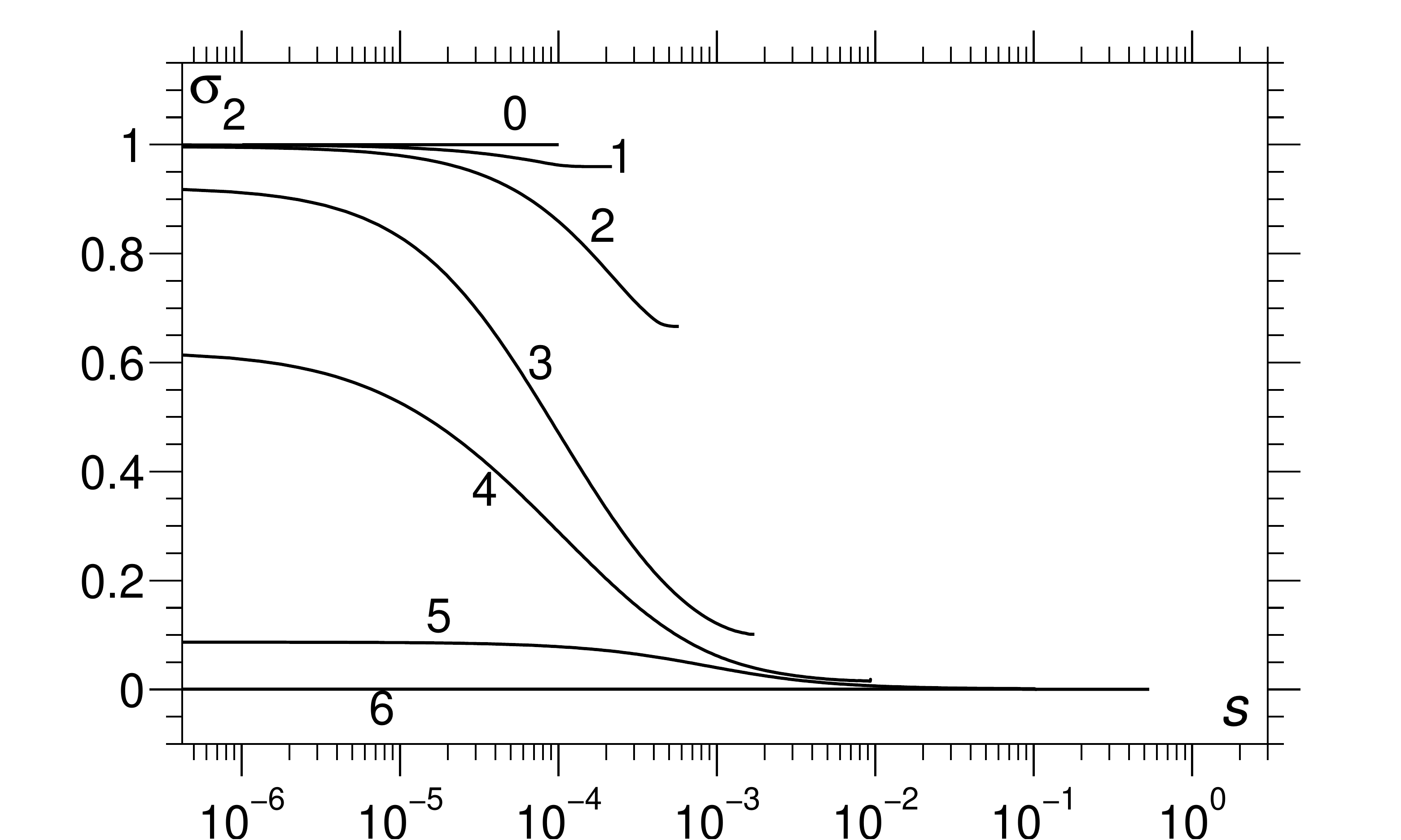}
 \caption{\label{F:sigma_evo} Evolution of the surface tension
 distributions along the free surface, $\sigma_1$, and internal
 interface, $\sigma_2$, as a function of distance from the contact line $s$ at times 0: $t=0$; 1: $t=10^{-5}$; 2: $t=10^{-4}$; 3:
 $t=10^{-3}$; 4: $t=10^{-2}$; 5: $t=10^{-1}$; 6: $t=1$, for the highest
 viscosity fluid ($Re=2.9$), obtained using the interface formation
 model.}
 \end{figure}

In Figure~\ref{F:sigma_evo}, the distributions of the surface tensions
along both the free surface and the internal interface are shown at
different instances through the simulation. Notably, although the free surface is
in equilibrium ($\sigma_1=1$) both initially and at the end of the coalescence
process when one has a single body of fluid confined by a smooth free surface,
as the interface formation dynamics unfolds ($t>0$),
the surface tension distribution near the contact line becomes driven away from
equilibrium, with, in particular, $\sigma_1 = 0.63$ at the contact line when
$t=10^{-2}$, which is not far away from its minimum value of $\sigma_1=0.61$ reached
at $t=0.017$. As can be seen from Figure~\ref{F:230cP_fs}, it is at this time that the contact angle
rapidly decreases from its initial value of $\theta_d=180^\circ$, imposed by
the initial conditions, to its equilibrium value of $\theta_d=90^\circ$, which it
is close to achieving by $t=10^{-1}$. Consequently, the behaviour of
$\sigma_1$ is non-monotonic in time, with an initial decrease in its distribution near
the contact line followed by a relaxation back towards its equilibrium state.
As one would expect, when there is a separation of length scales between the drop
radius and the length scale of interface formation, the surface tension on the free surface
far away from the contact line, roughly $s > 10^{-2}$, remains in its
equilibrium state throughout the coalescence process. However, the
internal interface, which has length $s=10^{-4}$ at the start of the
simulation, is comparable with the length scale on which the interface
formation model acts, and hence, as one can see from Figure~\ref{F:sigma_evo},
it takes a finite time for the interface to form, and for this interface there is no `far-field'
where the interface is in equilibrium until around $t=10^{-1}$, at which point the length of the internal interface has increased significantly.

In our comparison of the two models with experiments, the following
two aspects can be highlighted. Firstly, it is apparent that the
conventional model considerably overpredicts the speed at which
coalescence occurs. This is consistent with the fact that this model
introduces unphysical singular velocities at the start of the
process, as the cusp in the free surface shape is instantaneously
rounded. In our computations, this unphysicality is moderated by our
use of the zero velocity initial condition (\ref{ic-u}) but the
influence of this initial condition quickly dies out, and one ends
up with the rate of the widening of the bridge connecting the two drops
well above what is actually observed. In contrast, the interface
formation model predicts that the angle at which the free surface
meets the plane of symmetry will relax from its initial value of
$180^\circ$ to its eventual value of $90^\circ$ gradually, over some
characteristic time scale. In Figure~\ref{F:230cP_fs}, this
behaviour is observed, where the angle remains high for a
considerable amount of time, being greater than $170^\circ$ until
$t=10^{-2}$, gradually relaxing to $90^\circ$ and reaching
this value at around $t=10^{-1}$. What is unexpected, is that the
non-dimensional relaxation time of the interface
$\tau_{nd}=\tau/(R\mu/\sigma)=\tau_{\mu}\sigma/R=O(10^{-4})$ is not
a good approximation for the period in which the interface is out of
equilibrium, i.e.\ the free surface is not smooth; in fact, the time
scale over which interface formation acts is much larger, which
suggests that the influence of these effects could extend outside
the parameter space previously identified.

The second aspect, which is perhaps more important, is the trends
observed in experiments and predicted by the two models. In
experiment and in what the interface formation model predicts, one
can see what looks like two different regimes, roughly corresponding, coincidentally,
to the ranges of the electric and optical measurements, whereas the
conventional model describes the process as `more of the same', with
no qualitative difference between the early stages of the process
and the subsequent dynamics. This is consistent with the fact that
the conventional model assumes that coalescence as such occurs
instantly, resulting in a single body of fluid whose subsequent
evolution can be described in the standard way, as in the drop
oscillation problem, whilst the interface formation model suggests
that the formation of a single body of fluid is the result of a {\it
process\/} and hence presumes that this process has a dynamics
different from that of the drop oscillations. These differences
between the two models can be of great significance, for example,
for the modelling of microfluidics, and they indicate a promising
direction of experimental research.


\section{The influence of viscosity}\label{viscosity}

\begin{figure}
     \centering
     \includegraphics[scale=0.35]{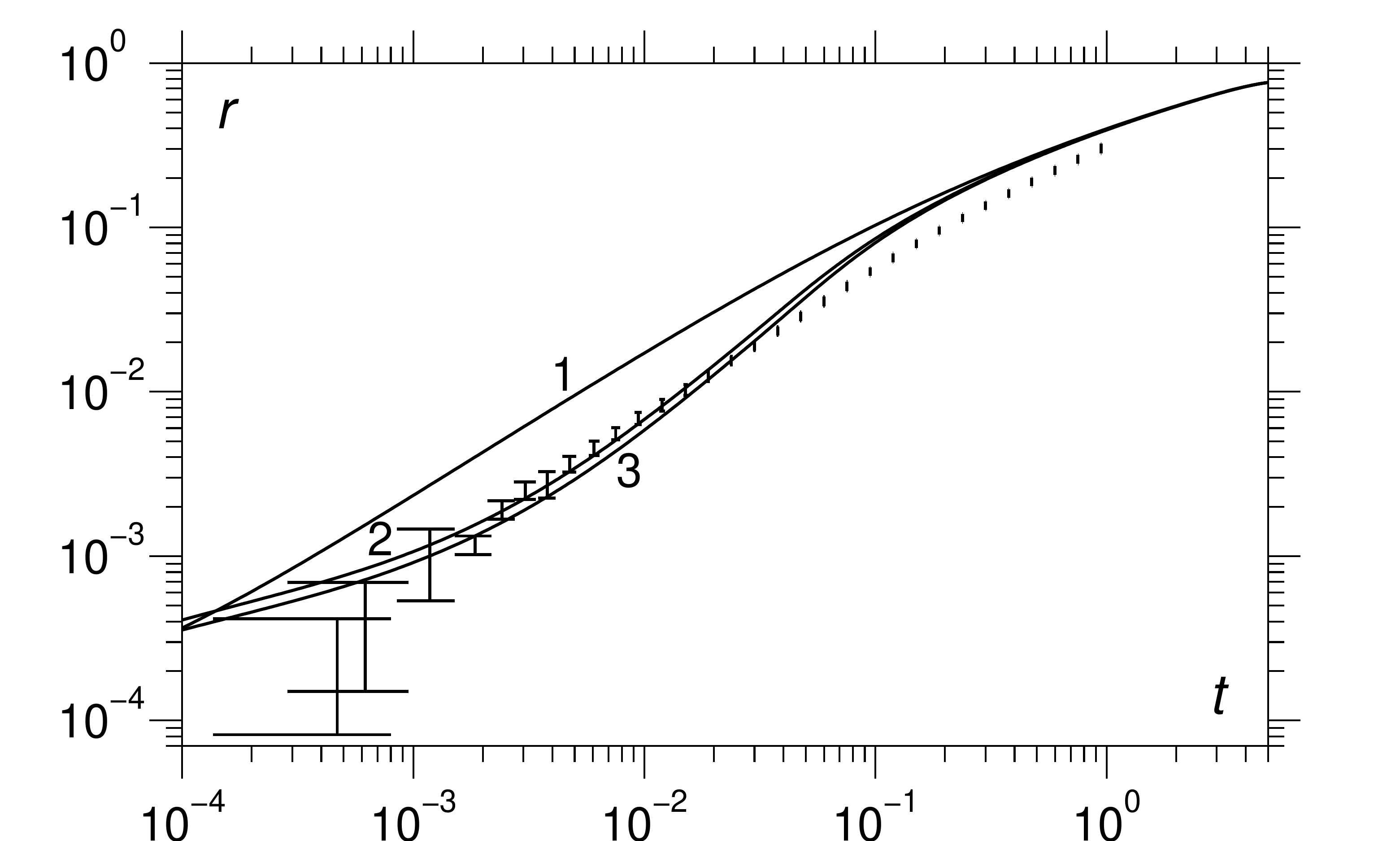}
\caption{\label{F:48cP_cl} Bridge radius as a function of time for viscosity $\mu=48$~mPa~s ($Re=68$) obtained using different models compared to experiments from \citep{paulsen11} (with error bars). Curve 1: the conventional model; curve 2: the interface formation model with $\rho^s_{1e}=0.45$ and other parameters from (\ref{base_parameters}); curve 3: the interface formation model with parameters from (\ref{base_parameters}).}
\end{figure}

\begin{figure}
     \centering
\includegraphics[scale=0.35]{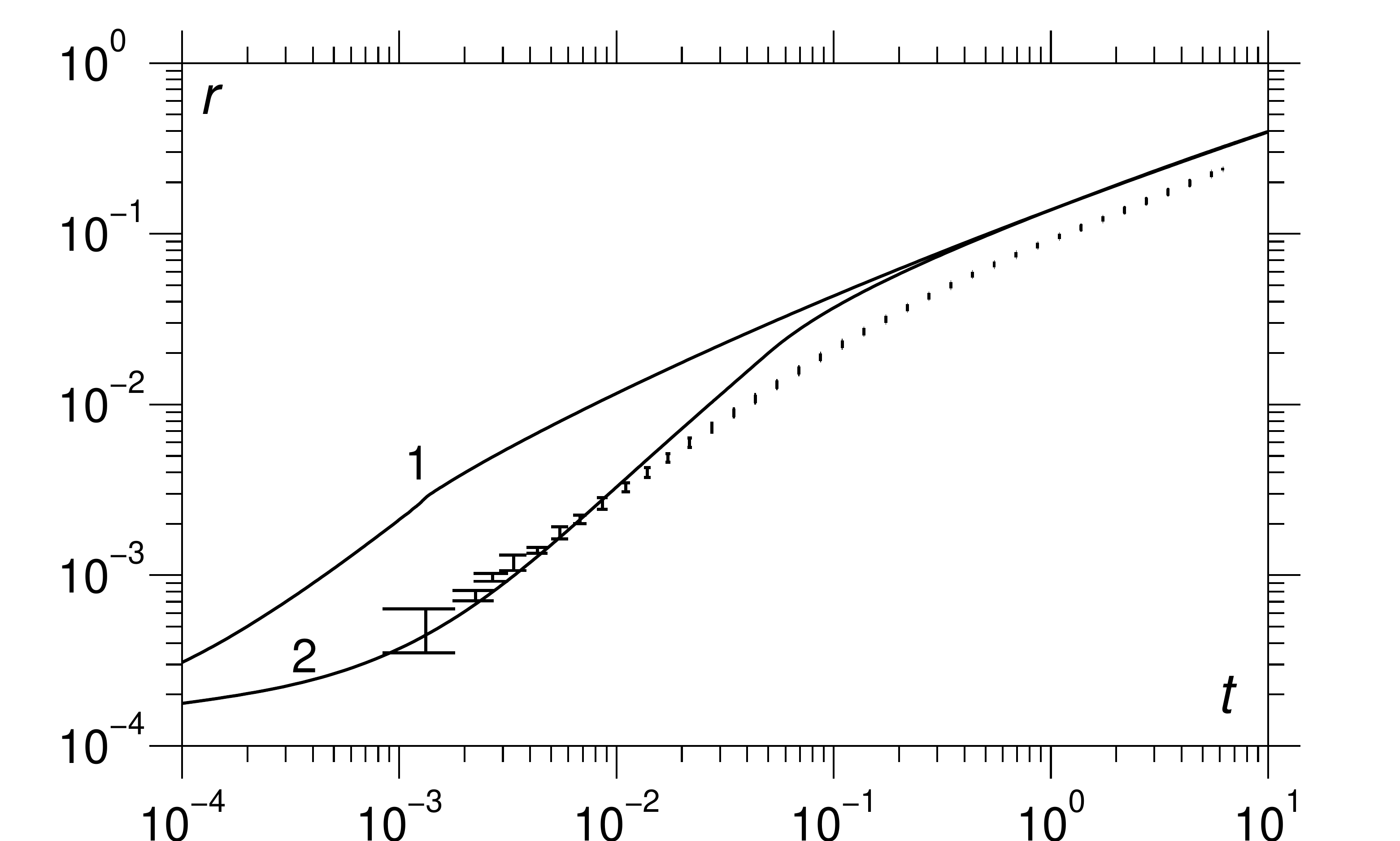}
\caption{\label{F:3p3cP_cl} Bridge radius as a function of time for viscosity $\mu=48$~mPa~s ($Re=1.4\times 10^4$) obtained using different models compared to experiments from \citep{paulsen11} (with error bars). Curve 1: the conventional model and curve 2: the interface formation model with parameters from (\ref{base_parameters}).}
\end{figure}

In Figures~\ref{F:48cP_cl} and \ref{F:3p3cP_cl} the influence
of decreasing the fluid's viscosity is explored by computing curves
for the $Re=68$ and $Re=1.4\times 10^4$ cases, respectively.  In both
figures, the interface formation model provides a considerably
better approximation of the initial stages of the drops' evolution.
In the $Re=68$ case we see a slightly better agreement with the
experimental data by taking $\rho^s_{1e}=0.45$ whilst little
improvement is achieved by altering any of the parameters for the
lowest viscosity.  Notably, it is apparent that the curves provided
by both models deviate from the experimental results at later times,
with a more significant error seen at lower viscosities. Given that
the predictions of the two theories have begun to coincide, this is
the region in which the interface formation is completed, so that
the surface parameters take their equilibrium values and the free
surface is smooth.  In other words, in terms of the interface
formation model, this deviation corresponds to the period after
coalescence has happened, a single body of fluid formed and it is
the physics incorporated in the conventional model that determines
the subsequent dynamics.

The deviation of both theories\footnote{More precisely, it is the conventional model as at this stage the interface formation/disappearance dynamics have ended, so that the interfaces are in equilibrium and the interface formation model becomes equivalent to the conventional one.} from experiment in the later stage of
the process seen in Figures~\ref{F:48cP_cl} and \ref{F:3p3cP_cl} cannot, as shown in the Appendix, be attributed to the
influence of gravity deforming the drops' shape, or to an incomplete
description of the overall flow geometry; these effects only
influence the drops' evolution on an even longer time scale.
Therefore, it seems most likely that the additional resistance to
the drops' motion near the bridge is coming from the influence of
air, which begins to resist the bridge's propagation more as the
radius of the bridge, i.e. the surface area of the bridge region,
increases. This is consistent with the fact that the deviation
becomes more pronounced as the air-to-liquid viscosity ratio
increases, i.e.\ the liquid's viscosity goes down. This effect only
kicks-in during the mid-stages of the drops' evolution, so that our
conclusions about the initial stages are not affected.  An
investigation into the role played by the ambient air in the process
of what is, strictly speaking, the post-coalescence evolution of a
strongly deformed single body of fluid is of considerable interest
and will be the subject of future research. One also might be
interested in proposing a new scaling law for this effect to provide
a simplified analytic description that could be validated by the
full numerical solution.


\section{Theory-guided experiments}\label{theory_driven}

\subsection{Free-surface shape}

Having found from the interface formation model that the
free-surface shape is non-smooth for a considerable
amount of time, it is reasonable to ask why this has not been
reported from experiments and how this effect can be brought to light. Taking the largest viscosity used in these experiments
($230$~mPa~s), we see from Figure~\ref{F:230cP_fs} that the contact
angle varies over the time period $10^{-2}<t<10^{-1}$ during which,
from Figure~\ref{F:230cP_cl}, the bridge radius varies in the range
$10^{-2}<r<10^{-1}$.  In other words, whilst the bridge evolves from
around $1$\% to $10$\% of the drop's total radius, the free-surface
profile is non-smooth.  As can be seen in Figure~\ref{F:230cP_cl},
some data points from the experiments of \citep{thoroddsen05} exist
in this regime, so that, in principle, this regime is within the
range of optical experiments. In fact, in \citep{thoroddsen05}, it is
noted (see their Figure~20) that, as the viscosity increases, for a
given bridge radius ($280$~$\mu$m) the curvature of the bridge's profile increases
rapidly as a function of viscosity. This is based on fitting circles
to the free surface images to extract a radius of curvature, a
process which (a) \emph{presumes} that the free surface is smooth and (b)
involves, as the authors admit, ``some subjectivity''. In fact, our
results obtained in the framework of the interface formation model
suggest that, for the highest viscosity which we consider, when
$r_{dim}=2.8\times 10^{-4}$~m, so that $r \approx 10^{-1}$, the free
surface will indeed be almost smooth. However, if instead one
considers $t=r=0.04$, which corresponds to a dimensional bridge
radius of $80$~$\mu$m, then our results suggest that the contact
angle should be measurable, at around $115^\circ$. This is
apparently within the optical range. Furthermore, if one goes to
higher viscosities, there is the possibility of making the contact
angle even more pronounced.

It is interesting to note from Figure~\ref{F:230cP_fs} that, when the
angle is already not too large, i.e.\ $\theta_d<120^\circ$, the
profiles obtained using the interface formation model (lower curves)
do not look very sharp where they meet $z=0$, and one can easily see
how, without allowing for the possibility that the free surface can
be non-smooth, these angles could easily be attributed to the errors
associated with the optical resolution.

\begin{figure}
     \centering
\includegraphics[scale=0.35]{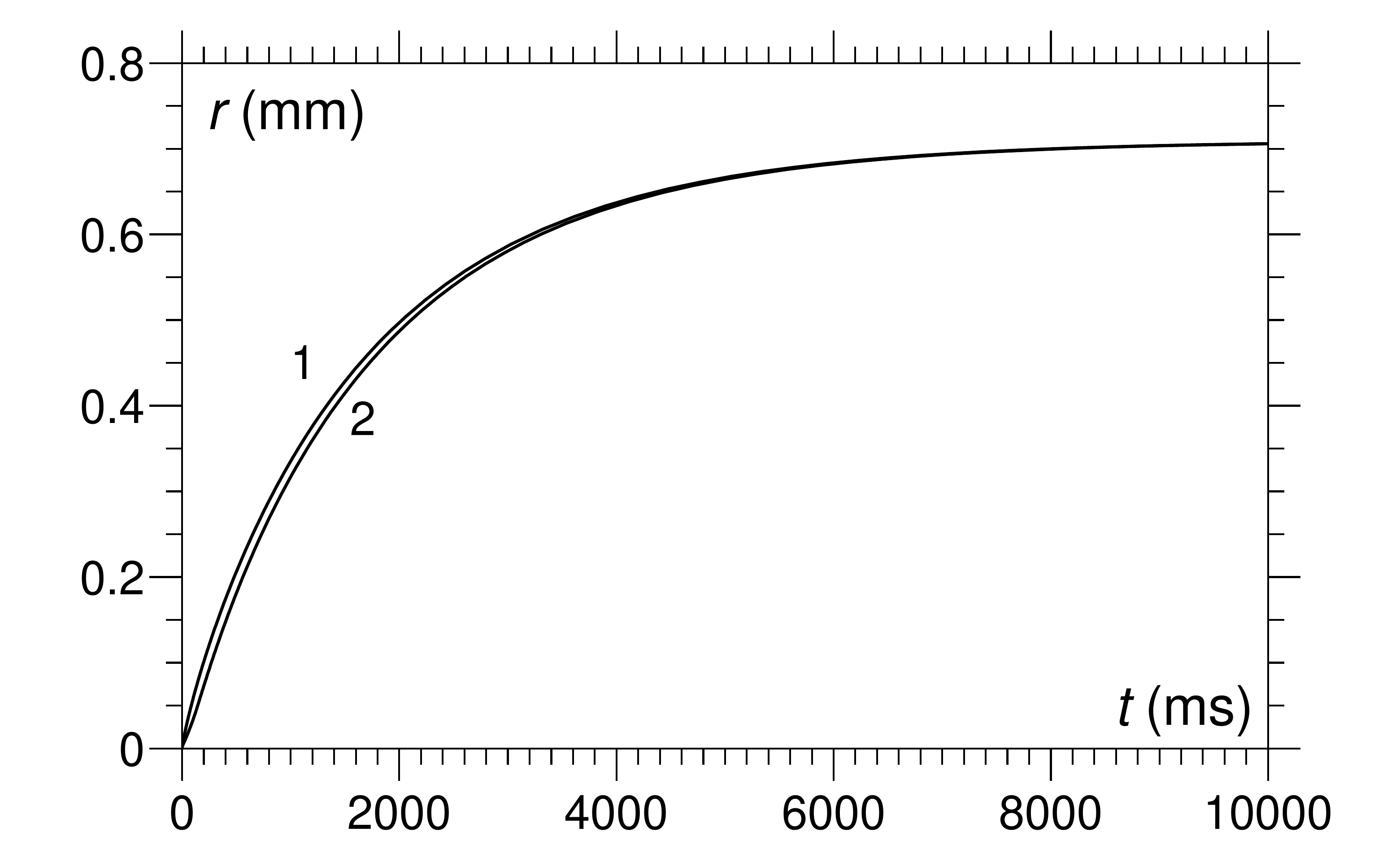}
\includegraphics[scale=0.35]{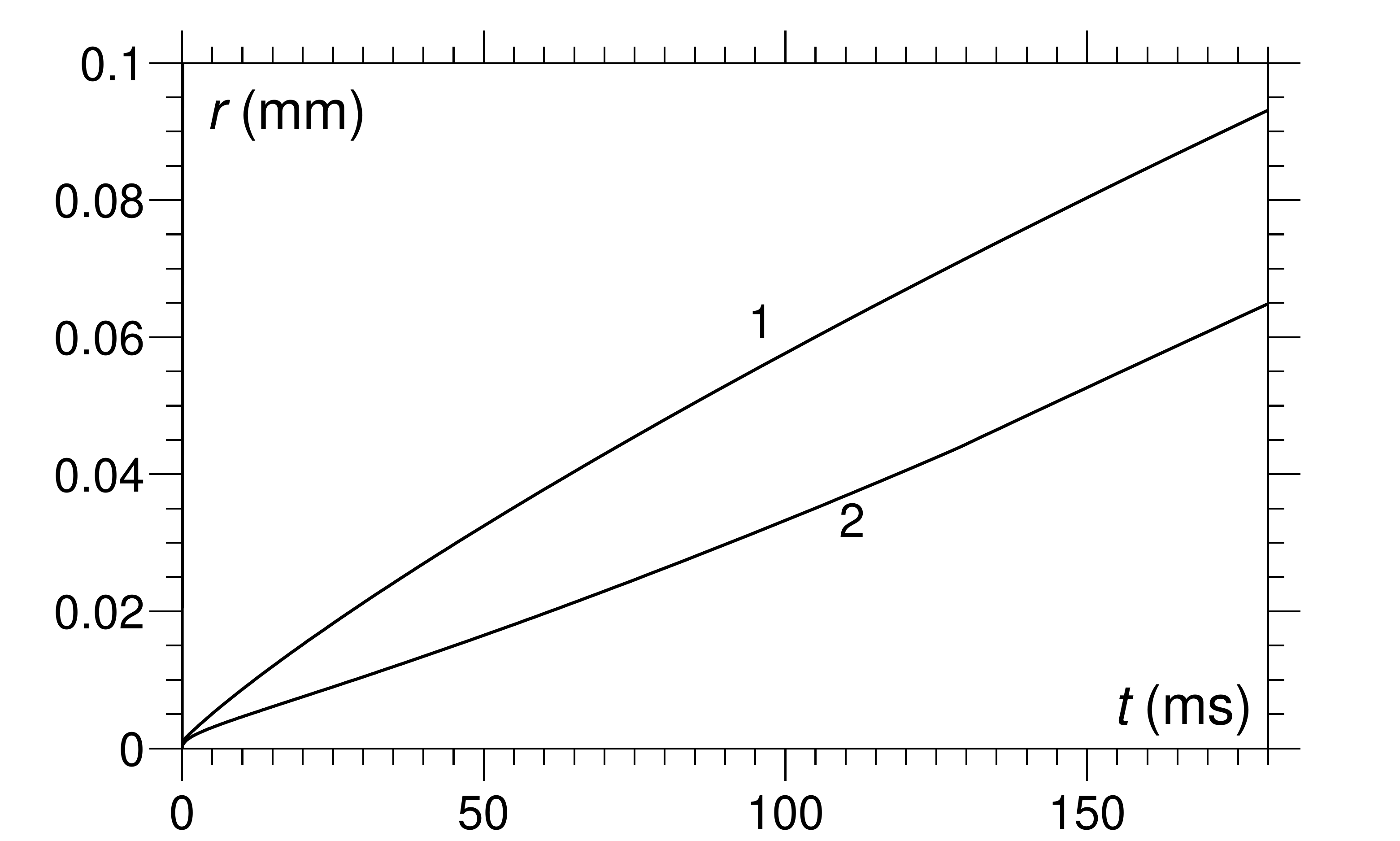}
\caption{\label{F:comparison} Example illustrating the dependence of the dimensional radius of the bridge on dimensional time for the coalescence of two liquid spheres of radius $1$~mm, viscosity $58000$~mPa~s and surface tension $20$~mN~m$^{-1}$ simulated using the conventional model (curve 1) and the interface formation model (curve 2). Although on a large time scale the two curves are similar, on a shorter, and yet easily measurable, time scale there are experimentally verifiable differences between the predictions of the two models.}
\end{figure}

Here, we are interested in suggesting theory-guided experiments,
which would allow experimentally obtained data to be interpreted in
terms of the concepts that the interface formation model adds to our
conventional understanding of fluid mechanical phenomena, such as,
in this particular case, describing how non-smooth free surface
profiles can be sustained. With the aforementioned estimates in mind, we return to the highest
viscosity fluid ($58000$~mPa~s) used in \citep{paulsen12}, and
consider whether one can bring the differences between the
conventional model and the interface formation model into the
optical range for these parameters. In Figure~\ref{F:comparison}, we
give an example showing that this is indeed possible. In particular, we see that with
the time of the order of $100$~ms and the bridge radius of the order
of $100$~$\mu$m, so that we are well within the optical range, there
is a clearly verifiable difference between the two models'
predictions. It should be pointed out that we have not been able to
ascertain the precise parameters which should be used for the
interface formation model for this particular fluid, and so have
used the parameters (\ref{base_parameters}) mentioned earlier. The key point is
that this is a perfect test case with which the use of the interface
formation model for the coalescence process could be scrutinized.

\subsection{Kinematics}

Another aspect of the interface formation model that lends itself to
experimental verification is the fact that the flow kinematics
produced in the framework of this model indicates that the fluid
particles initially belonging to the free surface move across the
contact line to become the fluid particles forming, first, the internal
interface and then the `ordinary' bulk particles. In other words,
there is a qualitative difference with what one has in the
conventional model where the fluid particles once forming the free
surface stay on the free surface at all time. From an experimental
viewpoint, this difference suggests `marking' the fluid particles of
the free surface with microscopic `markers', e.g.\ the molecules of
a surfactant with a sufficiently low concentration so that the
surfactant remains a `marker', as opposed to influencing the fluid's
dynamics. Then, one could monitor the percentage of the `markers'
that find themselves in the bulk of the fluid when the drops
coalesce to form a single body of fluid.

Notably, the kind of kinematics outlined above has been observed in
the experiments on the steady free-surface `cusps' forming in
convergent flow\citep{jeong92},
albeit the `markers' used in these experiments (particles of a
powder) were rather crude.  It is also worth mentioning that, as it has subsequently been shown, the `cusps' themselves, first discovered in \citep{joseph91}, turned out to be
corners\citep{shik05a}, so that the `contact angle' in the coalescence phenomenon is actually the unsteady version of the corners observed in steady convergent flows. The similarity between the
flow kinematics in the steady convergent flows and the coalescence
process indicates that the appearance of singularities in the
free-surface curvature and the corresponding qualitative change in
the flow kinematics could be a generic phenomenon with profound
implications.

\section{Concluding remarks}\label{conclusion}

Much literature on the coalescence of liquid drops has been
concerned with producing and testing various `scaling laws', which,
with the proper choice of constants, are expected to approximate the
actual solution one would obtain in the framework of the
conventional model. Here, we have used our computational platform to
show that in many cases these scaling laws indeed provide a fairly
good fit to the predictions of the conventional model and in some
cases appear to work even outside their `nominal' limits of
applicability. However, we have also shown that the conventional
model itself is unable to describe the coalescence phenomenon whose
details have come to light with the new experimental data. In fact,
for the three viscosities considered, even on a $\log$-$\log$ plot
there is a clear discrepancy between the predictions of the
conventional model and experimental data in all cases except the
late stages of coalescence of the most viscous drop, i.e.\ the
stages where coalescence as such is already over. Clearly then, the
scaling laws so often used in the literature are also ineffective at
describing these flows and any attempts to fit the data with
different coefficients will merely result in the dependencies that
are no longer close to the solution of the equations they are
supposed to represent.

The mathematical complexity of the interface formation model has
often been cited \citep{attane07,bayer06} as its drawback, although there is no reason to expect that intricate experimental
effects will be describable by simple mathematics. We have overcome
the mathematical difficulties of incorporating the interface
formation model into a numerical platform in our previous work \citep{sprittles_jcp}, which allowed us to use and compare both the
conventional and the interface formation model in the context of
dynamic wetting processes. In the present work, we have shown that
the interface formation model provides a natural description, as
well as a considerably easier numerical implementation compared to the conventional model, for the coalescence phenomena. The reason
for this is that the interface formation model is able to cope with the coalescence
event in a singularity-free manner, which makes computation far
easier and actually means that less resolution is required with this
model than the conventional one. The results of using the interface
formation model agree well with all experimental data apart from the
late stages of low viscosity drops, in which coalescence as such,
i.e.\ the formation of a single body of fluid with a smooth free
surface, has actually occurred already.

As previously mentioned, it seems most likely that the influence of
the surrounding air, which is neglected in our description, is
responsible for the above discrepancy between the theories and
experiment. The evidence in favour of this reason is that at the
highest viscosity, where the liquid-to-air viscosity ratio is large,
$\mu/\mu_{air}\sim10^4$, there is no discrepancy whereas at the
lowest liquid-to-air viscosity ratio $\mu/\mu_{air}\sim10^2$, i.e.\
where the viscosities are more comparable, an influence is seen.
Including the ambient gas dynamics will be the subject of future
work where we will consider both the possibility of using
lubrication theory to determine the forces acting on the free
surface from the gas, as well as extending our computational
framework to describe the gas flow.

Our computations have confirmed previous predictions that for
low-viscosity fluids, toroidal bubbles are to be expected. Such
bubbles are particularly prevalent when one uses the conventional
model to describe coalescence as it introduces a stronger capillary
wave that leads to the trapping of the bubbles.  Therefore, a potential
test case for the two models would be to predict when such bubbles
exist and what the size distribution of the bubbles will be. The
problem of describing the dynamics of the trapped bubbles and, in
particular, their stability with respect to azimuthal disturbances,
requires the development of more powerful computer codes which would
be capable of handling multiple topological changes to the fluid's
domain. This is the subject of current work. From the
theoretical standpoint, it is yet unclear even how accounting for
the ambient gas' viscosity will affect the formation of the bubbles,
and a natural approach to this problem is to include the gas
dynamics into the computational framework.  Ultimately, it will be
for the experiments to ascertain the appearance of the toroidal
bubbles and the conditions that promote this effect.  In this
regard, experiments in vacuum/low-pressure chambers are a
particularly promising line of enquiry as they could help to
elucidate several aspects associated with the role of the gas.

Much debate exists in the literature as to whether the conventional
model and its known extensions are able to describe a variety of
flow configurations in which, as suggested by qualitative analysis,
interfaces form or disappear. These
flows are often characterized by the conventional model predicting
singularities of various kind, as is the case for coalescence
\citep{shik00}, or not allowing a solution to exist at all, as in the
case of dynamic wetting \citep{shik07}. The advantage of using the coalescence phenomena to
investigate the possibility of dynamic interfacial effects is that,
in contrast to dynamic wetting experiments, there is no solid
surface involved; the solid's properties, such as roughness and chemical
inhomogeneity, are usually poorly defined, which creates room for
different interpretations of the experimental outcome.  If viewed
through the prism of the interface formation model, the coalescence
process considered here can be regarded essentially as the `dynamic
wetting' of a geometric surface (plane of symmetry), where the
`equilibrium contact angle' is $90^\circ$. In other words, in
coalescence, any observed non-smoothness of the free surface is
evidence in favour of the interface formation model. Furthermore,
the known effect of `hydrodynamic assist of dynamic wetting' \citep{blake99,blake94} suggests that in the coalescence process, for the
same liquid, the dynamic contact angle versus contact-line speed
curves will depend on the drops' size, and a close investigation of
this effect could provide valuable information about the interfacial
dynamics.

Our results suggest that, as drops' size decreases, the deviation
between the conventional and the interface formation model will
become more pronounced as the relative size of the trapped `internal
interface' will increase, which is particularly the case for
high-viscosity liquids. However, as the size of the system goes
down, one runs into the limitations of what can be measured using
the conventional optical techniques. To a certain extent, this catch
twenty-two situation has been resolved by the pioneering experiments
from the Chicago Group, e.g.\ \citep{paulsen11}, which allow, for the
first time, sub-optical measurements to be made reliably and
accurately. It would be interesting to see if a similar method can
be applied to wetting experiments to allow a similar resolution to
be achieved there, i.e.\ to determine the radius of the wetted area
for a drop impact and spreading onto a solid substrate as a function
of time from the resistance which this area produces. Such a method
could uncover the new effects predicted in \citep{sprittles_jcp},
which are similar to those observed in coalescence, namely that, as
the interface formation model indicates, the onset of spreading
corresponds to a much slower initial motion of the wetting line than
what the conventional models suggest.  Of particular importance is the predicted decrease of the dynamic contact angle as the contact-line speed increases, which is a specific feature of unsteady dynamic wetting.

It was interesting to see that, with regard to the coalescence
experiments, a better agreement between theory and experiment was
obtained by using a lower value of $\rho^s_{1e}$ as the
concentration of glycerol was increased in the mixture.  As both
water and glycerol have a similar density, this may seem somewhat
surprising; however, the hygroscopic nature of glycerol suggests
that at high concentrations often the interface of the drops can
consist of just one of the liquids, which then essentially acts as a
kind of surfactant to the whole mixture.  We can speculate that this
may be the nature of the observed effect, but the best way to
confirm that this is the case would be to use a different liquid,
such as a silicon oil, which does not suffer from such effects, in
order to conduct similar experiments and then, by checking the
results against the interface formation model's simulations,
determine whether there is a variation of $\rho^s_{1e}$ with, say,
viscosity.

\section*{Appendix: Influence of initial gravity and initial shape}

To simulate the coalescence of two drops, which retain their axisymmetry but, due to, say, gravity, lose their symmetry about the $z=0$ plane is a computationally tractable problem.  Here we have assumed that such asymmetry will not have a significant influence on the very initial stages of coalescence and, in particular, will not alter the conclusions of our comparison between theory and experiment. Once gravity is included, it will act to elongate/squash the upper/lower drop so that the radius of curvature of the upper/lower drop at the point where the two drops meet is decreased/increased.  Then, crudely, one could argue that these two opposite influences, which will act to decrease/increase the speed of coalescence, will neutralize each other.  To provide bounds on the effects which gravity could have, whilst retaining the plane of symmetry, we consider a body force which acts towards the $z=0$ plane, so that the drop, and its image, is elongated and the opposite case where the body force is away from the $z=0$ plane, acting to squash the drop. These tests, which provide the worst case scenario where a elongated/squashed drop coalesces with a copy of itself, so that there is no cancelling of effects, will provide a useful bounds on the influence that correctly incorporating gravity into our framework would have.

\begin{figure}
     \centering
\includegraphics[scale=0.305]{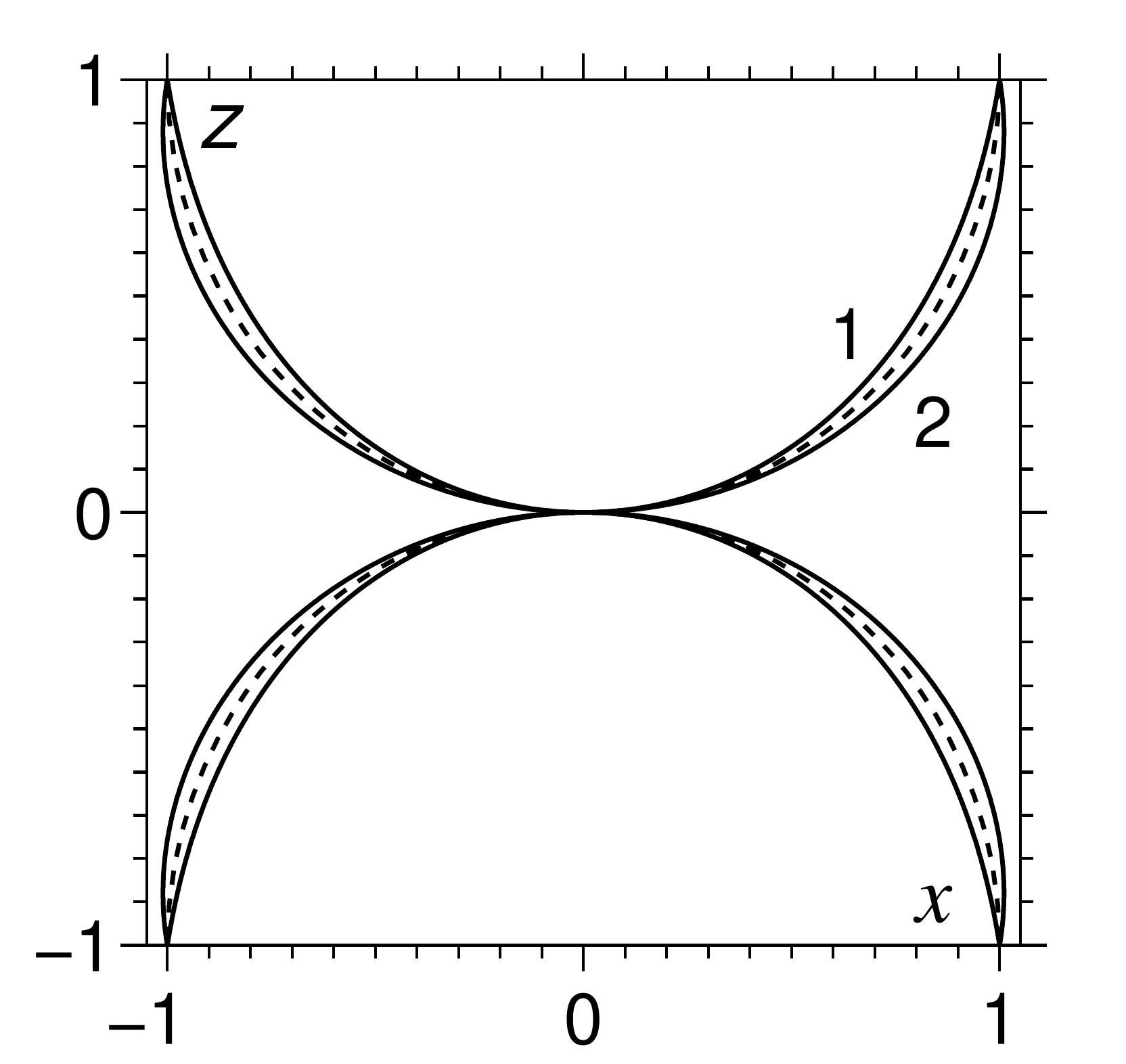}
\includegraphics[scale=0.305]{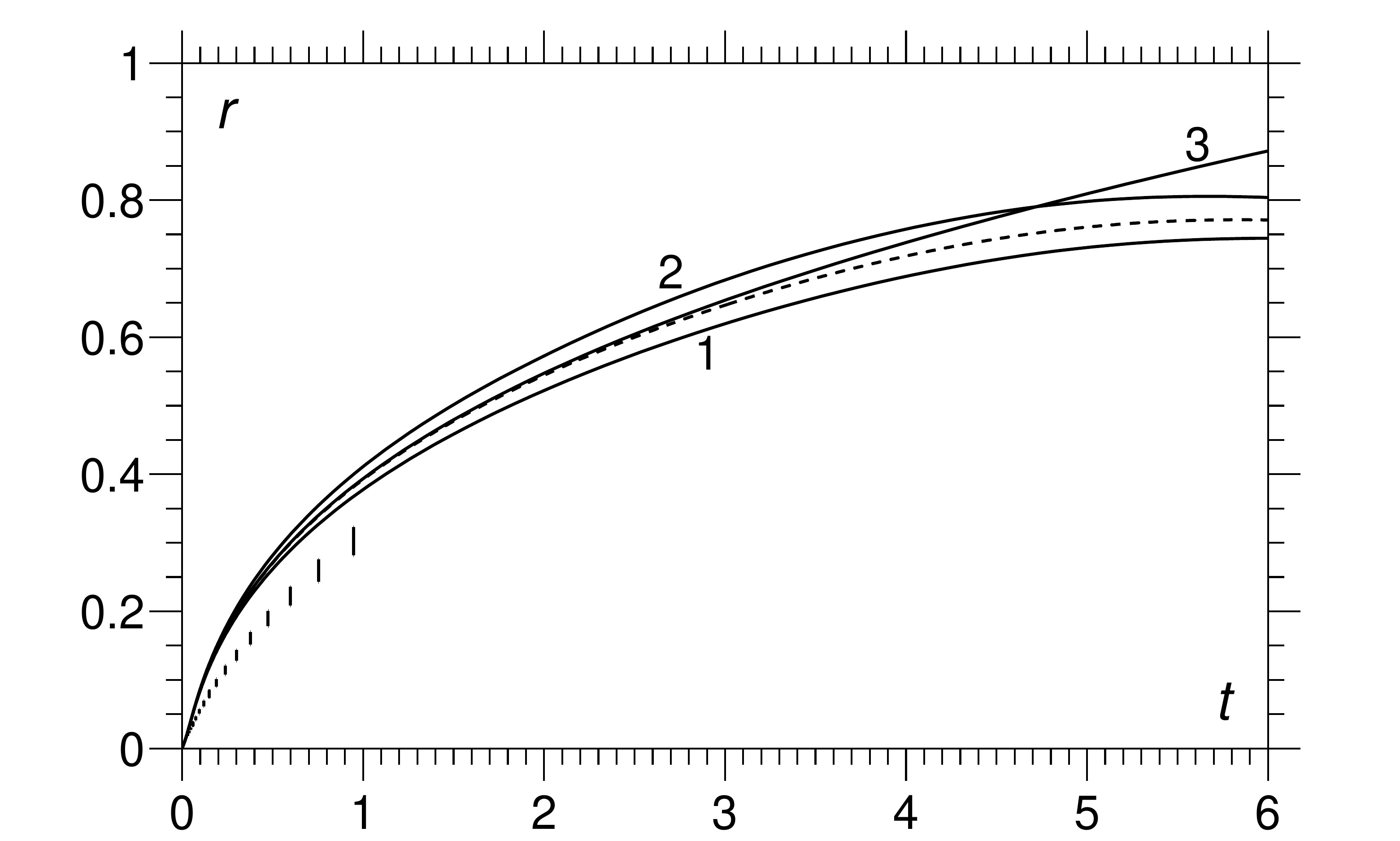}
\caption{\label{F:gravity} Left: Different initial shapes dependent on gravity for $2$~mm drops.  Right: Bridge radius as a function of time for the drops with $Re=68$ compared to error bars from \citep{paulsen11}, with curve 1 corresponding to $Bo=0.74$, curve 2 to $Bo=-0.74$ and the dashed curve is the hemispheres obtained for $Bo=0$.  Curve 3 is for the coalescence of free spheres as opposed to pinned hemispheres.}
\end{figure}

In Figure~\ref{F:gravity}, we see the influence which gravity has on the initial shapes and the subsequent evolution of the drops considered in \citep{paulsen11}, which are taken for the $Re=68$ case.  The bridge radius is plotted for simulations using the conventional model with the two elongated drops (curve 1) and the two squashed drops (curve 2) compared to the zero gravity case (dashed curve), where perfect hemispheres coalesce.  Also in the figure are the experimental error bars from \citep{paulsen11} and, most importantly, one can see that over the range $0<t<1$, the effect of the initial shape, has very little influence on the drops' dynamics.

Additionally, in Figure~\ref{F:gravity}, we show that the non-local effect of the different flow geometries used, i.e. free spheres and pinned hemispheres, have no influence on the very initial stages of coalescence, where the comparison to experiment has been made. As one can see, in this region the result for coalescing hemispheres (dashed line) is graphically indistinguishable from that obtained using spheres (curve 3). Notably, the two equilibrium bridge radii will differ, with the sphere obtaining a larger equilibrium radius.

\section*{Acknowledgements}

This publication is based on work supported by Award No KUK-C1-013-04 , made by King Abdullah University of Science and Technology (KAUST).\\

The authors would like to thank Dr J.D.~Paulsen, Dr J.C.~Burton and Professor S.R.~Nagel for providing us with the data from their experiments published in \citep{paulsen11} and Professor S.T.~Thoroddsen for useful discussions regarding the coalescence phenomena.

\bibliography{Bibliography}

\begin{thebibliography}{10}%
\makeatletter
\providecommand \@ifxundefined [1]{%
 \ifx #1\undefined \expandafter \@firstoftwo
 \else \expandafter \@secondoftwo
\fi
}%
\providecommand \@ifnum [1]{%
 \ifnum #1\expandafter \@firstoftwo
 \else \expandafter \@secondoftwo
\fi
}%
\providecommand \enquote [1]{``#1''}%
\providecommand \bibnamefont  [1]{#1}%
\providecommand \bibfnamefont [1]{#1}%
\providecommand \citenamefont [1]{#1}%
\providecommand\href[0]{\@sanitize\@href}%
\providecommand\@href[1]{\endgroup\@@startlink{#1}\endgroup\@@href}%
\providecommand\@@href[1]{#1\@@endlink}%
\providecommand \@sanitize [0]{\begingroup\catcode`\&12\catcode`\#12\relax}%
\@ifxundefined \pdfoutput {\@firstoftwo}{%
 \@ifnum{\z@=\pdfoutput}{\@firstoftwo}{\@secondoftwo}%
}{%
 \providecommand\@@startlink[1]{\leavevmode\special{html:<a href="#1">}}%
 \providecommand\@@endlink[0]{\special{html:</a>}}%
}{%
 \providecommand\@@startlink[1]{%
  \leavevmode
  \pdfstartlink
   attr{/Border[0 0 1 ]/H/I/C[0 1 1]}%
   user{/Subtype/Link/A<</Type/Action/S/URI/URI(#1)>>}%
  \relax
 }%
 \providecommand\@@endlink[0]{\pdfendlink}%
}%
\providecommand \url  [0]{\begingroup\@sanitize \@url }%
\providecommand \@url [1]{\endgroup\@href {#1}{\urlprefix}}%
\providecommand \urlprefix [0]{URL }%
\providecommand \Eprint[0]{\href }%
\@ifxundefined \urlstyle {%
  \providecommand \doi [1]{doi:\discretionary{}{}{}#1}%
}{%
  \providecommand \doi [0]{doi:\discretionary{}{}{}\begingroup
  \urlstyle{rm}\Url }%
}%
\providecommand \doibase [0]{http://dx.doi.org/}%
\providecommand \Doi[1]{\href{\doibase#1}}%
\providecommand \selectlanguage [0]{\@gobble}%
\providecommand \bibinfo [0]{\@secondoftwo}%
\providecommand \bibfield [0]{\@secondoftwo}%
\providecommand \translation [1]{[#1]}%
\providecommand \BibitemOpen[0]{}%
\providecommand \bibitemStop [0]{}%
\providecommand \bibitemNoStop [0]{.\EOS\space}%
\providecommand \EOS [0]{\spacefactor3000\relax}%
\providecommand \BibitemShut [1]{\csname bibitem#1\endcsname}%
\bibitem{bellehumeur04}%
  \BibitemOpen
  \bibfield{author}{%
  \bibinfo {author} {\bibfnamefont{C.~T.}\ \bibnamefont{Bellehumeur}}, \bibinfo
  {author} {\bibfnamefont{M.~K.}\ \bibnamefont{Biaria}},\ and\ \bibinfo
  {author} {\bibfnamefont{J.}~\bibnamefont{Vlachopoulos}},\ }%
  \bibfield{title}{%
  \enquote{\bibinfo {title} {An experimental study and model assessment of
  polymer sintering},}\ }%
  \bibfield{journal}{%
  \bibinfo {journal} {Polymer Engineering and Science}\ }%
  \textbf{\bibinfo {volume} {36}},\ \bibinfo {pages} {2198--2207} (\bibinfo
  {year} {2004})%
\bibitem{dreher99}%
  \BibitemOpen
  \bibfield{author}{%
  \bibinfo {author} {\bibfnamefont{T.~M.}\ \bibnamefont{Dreher}}, \bibinfo
  {author} {\bibfnamefont{J.}~\bibnamefont{Glass}}, \bibinfo {author}
  {\bibfnamefont{A.~J.}\ \bibnamefont{{O'Connor}}},\ and\ \bibinfo {author}
  {\bibfnamefont{G.~W.}\ \bibnamefont{Stevens}},\ }%
  \bibfield{title}{%
  \enquote{\bibinfo {title} {Effect of rheology on coalescence rates and
  emulsion stability},}\ }%
  \bibfield{journal}{%
  \bibinfo {journal} {AIChE Journal}\ }%
  \textbf{\bibinfo {volume} {45}},\ \bibinfo {pages} {1182--1190} (\bibinfo
  {year} {1999})%
\bibitem{grissom81}%
  \BibitemOpen
  \bibfield{author}{%
  \bibinfo {author} {\bibfnamefont{W.~M.}\ \bibnamefont{Grissom}}\ and\
  \bibinfo {author} {\bibfnamefont{F.~A.}\ \bibnamefont{Wierum}},\ }%
  \bibfield{title}{%
  \enquote{\bibinfo {title} {Liquid spray cooling of a heated surface},}\ }%
  \bibfield{journal}{%
  \bibinfo {journal} {International Journal of Heat and Mass Transfer}\ }%
  \textbf{\bibinfo {volume} {24}},\ \bibinfo {pages} {261--271} (\bibinfo
  {year} {1981})%
\bibitem{kovetz69}%
  \BibitemOpen
  \bibfield{author}{%
  \bibinfo {author} {\bibfnamefont{A.}~\bibnamefont{Kovetz}}\ and\ \bibinfo
  {author} {\bibfnamefont{B.}~\bibnamefont{Olund}},\ }%
  \bibfield{title}{%
  \enquote{\bibinfo {title} {The effect of coalescence and condensation on rain
  formation in a cloud of finite vertical extent},}\ }%
  \bibfield{journal}{%
  \bibinfo {journal} {Journal of Atmospheric Sciences}\ }%
  \textbf{\bibinfo {volume} {26}},\ \bibinfo {pages} {1060--1065} (\bibinfo
  {year} {1969})%
\bibitem{quake05}%
  \BibitemOpen
  \bibfield{author}{%
  \bibinfo {author} {\bibfnamefont{T.~M.}\ \bibnamefont{Squires}}\ and\
  \bibinfo {author} {\bibfnamefont{S.~R.}\ \bibnamefont{Quake}},\ }%
  \bibfield{title}{%
  \enquote{\bibinfo {title} {Microfluidics: Fluid physics at the nanoliter
  scale},}\ }%
  \bibfield{journal}{%
  \bibinfo {journal} {Reviews of Modern Physics}\ }%
  \textbf{\bibinfo {volume} {77}},\ \bibinfo {pages} {977--1026} (\bibinfo
  {year} {2005})%
\bibitem{derby10}%
  \BibitemOpen
  \bibfield{author}{%
  \bibinfo {author} {\bibfnamefont{B.}~\bibnamefont{Derby}},\ }%
  \bibfield{title}{%
  \enquote{\bibinfo {title} {Inkjet printing of functional and structural
  materials: Fluid property requirements, feature stability and resolution},}\
  }%
  \bibfield{journal}{%
  \bibinfo {journal} {Annual Review of Materials Research}\ }%
  \textbf{\bibinfo {volume} {40}},\ \bibinfo {pages} {395--414} (\bibinfo
  {year} {2010})%
\bibitem{singh10}%
  \BibitemOpen
  \bibfield{author}{%
  \bibinfo {author} {\bibfnamefont{M.}~\bibnamefont{Singh}}, \bibinfo {author}
  {\bibfnamefont{H.}~\bibnamefont{Haverinen}}, \bibinfo {author}
  {\bibfnamefont{P.}~\bibnamefont{Dhagat}},\ and\ \bibinfo {author}
  {\bibfnamefont{G.}~\bibnamefont{Jabbour}},\ }%
  \bibfield{title}{%
  \enquote{\bibinfo {title} {Inkjet printing process and its applications},}\
  }%
  \bibfield{journal}{%
  \bibinfo {journal} {Advanced Materials}\ }%
  \textbf{\bibinfo {volume} {22}},\ \bibinfo {pages} {673--685} (\bibinfo
  {year} {2010})%
\bibitem{richardson68}%
  \BibitemOpen
  \bibfield{author}{%
  \bibinfo {author} {\bibfnamefont{S.}~\bibnamefont{Richardson}},\ }%
  \bibfield{title}{%
  \enquote{\bibinfo {title} {Two-dimensional bubbles in slow viscous flows},}\
  }%
  \bibfield{journal}{%
  \bibinfo {journal} {Journal of Fluid Mechanics}\ }%
  \textbf{\bibinfo {volume} {33}},\ \bibinfo {pages} {475--493} (\bibinfo
  {year} {1968})%
\bibitem{thoroddsen05}%
  \BibitemOpen
  \bibfield{author}{%
  \bibinfo {author} {\bibfnamefont{S.~T.}\ \bibnamefont{Thoroddsen}}, \bibinfo
  {author} {\bibfnamefont{K.}~\bibnamefont{Takehara}},\ and\ \bibinfo {author}
  {\bibfnamefont{T.~G.}\ \bibnamefont{Etoh}},\ }%
  \bibfield{title}{%
  \enquote{\bibinfo {title} {The coalescence speed of a pendent and sessile
  drop},}\ }%
  \bibfield{journal}{%
  \bibinfo {journal} {Journal of Fluid Mechanics}\ }%
  \textbf{\bibinfo {volume} {527}},\ \bibinfo {pages} {85--114} (\bibinfo
  {year} {2005})%
\bibitem{paulsen11}%
  \BibitemOpen
  \bibfield{author}{%
  \bibinfo {author} {\bibfnamefont{J.~D.}\ \bibnamefont{Paulsen}}, \bibinfo
  {author} {\bibfnamefont{J.~C.}\ \bibnamefont{Burton}},\ and\ \bibinfo
  {author} {\bibfnamefont{S.~R.}\ \bibnamefont{Nagel}},\ }%
  \bibfield{title}{%
  \enquote{\bibinfo {title} {Viscous to inertial crossover in liquid drop
  coalescence},}\ }%
  \bibfield{journal}{%
  \bibinfo {journal} {Physical Review Letters}\ }%
  \textbf{\bibinfo {volume} {106}},\ \bibinfo {pages} {114501} (\bibinfo {year}
  {2011})%
\bibitem{frenkel45}%
  \BibitemOpen
  \bibfield{author}{%
  \bibinfo {author} {\bibfnamefont{J.}~\bibnamefont{Frenkel}},\ }%
  \bibfield{title}{%
  \enquote{\bibinfo {title} {Viscous flow of crystalline bodies under the
  action of surface tension},}\ }%
  \bibfield{journal}{%
  \bibinfo {journal} {Journal of Physics (USSR)}\ }%
  \textbf{\bibinfo {volume} {9}},\ \bibinfo {pages} {385--391} (\bibinfo {year}
  {1945})%
\bibitem{hopper84}%
  \BibitemOpen
  \bibfield{author}{%
  \bibinfo {author} {\bibfnamefont{R.~W.}\ \bibnamefont{Hopper}},\ }%
  \bibfield{title}{%
  \enquote{\bibinfo {title} {Coalescence of two equal cylinders: exact results
  for creeping viscous plane flow driven by capillarity},}\ }%
  \bibfield{journal}{%
  \bibinfo {journal} {Journal of the American Ceramic Society}\ }%
  \textbf{\bibinfo {volume} {67}},\ \bibinfo {pages} {262--264} (\bibinfo
  {year} {1984})%
\bibitem{hopper90}%
  \BibitemOpen
  \bibfield{author}{%
  \bibinfo {author} {\bibfnamefont{R.~W.}\ \bibnamefont{Hopper}},\ }%
  \bibfield{title}{%
  \enquote{\bibinfo {title} {Plane {Stokes} flow driven by capillarity on a
  free surface},}\ }%
  \bibfield{journal}{%
  \bibinfo {journal} {Journal of Fluid Mechanics}\ }%
  \textbf{\bibinfo {volume} {213}},\ \bibinfo {pages} {349--375} (\bibinfo
  {year} {1990})%
\bibitem{hopper93a}%
  \BibitemOpen
  \bibfield{author}{%
  \bibinfo {author} {\bibfnamefont{R.~W.}\ \bibnamefont{Hopper}},\ }%
  \bibfield{title}{%
  \enquote{\bibinfo {title} {Coalescence of two viscous cylinders by
  capillarity: {Part 1}. {Theory}},}\ }%
  \bibfield{journal}{%
  \bibinfo {journal} {Journal of the American Ceramic Society}\ }%
  \textbf{\bibinfo {volume} {76}},\ \bibinfo {pages} {2947--2952} (\bibinfo
  {year} {1993})%
\bibitem{hopper93b}%
  \BibitemOpen
  \bibfield{author}{%
  \bibinfo {author} {\bibfnamefont{R.~W.}\ \bibnamefont{Hopper}},\ }%
  \bibfield{title}{%
  \enquote{\bibinfo {title} {Coalescence of two viscous cylinders by
  capillarity: {Part 2}. shape evolution},}\ }%
  \bibfield{journal}{%
  \bibinfo {journal} {Journal of the American Ceramic Society}\ }%
  \textbf{\bibinfo {volume} {76}},\ \bibinfo {pages} {2953--2960} (\bibinfo
  {year} {1993})%
\bibitem{richardson92}%
  \BibitemOpen
  \bibfield{author}{%
  \bibinfo {author} {\bibfnamefont{S.}~\bibnamefont{Richardson}},\ }%
  \bibfield{title}{%
  \enquote{\bibinfo {title} {Two-dimensional slow viscous flows with
  time-dependent free boundaries driven by surface tension},}\ }%
  \bibfield{journal}{%
  \bibinfo {journal} {European Journal of Applied Mathematics}\ }%
  \textbf{\bibinfo {volume} {3}},\ \bibinfo {pages} {193--207} (\bibinfo {year}
  {1992})%
\bibitem{eggers99}%
  \BibitemOpen
  \bibfield{author}{%
  \bibinfo {author} {\bibfnamefont{J.}~\bibnamefont{Eggers}}, \bibinfo {author}
  {\bibfnamefont{J.~R.}\ \bibnamefont{Lister}},\ and\ \bibinfo {author}
  {\bibfnamefont{H.~A.}\ \bibnamefont{Stone}},\ }%
  \bibfield{title}{%
  \enquote{\bibinfo {title} {Coalescence of liquid drops},}\ }%
  \bibfield{journal}{%
  \bibinfo {journal} {Journal of Fluid Mechanics}\ }%
  \textbf{\bibinfo {volume} {401}},\ \bibinfo {pages} {293--310} (\bibinfo
  {year} {1999})%
\bibitem{duchemin03}%
  \BibitemOpen
  \bibfield{author}{%
  \bibinfo {author} {\bibfnamefont{L.}~\bibnamefont{Duchemin}}, \bibinfo
  {author} {\bibfnamefont{J.}~\bibnamefont{Eggers}},\ and\ \bibinfo {author}
  {\bibfnamefont{C.}~\bibnamefont{Josserand}},\ }%
  \bibfield{title}{%
  \enquote{\bibinfo {title} {Inviscid coalescence of drops},}\ }%
  \bibfield{journal}{%
  \bibinfo {journal} {Journal of Fluid Mechanics}\ }%
  \textbf{\bibinfo {volume} {487}},\ \bibinfo {pages} {167--178} (\bibinfo
  {year} {2003})%
\bibitem{menchacarocha01}%
  \BibitemOpen
  \bibfield{author}{%
  \bibinfo {author} {\bibfnamefont{A.}~\bibnamefont{Menchaca-Rocha}}, \bibinfo
  {author} {\bibfnamefont{A.}~\bibnamefont{Mart\'{\i}nez-D\'{a}valos}},
  \bibinfo {author} {\bibfnamefont{R.}~\bibnamefont{N\'{u}\'{n}ez}}, \bibinfo
  {author} {\bibfnamefont{S.}~\bibnamefont{Popinet}},\ and\ \bibinfo {author}
  {\bibfnamefont{S.}~\bibnamefont{Zaleski}},\ }%
  \bibfield{title}{%
  \enquote{\bibinfo {title} {Coalescence of liquid drops by surface tension},}\
  }%
  \bibfield{journal}{%
  \bibinfo {journal} {Physical Review E}\ }%
  \textbf{\bibinfo {volume} {63}},\ \bibinfo {pages} {046309} (\bibinfo {year}
  {2001})%
\bibitem{paulsen12}%
  \BibitemOpen
  \bibfield{author}{%
  \bibinfo {author} {\bibfnamefont{J.~D.}\ \bibnamefont{Paulsen}}, \bibinfo
  {author} {\bibfnamefont{J.~C.}\ \bibnamefont{Burton}}, \bibinfo {author}
  {\bibfnamefont{S.~R.}\ \bibnamefont{Nagel}}, \bibinfo {author}
  {\bibfnamefont{S.}~\bibnamefont{Appathurai}}, \bibinfo {author}
  {\bibfnamefont{M.~T.}\ \bibnamefont{Harris}},\ and\ \bibinfo {author}
  {\bibfnamefont{O.}~\bibnamefont{Basaran}},\ }%
  \bibfield{title}{%
  \enquote{\bibinfo {title} {The inexorable resistance of inertia determines
  the initial regime of drop coalescence},}\ }%
  \bibfield{journal}{%
  \bibinfo {journal} {Proceedings of the National Academy of Science}\ }%
  \textbf{\bibinfo {volume} {109}},\ \bibinfo {pages} {6857--6861} (\bibinfo
  {year} {2012})%
\bibitem{oguz89}%
  \BibitemOpen
  \bibfield{author}{%
  \bibinfo {author} {\bibfnamefont{H.~N.}\ \bibnamefont{Oguz}}\ and\ \bibinfo
  {author} {\bibfnamefont{A.}~\bibnamefont{Prosperetti}},\ }%
  \bibfield{title}{%
  \enquote{\bibinfo {title} {Surface-tension effects in the contact of liquid
  surfaces},}\ }%
  \bibfield{journal}{%
  \bibinfo {journal} {Journal of Fluid Mechanics}\ }%
  \textbf{\bibinfo {volume} {203}},\ \bibinfo {pages} {149--171} (\bibinfo
  {year} {1989})%
\bibitem{jagota88}%
  \BibitemOpen
  \bibfield{author}{%
  \bibinfo {author} {\bibfnamefont{A.}~\bibnamefont{Jagota}}\ and\ \bibinfo
  {author} {\bibfnamefont{P.~R.}\ \bibnamefont{Dawson}},\ }%
  \bibfield{title}{%
  \enquote{\bibinfo {title} {Micromechanical modeling of powder compacts - {I.}
  {Unit} problems for sintering and traction induced deformation},}\ }%
  \bibfield{journal}{%
  \bibinfo {journal} {Acta Metallurgica}\ }%
  \textbf{\bibinfo {volume} {36}},\ \bibinfo {pages} {2551--2561} (\bibinfo
  {year} {1988})%
\bibitem{martinez95}%
  \BibitemOpen
  \bibfield{author}{%
  \bibinfo {author} {\bibfnamefont{J.~I.}\
  \bibnamefont{Mart\'{\i}nez-Herrera}}\ and\ \bibinfo {author}
  {\bibfnamefont{J.~J.}\ \bibnamefont{Derby}},\ }%
  \bibfield{title}{%
  \enquote{\bibinfo {title} {Viscous sintering of spherical particles via
  finite element analysis},}\ }%
  \bibfield{journal}{%
  \bibinfo {journal} {Journal of the American Ceramic Society}\ }%
  \textbf{\bibinfo {volume} {78}},\ \bibinfo {pages} {645--649} (\bibinfo
  {year} {1995})%
\bibitem{aarts05}%
  \BibitemOpen
  \bibfield{author}{%
  \bibinfo {author} {\bibfnamefont{D.~G. A.~L.}\ \bibnamefont{Aarts}}, \bibinfo
  {author} {\bibfnamefont{H.~N.~W.}\ \bibnamefont{Lekkerkerker}}, \bibinfo
  {author} {\bibfnamefont{H.}~\bibnamefont{Guo}}, \bibinfo {author}
  {\bibfnamefont{G.~H.}\ \bibnamefont{Wegdam}},\ and\ \bibinfo {author}
  {\bibfnamefont{D.}~\bibnamefont{Bonn}},\ }%
  \bibfield{title}{%
  \enquote{\bibinfo {title} {Hydrodynamics of droplet coalescence},}\ }%
  \bibfield{journal}{%
  \bibinfo {journal} {Physical Review Letters}\ }%
  \textbf{\bibinfo {volume} {05}},\ \bibinfo {pages} {164503} (\bibinfo {year}
  {2005})%
\bibitem{wu04}%
  \BibitemOpen
  \bibfield{author}{%
  \bibinfo {author} {\bibfnamefont{M.}~\bibnamefont{Wu}}, \bibinfo {author}
  {\bibfnamefont{T.}~\bibnamefont{Cubaud}},\ and\ \bibinfo {author}
  {\bibfnamefont{C.}~\bibnamefont{Ho}},\ }%
  \bibfield{title}{%
  \enquote{\bibinfo {title} {Scaling law in liquid drop coalescence driven by
  surface tension},}\ }%
  \bibfield{journal}{%
  \bibinfo {journal} {Physics of Fluids}\ }%
  \textbf{\bibinfo {volume} {16}},\ \bibinfo {pages} {51--54} (\bibinfo {year}
  {2004})%
\bibitem{Note1}%
  \BibitemOpen
  \bibinfo {note} {It should be pointed out here that the measurements are
  taken for relatively large bridge radii, so that a comparison with the
  inertial scaling (\ref {inertial_scaling}) is valid, but the theory of \cite
  {duchemin03} is well past its limits of applicability, so that one cannot
  expect good agreement for the proposed prefactor.}\BibitemShut{Stop}%
\bibitem{case08}%
  \BibitemOpen
  \bibfield{author}{%
  \bibinfo {author} {\bibfnamefont{S.~C.}\ \bibnamefont{Case}}\ and\ \bibinfo
  {author} {\bibfnamefont{S.~R.}\ \bibnamefont{Nagel}},\ }%
  \bibfield{title}{%
  \enquote{\bibinfo {title} {Coalescence in low-viscosity liquids},}\ }%
  \bibfield{journal}{%
  \bibinfo {journal} {Physical Review Letters}\ }%
  \textbf{\bibinfo {volume} {100}},\ \bibinfo {pages} {084503} (\bibinfo {year}
  {2008})%
\bibitem{case09}%
  \BibitemOpen
  \bibfield{author}{%
  \bibinfo {author} {\bibfnamefont{S.~C.}\ \bibnamefont{Case}},\ }%
  \bibfield{title}{%
  \enquote{\bibinfo {title} {Coalescence of low-viscosity fluids in air},}\ }%
  \bibfield{journal}{%
  \bibinfo {journal} {Physical Review E}\ }%
  \textbf{\bibinfo {volume} {79}},\ \bibinfo {pages} {026307} (\bibinfo {year}
  {2009})%
\bibitem{burton04}%
  \BibitemOpen
  \bibfield{author}{%
  \bibinfo {author} {\bibfnamefont{J.~C.}\ \bibnamefont{Burton}}, \bibinfo
  {author} {\bibfnamefont{J.~E.}\ \bibnamefont{Rutledge}},\ and\ \bibinfo
  {author} {\bibfnamefont{P.}~\bibnamefont{Taborek}},\ }%
  \bibfield{title}{%
  \enquote{\bibinfo {title} {Fluid pinch-off dynamics at nanometer length
  scales},}\ }%
  \bibfield{journal}{%
  \bibinfo {journal} {Physical Review Letters}\ }%
  \textbf{\bibinfo {volume} {92}},\ \bibinfo {pages} {244505} (\bibinfo {year}
  {2004})%
\bibitem{shik00}%
  \BibitemOpen
  \bibfield{author}{%
  \bibinfo {author} {\bibfnamefont{Y.~D.}\ \bibnamefont{Shikhmurzaev}},\ }%
  \bibfield{title}{%
  \enquote{\bibinfo {title} {Coalescence and capillary breakup of liquid
  volumes},}\ }%
  \bibfield{journal}{%
  \bibinfo {journal} {Physics of Fluids}\ }%
  \textbf{\bibinfo {volume} {12}},\ \bibinfo {pages} {2386--2396} (\bibinfo
  {year} {2000})%
\bibitem{joseph91}%
  \BibitemOpen
  \bibfield{author}{%
  \bibinfo {author} {\bibfnamefont{D.~D.}\ \bibnamefont{Joseph}}, \bibinfo
  {author} {\bibfnamefont{J.}~\bibnamefont{Nelson}}, \bibinfo {author}
  {\bibfnamefont{M.}~\bibnamefont{Renardy}},\ and\ \bibinfo {author}
  {\bibfnamefont{Y.}~\bibnamefont{Renardy}},\ }%
  \bibfield{title}{%
  \enquote{\bibinfo {title} {Two-dimensional cusped interfaces},}\ }%
  \bibfield{journal}{%
  \bibinfo {journal} {Journal of Fluid Mechanics}\ }%
  \textbf{\bibinfo {volume} {223}},\ \bibinfo {pages} {383--409} (\bibinfo
  {year} {1991})%
\bibitem{shik05a}%
  \BibitemOpen
  \bibfield{author}{%
  \bibinfo {author} {\bibfnamefont{Y.~D.}\ \bibnamefont{Shikhmurzaev}},\ }%
  \bibfield{title}{%
  \enquote{\bibinfo {title} {Singularity of free-surface curvature in
  convergent flow: Cusp or corner?}.}\ }%
  \bibfield{journal}{%
  \bibinfo {journal} {Physics Letters A}\ }%
  \textbf{\bibinfo {volume} {345--385}},\ \bibinfo {pages} {378} (\bibinfo
  {year} {2005})%
\bibitem{shik05b}%
  \BibitemOpen
  \bibfield{author}{%
  \bibinfo {author} {\bibfnamefont{Y.~D.}\ \bibnamefont{Shikhmurzaev}},\ }%
  \bibfield{title}{%
  \enquote{\bibinfo {title} {Capillary breakup of liquid threads: A
  singularity-free solution},}\ }%
  \bibfield{journal}{%
  \bibinfo {journal} {IMA Journal of Applied Mathematics}\ }%
  \textbf{\bibinfo {volume} {70}},\ \bibinfo {pages} {880--907} (\bibinfo
  {year} {2005})%
\bibitem{shik05c}%
  \BibitemOpen
  \bibfield{author}{%
  \bibinfo {author} {\bibfnamefont{Y.~D.}\ \bibnamefont{Shikhmurzaev}},\ }%
  \bibfield{title}{%
  \enquote{\bibinfo {title} {Macroscopic mechanism of rupture of free liquid
  films},}\ }%
  \bibfield{journal}{%
  \bibinfo {journal} {Comptes Rendus Mecanique}\ }%
  \textbf{\bibinfo {volume} {333}},\ \bibinfo {pages} {205--210} (\bibinfo
  {year} {2005})%
\bibitem{shik93}%
  \BibitemOpen
  \bibfield{author}{%
  \bibinfo {author} {\bibfnamefont{Y.~D.}\ \bibnamefont{Shikhmurzaev}},\ }%
  \bibfield{title}{%
  \enquote{\bibinfo {title} {The moving contact line on a smooth solid
  surface},}\ }%
  \bibfield{journal}{%
  \bibinfo {journal} {International Journal of Multiphase Flow}\ }%
  \textbf{\bibinfo {volume} {19}},\ \bibinfo {pages} {589--610} (\bibinfo
  {year} {1993})%
\bibitem{shik97}%
  \BibitemOpen
  \bibfield{author}{%
  \bibinfo {author} {\bibfnamefont{Y.~D.}\ \bibnamefont{Shikhmurzaev}},\ }%
  \bibfield{title}{%
  \enquote{\bibinfo {title} {Moving contact lines in liquid/liquid/solid
  systems},}\ }%
  \bibfield{journal}{%
  \bibinfo {journal} {Journal of Fluid Mechanics}\ }%
  \textbf{\bibinfo {volume} {334}},\ \bibinfo {pages} {211--249} (\bibinfo
  {year} {1997})%
\bibitem{shik06}%
  \BibitemOpen
  \bibfield{author}{%
  \bibinfo {author} {\bibfnamefont{Y.~D.}\ \bibnamefont{Shikhmurzaev}},\ }%
  \bibfield{title}{%
  \enquote{\bibinfo {title} {{Singularities at the moving contact line.
  Mathematical, physical and computational aspects}},}\ }%
  \bibfield{journal}{%
  \bibinfo {journal} {Physica D}\ }%
  \textbf{\bibinfo {volume} {217}},\ \bibinfo {pages} {121--133} (\bibinfo
  {year} {2006})%
\bibitem{shik97a}%
  \BibitemOpen
  \bibfield{author}{%
  \bibinfo {author} {\bibfnamefont{Y.~D.}\ \bibnamefont{Shikhmurzaev}},\ }%
  \bibfield{title}{%
  \enquote{\bibinfo {title} {Spreading of drops on solid surfaces in a
  quasi-static regime},}\ }%
  \bibfield{journal}{%
  \bibinfo {journal} {Physics of Fluids}\ }%
  \textbf{\bibinfo {volume} {9}},\ \bibinfo {pages} {266--275} (\bibinfo {year}
  {1996})%
\bibitem{shik07}%
  \BibitemOpen
  \bibfield{author}{%
  \bibinfo {author} {\bibfnamefont{Y.~D.}\ \bibnamefont{Shikhmurzaev}},\ }%
  \emph{\bibinfo {title} {Capillary Flows with Forming Interfaces}}\ (\bibinfo
  {publisher} {Chapman \& Hall/CRC, Boca Raton},\ \bibinfo {year}
  {2007})%
\bibitem{sprittles_jcp}%
  \BibitemOpen
  \bibfield{author}{%
  \bibinfo {author} {\bibfnamefont{J.~E.}\ \bibnamefont{Sprittles}}\ and\
  \bibinfo {author} {\bibfnamefont{Y.~D.}\ \bibnamefont{Shikhmurzaev}},\ }%
  \bibfield{title}{%
  \enquote{\bibinfo {title} {Finite element simulation of dynamic wetting flows
  as an interface formation process},}\ }%
  \bibfield{journal}{%
  \bibinfo {journal} {Journal of Computational Physics}\ }%
  \textbf{\bibinfo {volume} {233}},\ \bibinfo {pages} {34--65} (\bibinfo {year}
  {2013})%
\bibitem{blake02}%
  \BibitemOpen
  \bibfield{author}{%
  \bibinfo {author} {\bibfnamefont{T.~D.}\ \bibnamefont{Blake}}\ and\ \bibinfo
  {author} {\bibfnamefont{Y.~D.}\ \bibnamefont{Shikhmurzaev}},\ }%
  \bibfield{title}{%
  \enquote{\bibinfo {title} {Dynamic wetting by liquids of different
  viscosity},}\ }%
  \bibfield{journal}{%
  \bibinfo {journal} {Journal of Colloid and Interface Science}\ }%
  \textbf{\bibinfo {volume} {253}},\ \bibinfo {pages} {196--202} (\bibinfo
  {year} {2002})%
\bibitem{young05}%
  \BibitemOpen
  \bibfield{author}{%
  \bibinfo {author} {\bibfnamefont{T.}~\bibnamefont{Young}},\ }%
  \bibfield{title}{%
  \enquote{\bibinfo {title} {An essay on the cohesion of fluids},}\ }%
  \bibfield{journal}{%
  \bibinfo {journal} {Philosophical Transactions of the Royal Society
  (London)}\ }%
  \textbf{\bibinfo {volume} {95}},\ \bibinfo {pages} {65--87} (\bibinfo {year}
  {1805})%
\bibitem{sprittles11c}%
  \BibitemOpen
  \bibfield{author}{%
  \bibinfo {author} {\bibfnamefont{J.~E.}\ \bibnamefont{Sprittles}}\ and\
  \bibinfo {author} {\bibfnamefont{Y.~D.}\ \bibnamefont{Shikhmurzaev}},\ }%
  \bibfield{title}{%
  \enquote{\bibinfo {title} {A finite element framework for describing dynamic
  wetting phenomena},}\ }%
  \bibfield{journal}{%
  \bibinfo {journal} {International Journal for Numerical Methods in Fluids}\
  }%
  \textbf{\bibinfo {volume} {68}},\ \bibinfo {pages} {1257--1298} (\bibinfo
  {year} {2012})%
\bibitem{sprittles_pof}%
  \BibitemOpen
  \bibfield{author}{%
  \bibinfo {author} {\bibfnamefont{J.~E.}\ \bibnamefont{Sprittles}}\ and\
  \bibinfo {author} {\bibfnamefont{Y.~D.}\ \bibnamefont{Shikhmurzaev}},\ }%
  \bibfield{title}{%
  \enquote{\bibinfo {title} {The dynamics of liquid drops and their interaction
  with solids of varying wettabilities},}\ }%
  \bibfield{journal}{%
  \bibinfo {journal} {Physics of Fluids}\ }%
  \textbf{\bibinfo {volume} {24}},\ \bibinfo {pages} {082001} (\bibinfo {year}
  {2012})%
\bibitem{kistler83}%
  \BibitemOpen
  \bibfield{author}{%
  \bibinfo {author} {\bibfnamefont{S.~F.}\ \bibnamefont{Kistler}}\ and\
  \bibinfo {author} {\bibfnamefont{L.~E.}\ \bibnamefont{Scriven}},\ }%
  \enquote{\bibinfo {title} {{Coating flows}},}\ in\ \emph{\bibinfo {booktitle}
  {Computational Analysis of Polymer Processing}},\ \bibinfo {editor} {edited
  by\ \bibinfo {editor} {\bibfnamefont{J.~R.~A.}\ \bibnamefont{Pearson}}\ and\
  \bibinfo {editor} {\bibfnamefont{S.~M.}\ \bibnamefont{Richardson}}}\
  (\bibinfo {publisher} {Applied Science Publishers London and New York},\
  \bibinfo {year} {1983})\ pp.\ \bibinfo {pages}
  {243--299}%
\bibitem{heil04}%
  \BibitemOpen
  \bibfield{author}{%
  \bibinfo {author} {\bibfnamefont{M.}~\bibnamefont{Heil}},\ }%
  \bibfield{title}{%
  \enquote{\bibinfo {title} {An efficient solver for the fully-coupled solution
  of large displacement fluid-structure interaction problems},}\ }%
  \bibfield{journal}{%
  \bibinfo {journal} {Computer Methods in Applied Mechanics and Engineering}\
  }%
  \textbf{\bibinfo {volume} {193}},\ \bibinfo {pages} {1--23} (\bibinfo {year}
  {2004})%
\bibitem{wilson06}%
  \BibitemOpen
  \bibfield{author}{%
  \bibinfo {author} {\bibfnamefont{M.~C.~T.}\ \bibnamefont{Wilson}}, \bibinfo
  {author} {\bibfnamefont{J.~L.}\ \bibnamefont{Summers}}, \bibinfo {author}
  {\bibfnamefont{Y.~D.}\ \bibnamefont{Shikhmurzaev}}, \bibinfo {author}
  {\bibfnamefont{A.}~\bibnamefont{Clarke}},\ and\ \bibinfo {author}
  {\bibfnamefont{T.~D.}\ \bibnamefont{Blake}},\ }%
  \bibfield{title}{%
  \enquote{\bibinfo {title} {Nonlocal hydrodynamic influence on the dynamic
  contact angle: Slip models versus experiment},}\ }%
  \bibfield{journal}{%
  \bibinfo {journal} {Physical Review E}\ }%
  \textbf{\bibinfo {volume} {83}},\ \bibinfo {pages} {041606} (\bibinfo {year}
  {2006})%
\bibitem{lotstedt86}%
  \BibitemOpen
  \bibfield{author}{%
  \bibinfo {author} {\bibfnamefont{P.}~\bibnamefont{L\"{o}tstedt}}\ and\
  \bibinfo {author} {\bibfnamefont{L.}~\bibnamefont{Petzold}},\ }%
  \bibfield{title}{%
  \enquote{\bibinfo {title} {Numerical solution of nonlinear differential
  equations with an algebraic constraints 1: Convergence results for backward
  differentiation formulas},}\ }%
  \bibfield{journal}{%
  \bibinfo {journal} {Mathematics of Computation}\ }%
  \textbf{\bibinfo {volume} {46}},\ \bibinfo {pages} {491--516} (\bibinfo
  {year} {1986})%
\bibitem{gresho2}%
  \BibitemOpen
  \bibfield{author}{%
  \bibinfo {author} {\bibfnamefont{P.~M.}\ \bibnamefont{Gresho}}\ and\ \bibinfo
  {author} {\bibfnamefont{R.~L.}\ \bibnamefont{Sani}},\ }%
  \emph{\bibinfo {title} {Incompressible Flow and the Finite Element Method.
  Volume 2. Isothermal Laminar Flow}}\ (\bibinfo {publisher} {John Wiley \&
  Sons, LTD, New York},\ \bibinfo {year} {1999})%
\bibitem{basaran92}%
  \BibitemOpen
  \bibfield{author}{%
  \bibinfo {author} {\bibfnamefont{O.~A.}\ \bibnamefont{Basaran}},\ }%
  \bibfield{title}{%
  \enquote{\bibinfo {title} {Nonlinear oscillations of viscous liquid drops},}\
  }%
  \bibfield{journal}{%
  \bibinfo {journal} {Journal of Fluid Mechanics}\ }%
  \textbf{\bibinfo {volume} {241}},\ \bibinfo {pages} {169--198} (\bibinfo
  {year} {1992})%
\bibitem{Note2}%
  \BibitemOpen
  \bibinfo {note} {We can see this from the data provided to us by Dr
  J.D.~Paulsen, Dr J.C.~Burton and Professor S.R.~Nagel, which was published in
  \cite {paulsen11}.}\BibitemShut{Stop}%
\bibitem{Note3}%
  \BibitemOpen
  \bibinfo {note} {Despite the drops in \cite {thoroddsen05} having a different
  radius, $R=1.5$~mm, and very slightly different viscosity, $\mu =220$~mPa~s,
  our simulations show that at such high viscosity these alterations can be
  scaled out by an appropriate choice of viscous time-scale, as we
  use.}\BibitemShut{Stop}%
\bibitem{fordham48}%
  \BibitemOpen
  \bibfield{author}{%
  \bibinfo {author} {\bibfnamefont{S.}~\bibnamefont{Fordham}},\ }%
  \bibfield{title}{%
  \enquote{\bibinfo {title} {On the calculation of surface tension from
  measurements of pendant drops},}\ }%
  \bibfield{journal}{%
  \bibinfo {journal} {Proceedings of the Royal Society of London. Series A:
  Mathematical and Physical}\ }%
  \textbf{\bibinfo {volume} {194}},\ \bibinfo {pages} {1--16} (\bibinfo {year}
  {1948})%
\bibitem{Note4}%
  \BibitemOpen
  \bibinfo {note} {More precisely, it is the conventional model as at this
  stage the interface formation/disappearance dynamics have ended, so that the
  interfaces are in equilibrium and the interface formation model becomes
  equivalent to the conventional one.}\BibitemShut{Stop}%
\bibitem{jeong92}%
  \BibitemOpen
  \bibfield{author}{%
  \bibinfo {author} {\bibfnamefont{J.~{-T}.}\ \bibnamefont{Jeong}}\ and\
  \bibinfo {author} {\bibfnamefont{H.~K.}\ \bibnamefont{Moffatt}},\ }%
  \bibfield{title}{%
  \enquote{\bibinfo {title} {Free-surface cusps associated with flow at low
  {Reynolds} number},}\ }%
  \bibfield{journal}{%
  \bibinfo {journal} {Journal of Fluid Mechanics}\ }%
  \textbf{\bibinfo {volume} {241}},\ \bibinfo {pages} {1--22} (\bibinfo {year}
  {1992})%
\bibitem{attane07}%
  \BibitemOpen
  \bibfield{author}{%
  \bibinfo {author} {\bibfnamefont{P.}~\bibnamefont{Attan\'{e}}}, \bibinfo
  {author} {\bibfnamefont{F.}~\bibnamefont{Girard}},\ and\ \bibinfo {author}
  {\bibfnamefont{V.}~\bibnamefont{Morin}},\ }%
  \bibfield{title}{%
  \enquote{\bibinfo {title} {An energy balance approach of the dynamics of drop
  impact on a solid surface},}\ }%
  \bibfield{journal}{%
  \bibinfo {journal} {Physics of Fluids}\ }%
  \textbf{\bibinfo {volume} {19}},\ \bibinfo {pages} {012101} (\bibinfo {year}
  {2007})%
\bibitem{bayer06}%
  \BibitemOpen
  \bibfield{author}{%
  \bibinfo {author} {\bibfnamefont{I.~S.}\ \bibnamefont{Bayer}}\ and\ \bibinfo
  {author} {\bibfnamefont{C.~M.}\ \bibnamefont{Megaridis}},\ }%
  \bibfield{title}{%
  \enquote{\bibinfo {title} {Contact angle dynamics in droplets impacting on
  flat surfaces with different wetting characteristics},}\ }%
  \bibfield{journal}{%
  \bibinfo {journal} {Journal of Fluid Mechanics}\ }%
  \textbf{\bibinfo {volume} {558}},\ \bibinfo {pages} {415--449} (\bibinfo
  {year} {2006})%
\bibitem{blake99}%
  \BibitemOpen
  \bibfield{author}{%
  \bibinfo {author} {\bibfnamefont{T.~D.}\ \bibnamefont{Blake}}, \bibinfo
  {author} {\bibfnamefont{M.}~\bibnamefont{Bracke}},\ and\ \bibinfo {author}
  {\bibfnamefont{Y.~D.}\ \bibnamefont{Shikhmurzaev}},\ }%
  \bibfield{title}{%
  \enquote{\bibinfo {title} {Experimental evidence of nonlocal hydrodynamic
  influence on the dynamic contact angle},}\ }%
  \bibfield{journal}{%
  \bibinfo {journal} {Physics of Fluids}\ }%
  \textbf{\bibinfo {volume} {11}},\ \bibinfo {pages} {1995--2007} (\bibinfo
  {year} {1999})%
\bibitem{blake94}%
  \BibitemOpen
  \bibfield{author}{%
  \bibinfo {author} {\bibfnamefont{T.~D.}\ \bibnamefont{Blake}}, \bibinfo
  {author} {\bibfnamefont{A.}~\bibnamefont{Clarke}},\ and\ \bibinfo {author}
  {\bibfnamefont{K.~J.}\ \bibnamefont{Ruschak}},\ }%
  \bibfield{title}{%
  \enquote{\bibinfo {title} {Hydrodynamic assist of wetting},}\ }%
  \bibfield{journal}{%
  \bibinfo {journal} {AIChE Journal}\ }%
  \textbf{\bibinfo {volume} {40}},\ \bibinfo {pages} {229--242} (\bibinfo
  {year} {1994})%
\end{thebibliography}%

\end{document}